\documentclass[submission, Phys]{SciPost}

\usepackage{bbm,bm,amsthm,amsfonts,amssymb,amsopn,amsmath,mathtools}
\usepackage{pgf,tikz,ifthen}
\usetikzlibrary{arrows,calc,matrix,backgrounds,fit,positioning,decorations.pathmorphing,decorations.markings,patterns,decorations.pathreplacing,shapes.misc}

\renewcommand{\vec}[1]{{\boldsymbol{\mathbf{#1}}}}
\newcommand{\ave}[1]{{\langle #1\rangle}}
\newcommand{\vvec}[1]{{\underline{#1}}}
\def\bra#1{\mathinner{\langle{#1}|}}
\def\ket#1{\mathinner{|{#1}\rangle}}
\newcommand{\bbra}[1]{({#1}|}
\newcommand{\nbbra}[1]{(\!({#1}|}
\newcommand{\kket}[1]{|{#1})}
\newcommand{\nkket}[1]{|{#1})\!)}
\newcommand{\bbrakket}[2]{(#1 \vert #2 )}
\newcommand{\aave}[1]{{( #1)}}
\newcommand{\braket}[2]{\langle #1 \vert #2 \rangle}
\newcommand{\unit}{\mathbbm{1}}
\newcommand{\tr}{\mathrm{tr}}

\definecolor{full}{rgb}{0,0,0}
\definecolor{old}{rgb}{1,1,1}
\definecolor{halfborder}{rgb}{0.5,0.5,0.5}
\definecolor{half}{rgb}{0.9,0.9,0.9}
\definecolor{border}{rgb}{0.3,0.3,0.3}
\def\a{0.2}
\def\lc{0.3}
\def\ba{0.45}
\def\b{8}
\def\off{1.75}

\definecolor{colObs}{rgb}{1,0.39,0.28}
\definecolor{colMPA}{rgb}{0,0,0.55}
\definecolor{colvMPA}{rgb}{0.50,0.69,0.86}
\definecolor{colvMPAp}{rgb}{0.50,0.69,0.14}
\definecolor{colU}{rgb}{0.93,0.80,0.60}
\definecolor{colUt}{rgb}{0.93,0.80,0.60}
\definecolor{colP}{rgb}{0.69,0.50,0.86}
\definecolor{colPt}{rgb}{0.50,0.69,0.86}
\def\dx{0.15}
\def\dt{0.3}
\def\r{0.08}
\definecolor{colA}{rgb}{0.93,0.93,0.45}
\definecolor{colB}{rgb}{0.62,0.94,0.65}
\definecolor{colUtB}{rgb}{0.86,0.94,0.77}
\definecolor{colUtT}{rgb}{0.62,0.65,0.94}

\newcommand\redemptyrectangle[2]{
  \draw[red,thick] ({\a*(#1)},{\a*(#2-1)})  -- ({\a*(#1+1)},{\a*(#2)})  -- ({\a*(#1)},{\a*(#2+1)})  
  -- ({\a*(#1-1)},{\a*(#2)})  -- cycle;
}
\newcommand\redfullrectangle[2]{
  \draw[red,thick,fill=full] ({\a*(#1)},{\a*(#2-1)})  -- ({\a*(#1+1)},{\a*(#2)})  -- ({\a*(#1)},{\a*(#2+1)})  
  -- ({\a*(#1-1)},{\a*(#2)})  -- cycle;
}
\newcommand\redrectangle[3]{
  \ifthenelse{\equal{#3}{1}}{\redfullrectangle{#1}{#2}}{\redemptyrectangle{#1}{#2}};
}
\newcommand\emptyrectangle[2]{
  \draw[border] ({\a*(#1)},{\a*(#2-1)})  -- ({\a*(#1+1)},{\a*(#2)})  -- ({\a*(#1)},{\a*(#2+1)})  
  -- ({\a*(#1-1)},{\a*(#2)})  -- cycle;
}
\newcommand\halffullrectangle[2]{
\draw[opacity=0.] ({\a*(#1)},{\a*(#2+1)})  --
  ({\a*(#1+1)},{\a*(#2)})  -- ({\a*(#1-1)},{\a*(#2)})  -- cycle;
  \draw[opacity=0.,fill=full,fill opacity=1] ({\a*(#1)},{\a*(#2-1)})  --
  ({\a*(#1+1)},{\a*(#2)})  -- ({\a*(#1-1)},{\a*(#2)})  -- cycle;
  \draw[border] ({\a*(#1)},{\a*(#2-1)})  -- ({\a*(#1+1)},{\a*(#2)})  --
  ({\a*(#1)},{\a*(#2+1)})  -- ({\a*(#1-1)},{\a*(#2)})  -- cycle;
}
\newcommand\fullrectangle[2]{
  \draw[border,fill=full] ({\a*(#1)},{\a*(#2-1)})  -- ({\a*(#1+1)},{\a*(#2)})  -- ({\a*(#1)},{\a*(#2+1)})  
  -- ({\a*(#1-1)},{\a*(#2)})  -- cycle;
}
\newcommand\greyrectangle[2]{
  \draw[border,fill=halfborder] ({\a*(#1)},{\a*(#2-1)})  -- ({\a*(#1+1)},{\a*(#2)})  -- ({\a*(#1)},{\a*(#2+1)})  
  -- ({\a*(#1-1)},{\a*(#2)})  -- cycle;
}
\newcommand\rectangle[3]{
  \ifthenelse{\equal{#3}{1}}{\fullrectangle{#1}{#2}}{\ifthenelse{\equal{#3}{2}}{\greyrectangle{#1}{#2}}{\ifthenelse{\equal{#3}{3}}{\halffullrectangle{#1}{#2}}{\emptyrectangle{#1}{#2}}}};
}
\newcommand\bigdashedrectangle[2]{
  \draw[white,thick] ({\ba*(#1)},{\ba*(#2-1)})  -- ({\ba*(#1+1)},{\ba*(#2)})  -- ({\ba*(#1)},{\ba*(#2+1)})  
  -- ({\ba*(#1-1)},{\ba*(#2)})  -- cycle;
  \draw[border,thin,dashed] ({\ba*(#1)},{\ba*(#2-1)})  -- ({\ba*(#1+1)},{\ba*(#2)})  -- ({\ba*(#1)},{\ba*(#2+1)})  
  -- ({\ba*(#1-1)},{\ba*(#2)})  -- cycle;
}
\newcommand\bigtextrectangle[4]{
  \draw[border,fill=#4] ({\ba*(#1)},{\ba*(#2-1)})  -- ({\ba*(#1+1)},{\ba*(#2)})  -- ({\ba*(#1)},{\ba*(#2+1)})  
  -- ({\ba*(#1-1)},{\ba*(#2)})  -- cycle;
  \node at ({\ba*(#1)},{\ba*(#2)}) {\scalebox{0.7}{#3}}
}
\newcommand\textrectangle[3]{
  \draw[border,fill=half] ({\a*(#1)},{\a*(#2-1)})  -- ({\a*(#1+1)},{\a*(#2)})  -- ({\a*(#1)},{\a*(#2+1)})  
  -- ({\a*(#1-1)},{\a*(#2)})  -- cycle;
  \node at ({\a*(#1)},{\a*(#2)}) {\scalebox{0.7}{#3}}
}

\newcommand\configThreeLeft[3]{
  \rectangle{(0.5}{(-1)}{#1};
  \rectangle{(-0.5}{(0)}{#2};
  \rectangle{(0.5}{(1)}{#3};
}
\newcommand\configThreeRight[3]{
  \rectangle{(-0.5}{(-1)}{#1};
  \rectangle{(0.5}{(0)}{#2};
  \rectangle{(-0.5}{(1)}{#3};
}
\newcommand\configFourLeft[4]{
  \rectangle{(0.5}{(-1)}{#1};
  \rectangle{(-0.5}{(0)}{#2};
  \rectangle{(0.5}{(1)}{#3};
  \rectangle{(-0.5}{(2)}{#4};
}
\newcommand\configFourRight[4]{
  \rectangle{(-0.5}{(-1)}{#1};
  \rectangle{(0.5}{(0)}{#2};
  \rectangle{(-0.5}{(1)}{#3};
  \rectangle{(0.5}{(2)}{#4};
}
\newcommand\rightSaw[4]{
  \begin{tikzpicture}[baseline={([yshift=-0.5ex]current bounding box.center)}, scale=0.6] 
    \rectangle{0}{-0.5}{#1};
    \rectangle{1}{0.5}{#2};
    \rectangle{2}{-0.5}{#3};
    \rectangle{3}{0.5}{#4};
  \end{tikzpicture}%
}
\newcommand\leftSaw[4]{
  \begin{tikzpicture}[baseline={([yshift=-0.5ex]current bounding box.center)}, scale=0.6] 
    \rectangle{0}{0.5}{#1};
    \rectangle{1}{-0.5}{#2};
    \rectangle{2}{0.5}{#3};
    \rectangle{3}{-0.5}{#4};
  \end{tikzpicture}%
}
\newcommand\rightsSaw[2]{
  \begin{tikzpicture}[baseline={([yshift=-0.5ex]current bounding box.center)}, scale=0.6] 
    \rectangle{0}{-0.5}{#1};
    \rectangle{1}{0.5}{#2};
  \end{tikzpicture}%
}
\newcommand\leftsSaw[2]{
  \begin{tikzpicture}[baseline={([yshift=-0.5ex]current bounding box.center)}, scale=0.6] 
    \rectangle{0}{0.5}{#1};
    \rectangle{1}{-0.5}{#2};
  \end{tikzpicture}%
}
\newcommand\rcaTextRule[4]{
  \begin{tikzpicture}[baseline={([yshift=-0.1ex]current bounding box.center)},scale=1.6]
    \rectangle{0}{0}{#1};
    \rectangle{1}{-1}{#2};
    \rectangle{2}{0}{#3};
    \redrectangle{1}{1}{#4};
    \node[text=border] at ({\a*(0)},{\a*(0)}) {\scalebox{0.7}{$s_1$}};
    \node[text=border] at ({\a*(1)},{\a*(-1)}) {\scalebox{0.7}{$s_2$}};
    \node[text=border] at ({\a*(2)},{\a*(0)}) {\scalebox{0.7}{$s_3$}};
    \node[text=border] at ({\a*(1)},{\a*(1)}) {\scalebox{0.7}{$s_2^{\prime}$}};
  \end{tikzpicture}
}
\newcommand\rcaRule[4]{
  \begin{tikzpicture}[baseline={([yshift=-0.1ex]current bounding box.center)},scale=1.6]
    \rectangle{0}{0}{#1};
    \rectangle{1}{-1}{#2};
    \rectangle{2}{0}{#3};
    \redrectangle{1}{1}{#4};
  \end{tikzpicture}
}
\newcommand\threeConfigLeft[3]{
  \begin{tikzpicture}[baseline={([yshift=-0.1ex]current bounding box.center)},scale=1.6]
    \rectangle{1}{-1}{#1};
    \rectangle{0}{0}{#2};
    \rectangle{1}{1}{#3};
  \end{tikzpicture}
}
\newcommand\threeConfigRight[3]{
  \begin{tikzpicture}[baseline={([yshift=-0.1ex]current bounding box.center)},scale=1.6]
    \rectangle{1}{-1}{#1};
    \rectangle{2}{0}{#2};
    \rectangle{1}{1}{#3};
  \end{tikzpicture}
}
\newcommand\textRcaDualRule[8]{
  \begin{tikzpicture}[baseline={([yshift=-0.1ex]current bounding box.center)},scale=1.2]
    \rectangle{0}{0}{#1};
    \rectangle{-1}{1}{#2};
    \rectangle{0}{2}{#3};
    \rectangle{-1}{3}{#4};
    \rectangle{0}{4}{#5};
    \rectangle{-1}{5}{#6};
    \rectangle{0}{6}{#7};
    \redrectangle{1}{3}{#8};
    \node[text=border] at ({\a*(0)},{\a*(0)}) {\scalebox{0.6}{$s_1$}};
    \node[text=border] at ({\a*(-1)},{\a*(1)}) {\scalebox{0.6}{$s_2$}};
    \node[text=border] at ({\a*(0)},{\a*(2)}) {\scalebox{0.6}{$s_3$}};
    \node[text=border] at ({\a*(-1)},{\a*(3)}) {\scalebox{0.6}{$s_4$}};
    \node[text=border] at ({\a*(0)},{\a*(4)}) {\scalebox{0.6}{$s_5$}};
    \node[text=border] at ({\a*(-1)},{\a*(5)}) {\scalebox{0.6}{$s_6$}};
    \node[text=border] at ({\a*(0)},{\a*(6)}) {\scalebox{0.6}{$s_7$}};
    \node[text=border] at ({\a*(1)},{\a*(3)}) {\scalebox{0.6}{$s_4^{\prime}$}};
  \end{tikzpicture}
}
\newcommand\rcaDualShortRule[4]{
  \begin{tikzpicture}[baseline={([yshift=-0.1ex]current bounding box.center)},scale=1.6]
    \rectangle{0}{0}{#1};
    \rectangle{1}{-1}{#2};
    \rectangle{1}{1}{#4};
    \redrectangle{2}{0}{#3};
  \end{tikzpicture}
}
\newcommand\rcaVerticalThree[3]{
  \begin{tikzpicture}[baseline={([yshift=-0.1ex]current bounding box.center)},scale=1.6]
    \rectangle{0}{2}{#1};
    \rectangle{-1}{3}{#2};
    \rectangle{0}{4}{#3};
  \end{tikzpicture}
}
\newcommand\rcaVerticalFive[5]{
  \begin{tikzpicture}[baseline={([yshift=-0.1ex]current bounding box.center)},scale=1.6]
    \rectangle{0}{0}{#1};
    \rectangle{-1}{1}{#2};
    \rectangle{0}{2}{#3};
    \rectangle{-1}{3}{#4};
    \rectangle{0}{4}{#5};
  \end{tikzpicture}
}
\newcommand\rcaVerticalSeven[7]{
  \begin{tikzpicture}[baseline={([yshift=-0.1ex]current bounding box.center)},scale=1.6]
    \rectangle{0}{0}{#1};
    \rectangle{-1}{1}{#2};
    \rectangle{0}{2}{#3};
    \rectangle{-1}{3}{#4};
    \rectangle{0}{4}{#5};
    \rectangle{-1}{5}{#6};
    \rectangle{0}{6}{#7};
  \end{tikzpicture}
}
\newcommand\rcaDualRule[8]{
  \begin{tikzpicture}[baseline={([yshift=-0.1ex]current bounding box.center)},scale=1.2]
    \rectangle{0}{0}{#1};
    \rectangle{-1}{1}{#2};
    \rectangle{0}{2}{#3};
    \rectangle{-1}{3}{#4};
    \rectangle{0}{4}{#5};
    \rectangle{-1}{5}{#6};
    \rectangle{0}{6}{#7};
    \redrectangle{1}{3}{#8};
  \end{tikzpicture}
}
\newcommand\rcaVerticalSevenUpdatedTopMiddle[9]{
  \begin{tikzpicture}[baseline={([yshift=-0.1ex]current bounding box.center)},scale=1.6]
    \rectangle{0}{0}{#1};
    \rectangle{-1}{1}{#2};
    \rectangle{0}{2}{#3};
    \rectangle{-1}{3}{#4};
    \rectangle{0}{4}{#5};
    \rectangle{-1}{5}{#6};
    \rectangle{0}{6}{#7};
    \rectangle{1}{5}{#8};
    \rectangle{1}{3}{#9};
  \end{tikzpicture}
}
\newcommand\rcaVerticalSevenUpdatedTop[8]{
  \begin{tikzpicture}[baseline={([yshift=-0.1ex]current bounding box.center)},scale=1.6]
    \rectangle{0}{0}{#1};
    \rectangle{-1}{1}{#2};
    \rectangle{0}{2}{#3};
    \rectangle{-1}{3}{#4};
    \rectangle{0}{4}{#5};
    \rectangle{-1}{5}{#6};
    \rectangle{0}{6}{#7};
    \rectangle{1}{5}{#8};
  \end{tikzpicture}
}

\newcommand\rect[5]{
  \draw[thick,fill=#4] ({\dx*(4*#3-1.5-1)},{\dt*(-#1+0.5)}) rectangle ({\dx*(4*#3+1.5-1)},{\dt*(-#2-0.5)});
  \node at ({\dx*(4*#3-1)},{-\dt*(#1+#2)/2.}) {$#5$};
}

\newcommand\proj[4]{
  \draw [very thick] ({\dx*(2*#3)},{-\dt*#1}) -- ({\dx*(2*#3)},{-\dt*#2});
  \foreach \t in {#1,...,#2}{
    \draw [thick,fill=#4] ({\dx*(2*#3)},{-\dt*\t}) circle (1.25pt);
  }
}
\newcommand\prop[4]{
  \draw [very thick] ({\dx*(2*#3)},{-\dt*#1}) -- ({\dx*(2*#3)},{-\dt*#2});
  \draw [thick,fill=#4] ({\dx*(2*#3)},{-\dt*#1}) circle (1.25pt);
  \draw [thick,fill=#4] ({\dx*(2*#3)},{-\dt*#2}) circle (1.25pt);
  \draw [thick,fill=#4] ({\dx*(2*#3)},{-\dt*(#1+#2)/2.}) circle (3pt);
}

\newcommand\obs[4]{
  \draw [thick,rounded corners=1,fill=#4] ({\dx*(2*#2)},{-\dt*#1}) +({-0.5*(#3)},{-0.5*(#3)}) rectangle +({0.5*(#3)},{0.5*(#3)});
}

\newcommand\vbvec[2]{
  \draw [thick] ({\dx*(2*#2)},{-\dt*#1}) +({0.75*\r},0) -- +({-0.75*\r},0);
}

\newcommand\vA[3]{
  \draw [thick,rounded corners=1,fill=#3] ({\dx*(2*#2)},{-\dt*#1})
  +({0.75*\r},{0.75*\r}) -- +(0,0) -- +({0.75*\r},{-0.75*\r}) -- +(0,0) -- +({-0.75*\r},{-0.75*\r}) -- +(0,0)
  -- +({-0.75*\r},{0.75*\r}) -- + (0,0) -- cycle;
}

\newcommand\vB[3]{
  \draw [thick,rounded corners=1,fill=#3] ({\dx*(2*#2)},{-\dt*#1})
  +({0.75*\r},{0.75*\r}) rectangle +({-0.75*\r},{-0.75*\r});
}

\newcommand\vC[4]{
  \draw [thick,rounded corners=1,fill=#4] ({\dx*(2*#3)-0.75*\r},{-\dt*#2-0.75*\r})
  rectangle ({\dx*(2*#3)+0.75*\r},{-0.5*\dt*(#1+#2)});
  \draw [thick,rounded corners=1,fill=#4] ({\dx*(2*#3)-0.75*\r},{-0.5*\dt*(#1+#2)})
  rectangle ({\dx*(2*#3)+0.75*\r},{-\dt*#1+0.75*\r});
}

\newcommand\mbvec[2]{
  \draw [thick,double,line cap=round] ({\dx*(2*#2)},{-\dt*#1}) +({0.75*\r},0) -- +({-0.75*\r},0);
}

\newcommand\bvec[2]{
  \draw [thick,double,line cap=round] ({\dx*(2*#2)},{-\dt*#1}) +(0,{0.75*\r}) -- +(0,{-0.75*\r});
}

\newcommand\mV[3]{
  \draw [thick,rounded corners=1,fill=#3] ({\dx*(2*#2)},{-\dt*#1})
  +({0.75*\r},{-0.8660254037844386*\r}) -- +({0.75*\r},{0.8660254037844386*\r}) 
  -- +({-0.75*\r},0) -- cycle;
}

\newcommand\mW[3]{
  \draw [thick,rounded corners=1,fill=#3] ({\dx*(2*#2)},{-\dt*#1})
  +({-0.75*\r},{-0.8660254037844386*\r}) -- +({-0.75*\r},{0.8660254037844386*\r}) 
  -- +({0.75*\r},0) -- cycle;
}

\newcommand\V[3]{
  \draw [thick,rounded corners=1,fill=#3] ({\dx*(2*#2)},{-\dt*#1})
  +({-0.8660254037844386*\r},{0.75*\r}) -- +({0.8660254037844386*\r},{0.75*\r}) 
  -- +(0,{-0.75*\r}) -- cycle;
}

\newcommand\W[3]{
  \draw [thick,rounded corners=1,fill=#3] ({\dx*(2*#2)},{-\dt*#1})
  +({-0.8660254037844386*\r},{-0.75*\r}) -- +({0.8660254037844386*\r},{-0.75*\r}) 
  -- +(0,{0.75*\r}) -- cycle;
}

\newcommand\mS[3]{
  \draw [thick,rounded corners=0.25,fill=#3] ({\dx*(2*#2)},{-\dt*#1}) +({-0.5*\r},{-0.5*\r}) rectangle +({0.5*\r},{0.5*\r});
}

\newcommand\addgrid[2]{
  \foreach \t in {1,...,#1}{
    \draw[gray,very thin] ({0.5*\dx},{-\t*\dt}) -- ({(#2*2+1.5)*\dx},{-\t*\dt});
  }
}
\newcommand\grid[2]{
  \foreach \t in {1,...,#1}{
    \draw[gray,very thin] (\dx,{-\t*\dt}) -- ({(#2*2+1)*\dx},{-\t*\dt});
  }
}

\newcommand{\rev}[1]{\textcolor{purple}{#1}}

\begin{document}

\begin{center}{\Large \textbf{
Space-like dynamics in a reversible cellular automaton
}}\end{center}

\begin{center}
K. Klobas\textsuperscript{*},
T. Prosen
\end{center}

\begin{center}
Department of Physics, Faculty of Mathematics and Physics,
University of Ljubljana, Ljubljana, Slovenia
\\
* katja.klobas@fmf.uni-lj.si
\end{center}

\begin{center}
\today
\end{center}


\section*{Abstract}
{\bf
  In this paper we study the space evolution in the \emph{Rule 54 reversible
  cellular automaton}, which is a paradigmatic example of a deterministic
  interacting lattice gas. We show that the spatial translation of time
  configurations of the automaton is given in terms of local deterministic maps
  with the support that is small but bigger than that of the time evolution.
  The model is thus an example of space-time dual reversible cellular
  automaton, i.e.\ its dual is also (in general different) reversible cellular automaton.
  We provide two equivalent interpretations of the result; the first one relies on
  the dynamics of quasi-particles and follows from an exhaustive check of all
  the relevant time configurations, while the second one relies on purely
  algebraic considerations based on the circuit representation of the dynamics.
  Additionally, we use the properties of the local space evolution maps to
  provide an alternative derivation of the matrix product representation of
  multi-time correlation functions of local observables positioned at the same
  spatial coordinate.
}

\vspace{10pt}
\noindent\rule{\textwidth}{1pt}
\tableofcontents\thispagestyle{fancy}
\noindent\rule{\textwidth}{1pt}
\vspace{10pt}

\section{Introduction}
\label{sec:intro}
Studying exactly solvable models has been traditionally a fruitful approach
towards explaining the emergence of macroscopic phenomena from
microscopics~\cite{baxter2016exactly,sutherland2004beautiful}. In
recent years, the many facets of integrability and solvability have been
explored outside equilibrium physics \cite{JSTAT2016}; for example, by studying
long-time asymptotics of the initial value problem for a many-body interacting
system --- the so-called quenches \cite{Calabrese}, by developing generalized
hydrodynamic description of integrable systems \cite{Doyon,Bertini}, or by
analysing  random matrix models with intrinsic spatial locality structure ---
e.g. random local quantum circuits \cite{Nahum1,Nahum2,Pollmann,Chalker}.

However, exact solutions of dynamical many-body problems for individual
interacting systems are extremely scarce.  A particularly interesting class of
local quantum circuits that are exactly solvable in the statistical sense, yet
they are not Bethe-ansatz or Yang-Baxter integrable, are \emph{dual unitary}
quantum circuits~\cite{bertini2019exact}. These are local interacting models in
discrete space and discrete time where the roles of space and time can be
exchanged while keeping dynamics unitary (a similar space-time duality has been
explored in integrable field
theories~\cite{avan2016lagrangian,findlay2019dual,doikou2019time}).  This
property implies a nontrivial structure that enables exact computation of
numerous physical quantities, such as local correlation
functions~\cite{bertini2019exact,gutkin2020local}, entanglement
spreading~\cite{bertini2019entanglement,piroli2019exact,gopalakrishnan4}, operator
entanglement~\cite{bertini2019operator1,bertini2019operator2}, and
OTOCs~\cite{claeys2020maximum}. However, dual unitarity restricts the growth of
correlations to the maximal speed, which enforces strictly ballistic transport
of conserved charges if present~\cite{bertini2019exact,bertini2019operator1}.
Therefore these models cannot describe generic behaviour of systems with
sub-ballistic transport of conserved quantities.

This motivates us to study the effect of exchanging time and space evolution in
other $1+1$ dimensional models, where the strict dual unitarity condition does
not hold in hope of finding a generalized space-time duality allowing for
potentially richer macroscopic physical properties, such as diffusive or
super-diffusive transport.  This question can be rephrased in the context of
classical deterministic interacting lattice systems, where the property analogous
to unitarity is the symplectic feature of the dynamical evolution law. An example
of dual symplectic classical lattice dynamics with continuous local degrees
of freedom that exhibits super-diffusive
transport in the Kardar-Parisi-Zhang universality class has been recently
proposed~\cite{Ziga1,Ziga2}.

However, one can consider an even simpler class of interacting lattice systems
where the local field variable takes only a discrete set of values, the
so-called cellular automata. There, the feature analogous to symplecticity is the
reversibility of dynamics, meaning that the local dynamical map over a discrete
set of configurations is always one-to-one. A particularly interesting solvable
example of such models is the Rule 54 reversible cellular automaton (RCA54)
introduced by Bobenko et.\ al.\ in~\cite{bobenko1993two}~\footnote{
  The model should not be mistaken for the cellular automaton given
    by rule 54 according to Wolfram's classification~\cite{wolfram2002new} which is not reversible. Even though
    both systems have the same local update rule, the way the update is implemented
    completely changes the dynamics. In particular, such reversible cellular automata
    can be understood as a caricature of 2nd order differential equations (2nd Newton's law)
    rather than 1st order (rate equation). This point is made more explicit
    by Takesue's classification~\cite{Takesue} under which RCA54 can be interpreted
  as ERCA250R.}
and studied extensively in the last years, both in
classical~\cite{prosenMejiaMonasterioCA54,inoueTakesueCA54,prosenBucaCA54,bucaetalLargeDev,TMPA2018,vMPA2019}
and in quantum
setting~\cite{gopalakrishnan1,gopalakrishnan2,gopalakrishnan3,alba2019RCA54}.
In particular, dynamical structure factor of this model has been computed
exactly \cite{TMPA2018} and shown to exhibit diffusive transport.

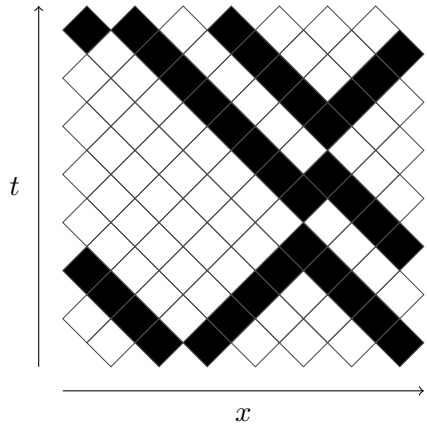
\begin{figure}
  \centering
  \begin{tikzpicture}[scale=1.6]
    \draw[->] ({-2*(\a)},{-2*(\a)}) -- ({-2*(\a)},{13*(\a)});
    \node at ({-3*\a},{5.5*(\a)}) {$t$};
    \node at ({6.5*\a},{-4*(\a)}) {$x$};
    \draw[->] ({-1*(\a)},{-3*(\a)}) -- ({14*(\a)},{-3*(\a)});
    \rectangle{0}{0}{0};
    \rectangle{1}{-1}{0};
    \rectangle{2}{0}{1};
    \rectangle{3}{-1}{1};
    \rectangle{4}{0}{0};
    \rectangle{5}{-1}{1};
    \rectangle{6}{0}{1};
    \rectangle{7}{-1}{0};
    \rectangle{8}{0}{0};
    \rectangle{9}{-1}{0};
    \rectangle{10}{0}{0};
    \rectangle{11}{-1}{0};
    \rectangle{12}{0}{1}
    \rectangle{13}{-1}{1}
    \rectangle{0}{2}{1};
    \rectangle{1}{1}{1};
    \rectangle{2}{2}{0};
    \rectangle{3}{1}{0};
    \rectangle{4}{2}{0};
    \rectangle{5}{1}{0};
    \rectangle{6}{2}{0};
    \rectangle{7}{1}{1};
    \rectangle{8}{2}{1};
    \rectangle{9}{1}{0};
    \rectangle{10}{2}{1};
    \rectangle{11}{1}{1};
    \rectangle{12}{2}{0};
    \rectangle{13}{1}{0};
    \rectangle{0}{4}{0};
    \rectangle{1}{3}{0};
    \rectangle{2}{4}{0};
    \rectangle{3}{3}{0};
    \rectangle{4}{4}{0};
    \rectangle{5}{3}{0};
    \rectangle{6}{4}{0};
    \rectangle{7}{3}{0};
    \rectangle{8}{4}{0};
    \rectangle{9}{3}{1};
    \rectangle{10}{4}{0};
    \rectangle{11}{3}{0};
    \rectangle{12}{4}{1};
    \rectangle{13}{3}{1};
    \rectangle{0}{6}{0};
    \rectangle{1}{5}{0};
    \rectangle{2}{6}{0};
    \rectangle{3}{5}{0};
    \rectangle{4}{6}{0};
    \rectangle{5}{5}{0};
    \rectangle{6}{6}{0};
    \rectangle{7}{5}{0};
    \rectangle{8}{6}{1};
    \rectangle{9}{5}{1};
    \rectangle{10}{6}{1};
    \rectangle{11}{5}{1};
    \rectangle{12}{6}{0};
    \rectangle{13}{5}{0};
    \rectangle{0}{8}{0};
    \rectangle{1}{7}{0};
    \rectangle{2}{8}{0};
    \rectangle{3}{7}{0};
    \rectangle{4}{8}{0};
    \rectangle{5}{7}{0};
    \rectangle{6}{8}{1};
    \rectangle{7}{7}{1};
    \rectangle{8}{8}{0};
    \rectangle{9}{7}{0};
    \rectangle{10}{8}{1};
    \rectangle{11}{7}{0};
    \rectangle{12}{8}{0};
    \rectangle{13}{7}{0};
    \rectangle{0}{10}{0};
    \rectangle{1}{9}{0};
    \rectangle{2}{10}{0};
    \rectangle{3}{9}{0};
    \rectangle{4}{10}{1};
    \rectangle{5}{9}{1};
    \rectangle{6}{10}{0};
    \rectangle{7}{9}{0};
    \rectangle{8}{10}{1};
    \rectangle{9}{9}{1};
    \rectangle{10}{10}{0};
    \rectangle{11}{9}{1};
    \rectangle{12}{10}{1};
    \rectangle{13}{9}{0};
    \rectangle{0}{12}{1};
    \rectangle{1}{11}{0};
    \rectangle{2}{12}{1};
    \rectangle{3}{11}{1};
    \rectangle{4}{12}{0};
    \rectangle{5}{11}{0};
    \rectangle{6}{12}{1};
    \rectangle{7}{11}{1};
    \rectangle{8}{12}{0};
    \rectangle{9}{11}{0};
    \rectangle{10}{12}{0};
    \rectangle{11}{11}{0};
    \rectangle{12}{12}{0};
    \rectangle{13}{11}{1};
  \end{tikzpicture}
  \caption{\label{fig:dynExample} Example of RCA54 time evolution. The
    configuration at the bottom is evolved upwards according to the time
    evolution rules~\eqref{eq:timePropRule}. Full sites (black rectangles) can
    be thought of as solitons that move with velocities $\pm 1$ and scattering
    displaces them one site backwards.  If the roles of space and time are
    exchanged, the dynamics can be still interpreted as solitons moving with
    velocities $\pm 1$, but when scattering their positions are moved one site
    \emph{forward} with respect to the original trajectories.
  }
\end{figure}

We argue that RCA54 is an excellent candidate for studying deterministic space evolution.
Indeed, a recent study revealed that probability distributions of time
configurations exhibit an efficient matrix-product description~\cite{vMPA2019},
suggesting that translating a given time configuration in space might be given
by a composition of deterministic maps with a finite support. Another indication
that a reversible space evolution formulation of RCA54 should be possible comes from the
quasi-particle interpretation of the dynamics; RCA54 rules describe solitons (kinks)
that move with fixed velocities and interact pairwise acquiring a delay for one
site after each scattering. Exchanging the roles of space and time results in
similar dynamics, the only difference is that the scattering now moves the
solitons one site forward with respect to the original trajectory. See
Figure~\ref{fig:dynExample} for a representative example.

In the paper we put this intuitive picture on formal grounds by expressing the space
evolution in terms of local deterministic maps, i.e. again as a reversible cellular automaton. 
In Section~\ref{sec:themodel} we define
the model and introduce the statistical states and observables.
In Section~\ref{sec:SpaceEvolution} we construct the local space evolution
rules. We demonstrate that the space evolution is indeed local and deterministic,
but it has to be defined on a reduced configuration space since not all of
the configurations can be realized in the time evolution. Therefore, the corresponding spatial cellular
automaton is configurationally constrained.
Furthermore, the map implementing the space evolution has a larger support than the local time
evolution map. In Section~\ref{sec:circuits} we provide an alternative
view of the problem by recasting the dynamics in terms of reversible logical circuits with three-site 
gates. This allows us to express the space evolution in an equivalent but simpler way.
In Section~\ref{sec:eqTS} we use the circuit representation to find an alternative
construction of \emph{time-states} (as introduced in~\cite{vMPA2019}) that does not
explicitly depend on the quasi-particle interpretation. Finally,
Section~\ref{sec:concl} contains some closing remarks.

\section{The model}\label{sec:themodel}
\subsection{Rule 54 dynamics}
The model is defined on a one-dimensional zig-zag lattice of even length $2n$ with each
site being either occupied or empty. A configuration at time~$t$
is given as a string of $2n$ binary digits,
$\vvec{s}^t=(\ldots,s_x^t,s_{x+1}^{t-1},s_{x+2}^{t},\ldots)$, where
the subscript denotes the position coordinate along the chain, superscript is the time
coordinate and the time and space coordinates have the same parity, $x+t\equiv 0\pmod{2}$.
Explicitly,
\begin{equation}\label{eq:configurations}
  \vvec{s}^{2 t}=(s_1^{2t-1},s_2^{2t},s_3^{2t-1},\ldots, s_{2n}^{2t}),\qquad
  \vvec{s}^{2 t+1}=(s_1^{2t+1},s_2^{2t},s_3^{2t+1},\ldots, s_{2n}^{2t}),
\end{equation}
where $s_x^t=1$ represents an occupied site and $s_x^t=0$ an empty one.
The time evolution is defined in discrete time and it is characterized by
a local three site update rule that changes the value of the middle bit (site)
depending on the configuration of the triple of neighbouring sites,
\begin{equation}\label{eq:timePropRule}
  s_2^{\prime}=\chi(s_1,s_2,s_3)= s_1+s_2+s_3+s_1 s_3 \pmod{2}.
\end{equation}
At every time step, the bits with the smaller time label are updated,
\begin{equation}\label{eq:timeProp}
  s_x^{t+1}=\chi(s_{x-1}^t,s_x^{t-1},s_{x+1}^t),
\end{equation}
where the {\em periodic boundaries} are assumed, $s_{2n+1}^t\equiv s_1^t$.
Geometrically, the time evolution can be imagined to update the bottom sites
of the zig-zag chain upwards while the previous top sites become the
bottom sites of the propagated chain corresponding to the new time step,
as schematically shown in Figure~\ref{fig:TEgeometry}.
\begin{figure}
  \begin{equation*}
    \begin{tikzpicture}[baseline={([yshift=0.5ex]current bounding box.center)}]
      \bigdashedrectangle{0}{2};
      \bigdashedrectangle{1}{1};
      \bigdashedrectangle{2}{2};
      \bigdashedrectangle{3}{1};
      \bigdashedrectangle{4}{2};
      \bigdashedrectangle{5}{1};
      \bigtextrectangle{0}{0}{$s_{x-3}^{t-1}$}{half};
      \bigtextrectangle{1}{-1}{$s_{x-2}^{t-2}$}{half};
      \bigtextrectangle{2}{0}{$s_{x-1}^{t-1}$}{half};
      \bigtextrectangle{3}{-1}{$s_{x}^{t-2}$}{half};
      \bigtextrectangle{4}{0}{$s_{x+1}^{t-1}$}{half};
      \bigtextrectangle{5}{-1}{$s_{x+2}^{t-2}$}{half};
      \node at ({\ba*(2.75)},{\ba*(-3.25)}) {$\vvec{s}^{t-1}$};
    \end{tikzpicture}\longrightarrow\ 
    \begin{tikzpicture}[baseline={([yshift=0.5ex]current bounding box.center)}]
      \bigdashedrectangle{1}{-1};
      \bigdashedrectangle{3}{-1};
      \bigdashedrectangle{5}{-1};
      \bigtextrectangle{0}{0}{$s_{x-3}^{t-1}$}{half};
      \bigtextrectangle{2}{0}{$s_{x-1}^{t-1}$}{half};
      \bigtextrectangle{4}{0}{$s_{x+1}^{t-1}$}{half};

      \bigdashedrectangle{0}{2};
      \bigdashedrectangle{2}{2};
      \bigdashedrectangle{4}{2};
      \bigtextrectangle{1}{1}{$s_{x-2}^{t}$}{half};
      \bigtextrectangle{3}{1}{$s_{x}^{t}$}{half};
      \bigtextrectangle{5}{1}{$s_{x+2}^{t}$}{half};

      \node at ({\ba*(2.75)},{\ba*(-3.25)}) {$\vvec{s}^{t}$};
    \end{tikzpicture}\longrightarrow\ 
    \begin{tikzpicture}[baseline={([yshift=0.5ex]current bounding box.center)}]
      \bigdashedrectangle{0}{0};
      \bigdashedrectangle{1}{-1};
      \bigdashedrectangle{2}{0};
      \bigdashedrectangle{3}{-1};
      \bigdashedrectangle{4}{0};
      \bigdashedrectangle{5}{-1};
      \bigtextrectangle{0}{2}{$s_{x-3}^{t+1}$}{half};
      \bigtextrectangle{1}{1}{$s_{x-2}^{t}$}{half};
      \bigtextrectangle{2}{2}{$s_{x-1}^{t+1}$}{half};
      \bigtextrectangle{3}{1}{$s_{x}^{t}$}{half};
      \bigtextrectangle{4}{2}{$s_{x+1}^{t+1}$}{half};
      \bigtextrectangle{5}{1}{$s_{x+2}^{t}$}{half};

      \node at ({\ba*(2.75)},{\ba*(-3.25)}) {$\vvec{s}^{t+1}$};
    \end{tikzpicture}
  \end{equation*}
  \caption{\label{fig:TEgeometry}Schematic representation of the lattice
    geometry and time evolution. At every time step, half of the sites
    are updated as expressed in~\eqref{eq:timeProp}.
    The new value depends on the values of the three consecutive sites
    as described in~\eqref{eq:timePropRule}.
  }
\end{figure}
Using this convention, the local time evolution rule~\eqref{eq:timePropRule}
can be represented graphically as 
\begin{equation}\label{eq:rca54Rules}
  \rcaTextRule{0}{0}{0}{0}\quad
  \rcaRule{0}{0}{1}{1}\quad
  \rcaRule{0}{1}{0}{1}\quad
  \rcaRule{0}{1}{1}{0}\quad
  \rcaRule{1}{0}{0}{1}\quad
  \rcaRule{1}{0}{1}{1}\quad
  \rcaRule{1}{1}{0}{0}\quad
  \rcaRule{1}{1}{1}{0}\enskip.
\end{equation}
This graphic representation of update rules immediately offers an alternative
interpretation of the dynamics. Black sites represent particles that move with
a constant velocity $1$ either to the left or right. When two oppositely moving
particles meet, they annihilate each other and reappear in the next time step
continuing with the same velocity. Or, alternatively speaking, they form a
virtual bound state which decays after one unit of time.  As a result, their
positions are shifted backwards by one site with respect to the original
trajectories.  This behaviour can be also observed by considering the example
of time evolution shown in Figure~\ref{fig:dynExample}. 

\subsection{Macroscopic states}\label{sec:MacroStates}
The statistical states of the system are probability distributions over the
set of configurations. They can be represented as  vectors from $\mathbb{R}^{2^{2n}}$
with an appropriate normalization,
\begin{equation}
  \vec{p}=\begin{bmatrix}p_{0}&p_{1}&\ldots&p_{2^{2n}-1}\end{bmatrix}^T,\qquad
  \sum_{s=0}^{2^{2n}-1} p_s=1,
\end{equation}
where the component $p_s \ge 0$ corresponds to the probability of the configuration
$(s_1,s_2,\ldots,s_{2n})$ given by the binary representation of $s$;
$s=\sum_{j=1}^{2n} 2^{2n-j}s_{j}$.\footnote{To simplify notation, we will interchangeably
use $p_{\underline{s}}$, $p_s$ or~$p_{s_1,s_2,\ldots,s_{2n}}$ to denote probability
of a configuration $(s_1,s_2,\ldots,s_{2n})$, depending on convenience.}
The time evolution of statistical states
is given in terms of a local three-site permutation operator $U$ that leaves the left
and right sites intact, while the middle site is changed according to
the update rule~\eqref{eq:timePropRule},
\begin{equation}\label{eq:defU}
  U=\begin{bmatrix}
    1& & & & & & & \\
    & & &1& & & & \\
    & &1& & & & & \\
    &1& & & & & & \\
    & & & & & &1& \\
    & & & & & & &1\\
    & & & &1& & & \\
    & & & & &1& & \\
  \end{bmatrix},\qquad
  U_{(s_1^{\prime},s_2^{\prime},s_3^{\prime}),(s_1,s_2,s_3)}=
  \delta_{s_1^{\prime},s_1}
  \delta_{s_2^{\prime},\chi(s_1,s_2,s_3)}
  \delta_{s_3^{\prime},s_3}.
\end{equation}
Due to the staggering, the full time evolution of states is given by
the alternation of even and odd time evolution operators $U^{\text{e/o}}$,
\begin{equation}\label{eq:TEStates}
  \vec{p}(t+1)=\begin{cases}
    U^{\text{e}}\vec{p}(t),\qquad&t\equiv 0\pmod{2},\\
    U^{\text{o}}\vec{p}(t),\qquad&t\equiv 1\pmod{2},
  \end{cases}
\end{equation}
where $U^{\text{e}}$ ($U^{\text{o}}$) are products of local operators
$U$ acting on even (odd) triples of sites,
\begin{equation}\label{eq:defTimeStep}
  U^{\text{e}}=\prod_{j=1}^{n} U_{2j},\qquad U^{\text{o}}=\prod_{j=1}^{n} U_{2j+1},
\end{equation}
with $U_k$ being the shorthand notation for the local operator $U$ 
that acts nontrivially on the sites $(k-1,k,k+1)$,
\begin{equation}
  U_k\equiv \unit^{\otimes k-2} \otimes U \otimes \unit^{\otimes 2n-k-1}.
\end{equation}

A distinguished set of statistical states are \emph{the stationary states}, which
are invariant under the time evolution. Due to the staggering, we require
these states to map into themselves after \emph{even} time steps and
therefore each stationary state is associated with two vectors, $\vec{p}$ and
$\vec{p}^{\prime}$, corresponding to even and odd time steps respectively,
\begin{equation}\label{eq:TinvCond}
  \vec{p}^{\prime}=U^{\text{e}}\vec{p},\qquad \vec{p}=U^{\text{o}}\vec{p}^\prime.
\end{equation}

We consider a simple class of $2$-parameter stationary states,
introduced in~\cite{prosenBucaCA54,bucaetalLargeDev}.
The state, denoted by $\vec{p}(\xi,\omega)$, exhibits an efficient matrix product
representation and is a simple example of a (\emph{generalized}) Gibbs state.
The two parameters $\xi,\omega$ are connected to the chemical potentials corresponding to the
densities of left and right moving solitons~\footnote{More concretely, $\log\xi$ and $\log\omega$
  are precisely the chemical potentials corresponding to the densities of left and right movers
respectively.}.
To express the state, we first define~$\vec{W}(\xi,\omega)$ and $\vec{W}^{\prime}(\xi,\omega)$
as vectors in the physical space,
\begin{equation}
  \vec{W}(\xi,\omega)=\begin{bmatrix}
    W_0(\xi,\omega)\\ W_1(\xi,\omega)
  \end{bmatrix},\qquad
  \vec{W}^{\prime}(\xi,\omega)=\begin{bmatrix}
    W^{\prime}_0(\xi,\omega)\\ W^{\prime}_1(\xi,\omega)
  \end{bmatrix},
\end{equation}
where the components~$W^{(\prime)}_s$ ($s=0,1$) are matrices, acting on a three-dimensional auxiliary space,
\begin{equation}
  W_0(\xi,\omega)=\begin{bmatrix}
    1&0&0\\
    \xi&0&0\\
    1&0&0
  \end{bmatrix}=W_0^{\prime}(\omega,\xi),\qquad
  W_1(\xi,\omega)=\begin{bmatrix}
    0&\xi&0\\
    0&0&1\\
    0&0&\omega
  \end{bmatrix}=W_1^{\prime}(\omega,\xi).
\end{equation}
Note that the pair of matrices~$W^{\prime}_s(\xi,\omega)$ is obtained
from~$W_s(\xi,\omega)$ by exchanging the roles of the parameters.
The stationary state~$\vec{p}(\xi,\omega)$ takes the following matrix product
form,
\begin{equation}\label{eq:eqStates}
  \vec{p}(\xi,\omega)=\frac{1}{Z_{2n}(\xi,\omega)} \tr\big(
    \vec{W}_1(\xi,\omega)\vec{W}^{\prime}_2(\xi,\omega)\vec{W}_3(\xi,\omega)\cdots
  \vec{W}_{2n}(\xi,\omega) \big),
\end{equation}
where the subscripts of bold vectors refer to the physical sites, and
$Z_{2n}(\xi,\omega)$ is the partition sum,
\begin{equation}
  Z_{2n}(\xi,\omega)=
  \tr\Big((W_0(\xi,\omega)+W_1(\xi,\omega))(W_0^{\prime}(\xi,\omega)
  +W_1^{\prime}(\xi,\omega))\Big)^n.
\end{equation}
To lighten the notation, when not ambiguous, we will suppress the
explicit dependence on the parameters.

The matrices~$W_s$, $W^{\prime}_s$ fulfill the following cubic algebraic
relation,
\begin{equation}\label{eq:cubicRel1PC}
  W_{s_1}W^{\prime}_{\chi(s_1,s_2,s_3)} W_{s_3} S
  = W_{s_1} S W_{s_2} W^{\prime}_{s_3},
  \qquad
  s_1,s_2,s_3=0,1,
\end{equation}
where we introduced the matrix $S$, 
\begin{equation}
  S=\begin{bmatrix}1&0&0\\0&0&1\\0&1&0\end{bmatrix},\qquad S^2=\unit.
\end{equation}
The relation~\eqref{eq:cubicRel1PC} can be compactly summarized as
\begin{equation}\label{eq:cubicRel1}
  U \vec{W}_1\vec{W}^{\prime}_2 \vec{W}_3 S
  = \vec{W}_1 S \vec{W}_2\vec{W}^{\prime}_3,
\end{equation}
where each of the $8$ physical components corresponds to one of the
combinations of~$(s_1,s_2,s_3)$ in~\eqref{eq:cubicRel1PC}. Defining the
odd-time version of the state as
\begin{equation}
  \vec{p}^{\prime}=\frac{1}{Z_{2n}}\tr\Big(\vec{W}^{\prime}_1\vec{W}_2
  \vec{W}^{\prime}_3\cdots \vec{W}_{2n}\Big),
\end{equation}
we can quickly see that time invariance condition~\eqref{eq:TinvCond} follows
by repeatedly applying the cubic relation~\eqref{eq:cubicRel1} and taking
into account that the local time evolution operator is its own
inverse~$U=U^{-1}$ (for details see\ \cite{prosenBucaCA54}).

\subsection{Local observables}
Observables are real valued functions over the set of configurations and form a
commutative algebra,
\begin{equation}
  \mathcal{A},\mathcal{B}: \mathbb{Z}_2^{2n} \to \mathbb{R},\qquad
  \left(\mathcal{A}\mathcal{B}\right)(\underline{s}) =
  \mathcal{A}(\underline{s})\mathcal{B}(\underline{s}).
\end{equation}
The space of observables can be thought of as a vector space that is dual to
the space of macroscopic states (probability vectors)~$\vec{p}$. It is
possible to define time evolution of observables via the following explicit
expression of expectation values,
\begin{equation}
  \ave{\mathcal{A}(t)}_{\vec{p}}=\sum_{\underline{s}^0}
  \mathcal{A}(\underline{s}^t) p_{\underline{s}^0}.
\end{equation}

In the paper, however, we will mostly deal with \emph{one-site observables}, which
only depend on the configuration at one site, $\mathcal{A}_x(\underline{s})=a(s_x)$,
where $a$ is as a real valued function from the one-site configuration
space, $a:\mathbb{Z}_2\to\mathbb{R}$.
Therefore, the expectation value of~$a$ at site $x$ and time $t$ takes the
following form,
\begin{equation}\label{eq:expectationValues1}
  \ave{a(x,t)}_{\vec{p}}=\sum_{\underline{s}^0} p_{\underline{s}^{0}}\, a(s_x^t).
\end{equation}
This expression can be also interpreted as an inner product by introducing a one-site
(unnormalized) maximum entropy vector~$\vec{\omega}$ and a diagonal matrix
representation of the observable,
\begin{equation}
  \vec{\omega} = \begin{bmatrix}1 & 1\end{bmatrix},\qquad
  \mathcal{O}_x(a) = \unit^{\otimes x-1} \otimes
  \begin{bmatrix} a(0) & 0 \\ 0 & a(1) \end{bmatrix}
  \otimes \unit^{\otimes 2n-x}.
\end{equation}
Using this notation, the expectation values~\eqref{eq:expectationValues1} 
can be expressed as
\begin{equation}\label{eq:expectationValues}
  \ave{a(x,t)}_{\vec{p}}
  = \vec{\omega}^{\otimes 2n} \mathcal{O}_x(a)\vec{p}(t).
\end{equation}

\subsection{Time states}\label{sec:timestates}
We proceed to define \emph{time configurations} as configurations of empty/full
sites observed at the same position $x$ and different times $t$, e.g.\
configurations of vertical zig-zag shaped chains from
Figure~\ref{fig:dynExample}.  Analogously to~\eqref{eq:configurations}, time
configurations $\vvec{s}_x$ are bit sequences
\begin{equation}
  (\ldots, s_x^{t-2}, s_{x-1}^{t-1}, s_x^t, s_{x-1}^{t+1},\ldots),
\end{equation}
where the space and time label
have the same parity, $x+t\equiv 0\pmod{2}$.  Explicitly,
\begin{equation}
  \vvec{s}_{2 x}=(s_{2x-1}^{1},s_{2x}^{2},s_{2x-1}^{3},\ldots, s_{2x}^{2m}),\qquad
  \vvec{s}_{2 x+1}=(s_{2x+1}^{1},s_{2x}^{2},s_{2x+1}^{3},\ldots,s_{2x}^{2m}).
\end{equation}
For simplicity we assume that the time label $t$ takes the values in a finite
range between $1$ and $2m$. In analogy with the statistical states, 
\emph{time states} are probability distributions over the space of time
configurations and can be represented as vectors from $\mathbb{R}^{2^{2m}}$,
\begin{equation}
  \vec{q}=\begin{bmatrix}
    q_0 & q_1 & q_2 & \ldots q_{2^{2m}-1}
  \end{bmatrix}^T,
  \qquad
  \sum_{s=0}^{2^{2m}-1} q_{s}=1,
\end{equation}
where each component $q_{s} \ge 0$ corresponds to the probability of the time configuration
given by the binary representation of $s$.

However, not every string of $2m$ binary digits represents a valid time
configuration.  In particular, the rules~\eqref{eq:rca54Rules}
imply that in a time configuration full sites always come in pairs,
while three consecutive full sites are forbidden. Therefore it makes sense to restrict the
discussion only to \emph{allowed} (also referred to as \emph{accessible}) time
configurations, where no substrings $(1,1,1)$ or $(0,1,0)$ appear. Accordingly,
the only nonzero components of time states should correspond to allowed
configurations.  This is equivalent to requiring the time states to be
invariant under the action of local projectors $P_k$,
\begin{equation}\label{eq:allowedTC}
  P_k \vec{q} = \vec{q},\qquad P_k\equiv \unit^{\otimes k-2}\otimes P\otimes \unit^{\otimes{2m-k-1}},
\end{equation}
where $P$ is the $3$ site projector to the allowed subspace of states,
\begin{equation}\label{eq:projector}
  P_{(s_1^{\prime},s_2^{\prime},s_3^{\prime}),(s_1,s_2,s_3)}=
  \delta_{s_1^{\prime},s_1}\delta_{s_2^{\prime},s_2}\delta_{s_3^{\prime},s_3}
  (1-\delta_{s_1,s_3}\delta_{s_2,1}).
\end{equation}

In analogy with stationary states one can define \emph{equilibrium
time-states}. These states correspond to probability distributions of time
configurations observed under the assumption of the system being in the equilibrium
state~$\vec{p}$ as introduced by Eq.~\eqref{eq:eqStates}. Explicitly, the
equilibrium time-states are uniquely determined by the expectation values of
\emph{multi-time correlation functions} of one-site observables at the same
position,~\footnote{Due to the geometry of the problem, the observables are
  technically positioned at one of the two neighbouring sites, depending on the
parity of the time-step.} $C_{a_1,a_2,\ldots,a_{2m}}(\vec{p})
=\lim_{n\to\infty}C^{(2n)}_{a_1,a_2,\ldots,a_{2m}}(\vec{p})$, defined as the
large system size limit of the following correlation function,
\begin{equation}
  \begin{aligned}
    C^{(2n)}_{a_1,a_2,a_3,\ldots,a_{2m}}(\vec{p})&=
    \ave{a_1(n^{\ast},0) a_2(n^{\ast}+1,1) a_3(n^{\ast},2)\cdots a_{2m}(n^{\ast}+1,2m-1)}_{\vec{p}},\\
    \textrm{where}\quad n^{\ast}&=2 \left\lfloor\frac{n}{2}\right\rfloor,
  \end{aligned}
\end{equation}
i.e.\ $n^{\ast}=n$ for even $n$ and $n^{\ast}=n-1$ for odd $n$.
Note that we choose~$n^{\ast}=n$ to denote an even site close to the middle
of the chain of length $2n$.
By definition~\eqref{eq:expectationValues}, the correlation
function can therefore be recast as the following inner product between the
vector~$\vec{\omega}^{\otimes 2n}$ and the equilibrium distribution
vector~$\vec{p}$ on which the appropriate sequential product of time evolution
operators~$U^{\text{e/o}}$ and matrix representations of local 
observables~$\mathcal{O}_{n^{\ast}/n^{\ast}+1}(a_j)$ is applied,
\begin{equation}\label{eq:multiTimeCorrelations}
  C^{(2n)}_{a_1,a_2,a_3,\ldots,a_{2m}}(\vec{p})=\vec{\omega}^{\otimes 2n}
  \,
  \mathcal{O}_{n^{\ast}+1}(a_{2m})
  U^{\text{e}}
  \cdots
  U^{\text{e}}
  \mathcal{O}_{n^{\ast}}(a_3)
  U^{\text{o}}
  \mathcal{O}_{n^{\ast}+1}(a_2)
  U^{\text{e}}
  \mathcal{O}_{n^{\ast}}(a_1) \, \vec{p}.
\end{equation}
The equilibrium time state~$\vec{q}$ is determined as the probability
distribution that uniquely fixes all the values of multi-time correlation
functions~$C_{a_1, a_2,\ldots a_{2m}}(\vec{p})$,
\begin{equation}
  C_{a_1,a_2,\ldots a_{2m}} =
  \smashoperator{\sum_{s_1,s_2,\ldots,s_{2m}}} q_{s_1 s_2\ldots s_{2m}}
  \prod_{j}^{2m} a_j(s_j) =
  \vec{\omega}^{\otimes 2m}
  \mathcal{O}_{2m}(a_{2m}) \cdots
  \mathcal{O}_{2}(a_{2})
  \mathcal{O}_{1}(a_{1})
  \vec{q},
\end{equation}
where the first equality follows from the definition and the second equality is
just the convenient vectorial representation. As was shown in
Ref.~\cite{vMPA2019}, these equilibrium time-states exhibit a simple matrix
product representation.

\section{Space evolution}\label{sec:SpaceEvolution}
\begin{figure}
  \begin{equation*}
    \begin{tikzpicture}[baseline={([yshift=0.5ex]current bounding box.center)}]
      \bigdashedrectangle{2}{0};
      \bigdashedrectangle{1}{1};
      \bigdashedrectangle{2}{2};
      \bigdashedrectangle{1}{3};
      \bigdashedrectangle{2}{4};
      \bigdashedrectangle{1}{5};
      \bigtextrectangle{0}{0}{$s_{x-1}^{t-3}$}{half};
      \bigtextrectangle{-1}{1}{$s_{x-2}^{t-2}$}{half};
      \bigtextrectangle{0}{2}{$s_{x-1}^{t-1}$}{half};
      \bigtextrectangle{-1}{3}{$s_{x-2}^{t}$}{half};
      \bigtextrectangle{0}{4}{$s_{x-1}^{t+1}$}{half};
      \bigtextrectangle{-1}{5}{$s_{x-2}^{t+2}$}{half};
      \node at ({\ba*(0.75)},{\ba*(-2)}) {$\vvec{s}_{x-1}$};
    \end{tikzpicture}\longrightarrow\ 
    \begin{tikzpicture}[baseline={([yshift=0.5ex]current bounding box.center)}]
      \bigdashedrectangle{-1}{1};
      \bigdashedrectangle{-1}{3};
      \bigdashedrectangle{-1}{5};
      \bigtextrectangle{0}{0}{$s_{x-1}^{t-3}$}{half};
      \bigtextrectangle{0}{2}{$s_{x-1}^{t-1}$}{half};
      \bigtextrectangle{0}{4}{$s_{x-1}^{t+1}$}{half};
      \bigdashedrectangle{2}{0};
      \bigdashedrectangle{2}{2};
      \bigdashedrectangle{2}{4};
      \bigtextrectangle{1}{1}{$s_{x}^{t-2}$}{half};
      \bigtextrectangle{1}{3}{$s_{x}^{t}$}{half};
      \bigtextrectangle{1}{5}{$s_{x}^{t+2}$}{half};
      \node at ({\ba*(0.75)},{\ba*(-2)}) {$\vvec{s}_{x\phantom{+1}}$};
    \end{tikzpicture}\longrightarrow\ 
    \begin{tikzpicture}[baseline={([yshift=0.5ex]current bounding box.center)}]
      \bigdashedrectangle{0}{0};
      \bigdashedrectangle{-1}{1};
      \bigdashedrectangle{0}{2};
      \bigdashedrectangle{-1}{3};
      \bigdashedrectangle{0}{4};
      \bigdashedrectangle{-1}{5};
      \bigtextrectangle{2}{0}{$s_{x+1}^{t-3}$}{half};
      \bigtextrectangle{1}{1}{$s_{x}^{t-2}$}{half};
      \bigtextrectangle{2}{2}{$s_{x+1}^{t-1}$}{half};
      \bigtextrectangle{1}{3}{$s_{x}^{t}$}{half};
      \bigtextrectangle{2}{4}{$s_{x+1}^{t+1}$}{half};
      \bigtextrectangle{1}{5}{$s_{x}^{t+2}$}{half};
      \node at ({\ba*(0.75)},{\ba*(-2)}) {$\vvec{s}_{x+1}$};
    \end{tikzpicture}
  \end{equation*}
  \caption{\label{fig:SEgeometry} Illustration of the geometry of space evolution.
    In analogy with the time evolution shown in Figure~\ref{fig:TEgeometry}, at
    every step the bits with the smaller space label deterministically
    change, while the others stay the same.
  }
\end{figure}
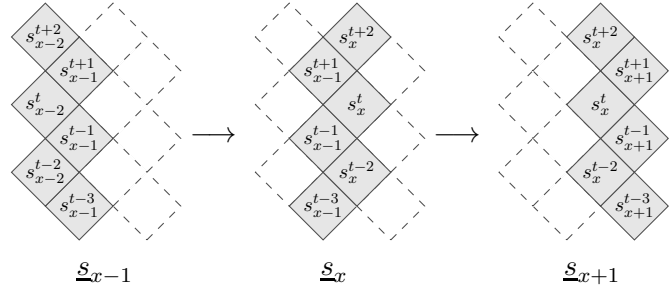
Our goal is to express the evolution of time configurations in the space direction
as schematically shown in Figure~\ref{fig:SEgeometry}. In general, there is no
guarantee that the space evolution can be expressed as a composition of local
deterministic maps. In our case, however, we expect this to be the case due to the
soliton description of the model: the dynamics in the space direction can be
understood as particles moving either upwards or downwards with velocity $1$.
When two oppositely moving particles meet, they get displaced one site forward
with respect to their original trajectories, mimicking repulsive interaction. 
This suggests the existence of a
deterministic local map,
\begin{equation}
  s_{x+1}^{t}=\phi(s_{x/x-1}^{t-r},\ldots,s_{x-1}^{t-2},s_x^{t-1},s_{x-1}^t,
  s_x^{t+1},s_{x-1}^{t+2},\ldots,s_{x/x-1}^{t+r}), 
  \label{eq:sm}
\end{equation}
where $r\in\mathbb{N}$ characterizes the support (of size $2r+1$) of the map.

The time evolution diagrams~\eqref{eq:rca54Rules} immediately imply that local space
propagators cannot be expressed in terms of maps with support $3$ (i.e.\ $r=1$).
Indeed, it is easy to see that the closest two neighbouring sites do not encode
enough information to deterministically propagate the state in space. In
particular, the last two pairs of diagrams have the same configurations of the
left three bits and different values of the right site. Therefore, the support
must be larger. Note that we additionally have to require that the local space
maps shifted by an even number of sites commute, i.e.\ the order in which we
apply (\ref{eq:sm}) on a given time configuration should not matter.

It is easy to see that the support $7$ (i.e.\ $r=3$) suffices to express the
deterministic space propagation rules. We start by observing that the first
four of the diagrams~\eqref{eq:rca54Rules} give the $3$-site deterministic
mapping also in the space direction,
\begin{equation}\label{eq:simpleDualRules}
  \rcaDualShortRule{0}{0}{0}{0}\quad
  \rcaDualShortRule{0}{0}{1}{1}\quad
  \rcaDualShortRule{0}{1}{0}{1}\quad
  \rcaDualShortRule{0}{1}{1}{0}.
\end{equation}
Now let us consider the subconfiguration $(0,1,1)$, which does not have
a unique $3$-site mapping and we add two neighbouring sites on the top. By avoiding the forbidden
subconfigurations, there are only two possibilities of how the configuration can
continue; either $(0,1,1,0,0)$ or $(0,1,1,0,1)$, which can be
explicitly visualised as
\begin{equation}
  \begin{tikzpicture}[baseline={([yshift=-0.5ex]current bounding box.center)},
    scale=1.2] 
    \rectangle{-2}{0}{0};
    \rectangle{-3}{1}{1};
    \rectangle{-2}{2}{1};

    \draw [thick,->] ({-0.5*\a},{1.5*\a}) -- ({1.5*\a},{3.5*\a});
    \draw [thick,->] ({-0.5*\a},{0.5*\a}) -- ({1.5*\a},{-1.5*\a});

    \rectangle{4.5}{-4.5}{0};
    \rectangle{3.5}{-3.5}{1};
    \rectangle{4.5}{-2.5}{1};
    \rectangle{3.5}{-1.5}{0};
    \rectangle{4.5}{-0.5}{1};

    \rectangle{4.5}{2.5}{0};
    \rectangle{3.5}{3.5}{1};
    \rectangle{4.5}{4.5}{1};
    \rectangle{3.5}{5.5}{0};
    \rectangle{4.5}{6.5}{0};
  \end{tikzpicture}.
\end{equation}
The top three sites in both configurations can be uniquely evolved by the 
$3$-site local maps~\eqref{eq:simpleDualRules}.
After applying these deterministic rules
we try to update the central bit to value $0$ or $1$,
while requiring that the updated configuration does not violate the
time-configuration restriction~\eqref{eq:allowedTC}. In both cases only one
configuration obeys the restriction,
\begin{equation}
  \begin{tikzpicture}[baseline={([yshift=-0.1ex]current bounding box.center)},
    scale=1.2]
    \rectangle{0}{0}{0};
    \rectangle{-1}{1}{1};
    \rectangle{0}{2}{1};
    \rectangle{-1}{3}{0};
    \rectangle{0}{4}{0};
    \draw [red,thick,rounded corners=1.5]
    (0,{5.5*\a}) -- ({1.5*\a},{4*\a}) -- ({0.5*\a},{3*\a})
    -- ({1.5*\a},{2*\a}) -- (0,{0.5*\a}) -- ({-2.5*\a},{3*\a}) -- cycle;
    \draw [thick,->] ({1.5*\a},{3*\a}) -- ({6.5*\a},{3*\a})
    node[midway,below] {\scriptsize{update}};
    \rectangle{9}{0}{0};
    \rectangle{8}{1}{1};
    \rectangle{9}{2}{1};
    \rectangle{8}{3}{0};
    \rectangle{9}{4}{0};
    \rectangle{10}{3}{1};
    \draw [red,thick,rounded corners=1.5]
    ({9*\a},{3.5*\a}) -- ({10.5*\a},{2*\a}) -- ({9.5*\a},{1*\a})
    -- ({10.5*\a},{0*\a}) -- ({9*\a},{-1.5*\a}) -- ({6.5*\a},{1*\a}) -- cycle;
    \draw [thick] ({10.5*\a},{1*\a}) -- ({15.5*\a},{1*\a})
    node[midway,above] {\scriptsize{\ \ update}};
    \draw [thick,->] ({15.5*\a},{\a}) -- ({17*\a},{2.5*\a});
    \draw [thick,->] ({15.5*\a},{\a}) -- ({17*\a},{-0.5*\a});
    \rectangle{18}{3}{0};
    \rectangle{17}{4}{1};
    \rectangle{18}{5}{1};
    \rectangle{17}{6}{0};
    \rectangle{18}{7}{0};
    \rectangle{19}{6}{1};
    \rectangle{19}{4}{0};

    \rectangle{18}{-5}{0};
    \rectangle{17}{-4}{1};
    \rectangle{18}{-3}{1};
    \rectangle{17}{-2}{0};
    \rectangle{18}{-1}{0};
    \rectangle{19}{-2}{1};
    \rectangle{19}{-4}{1};

    \draw[thick,red] ({20*\a},{-5*\a}) -- ({16*\a},{-1*\a});
    \draw[thick,red] ({20*\a},{-\a}) -- ({16*\a},{-5*\a});
  \end{tikzpicture},\qquad
  \begin{tikzpicture}[baseline={([yshift=-0.1ex]current bounding box.center)},
    scale=1.2]
    \rectangle{0}{0}{0};
    \rectangle{-1}{1}{1};
    \rectangle{0}{2}{1};
    \rectangle{-1}{3}{0};
    \rectangle{0}{4}{1};
    \draw [red,thick,rounded corners=1.5]
    (0,{5.5*\a}) -- ({1.5*\a},{4*\a}) -- ({0.5*\a},{3*\a})
    -- ({1.5*\a},{2*\a}) -- (0,{0.5*\a}) -- ({-2.5*\a},{3*\a}) -- cycle;
    \draw [thick,->] ({1.5*\a},{3*\a}) -- ({6.5*\a},{3*\a})
    node[midway,below] {\scriptsize{update}};
    \rectangle{9}{0}{0};
    \rectangle{8}{1}{1};
    \rectangle{9}{2}{1};
    \rectangle{8}{3}{0};
    \rectangle{9}{4}{1};
    \rectangle{10}{3}{0};
    \draw [red,thick,rounded corners=1.5]
    ({9*\a},{3.5*\a}) -- ({10.5*\a},{2*\a}) -- ({9.5*\a},{1*\a})
    -- ({10.5*\a},{0*\a}) -- ({9*\a},{-1.5*\a}) -- ({6.5*\a},{1*\a}) -- cycle;
    \draw [thick] ({10.5*\a},{1*\a}) -- ({15.5*\a},{1*\a})
    node[midway,above] {\scriptsize{\ \ update}};
    \draw [thick,->] ({15.5*\a},{\a}) -- ({17*\a},{2.5*\a});
    \draw [thick,->] ({15.5*\a},{\a}) -- ({17*\a},{-0.5*\a});
    \rectangle{18}{3}{0};
    \rectangle{17}{4}{1};
    \rectangle{18}{5}{1};
    \rectangle{17}{6}{0};
    \rectangle{18}{7}{1};
    \rectangle{19}{6}{0};
    \rectangle{19}{4}{0};
    \rectangle{18}{-5}{0};
    \rectangle{17}{-4}{1};
    \rectangle{18}{-3}{1};
    \rectangle{17}{-2}{0};
    \rectangle{18}{-1}{1};
    \rectangle{19}{-2}{0};
    \rectangle{19}{-4}{1};
    \draw[thick,red] ({20*\a},{3*\a}) -- ({16*\a},{7*\a});
    \draw[thick,red] ({20*\a},{7*\a}) -- ({16*\a},{3*\a});
  \end{tikzpicture},
\end{equation}
which provides a deterministic mapping corresponding to 5th and 6th
diagram of the time evolution rules~\eqref{eq:rca54Rules}. By
adding two undetermined sites to the bottom (denoted by grey squares),
the two (formally $7$-site) maps are graphically represented as
\begin{equation}\label{eq:rca54DualComplicated1}
  \rcaDualRule{2}{2}{0}{1}{1}{0}{0}{0}\quad
  \rcaDualRule{2}{2}{0}{1}{1}{0}{1}{1}\enskip.
\end{equation}
The last two rules, corresponding to 7th and 8th diagram of~\eqref{eq:rca54Rules},
are obtained by flipping~\eqref{eq:rca54DualComplicated1} upside down,
\begin{equation}\label{eq:rca54DualComplicated2}
  \rcaDualRule{0}{0}{1}{1}{0}{2}{2}{0}\quad
  \rcaDualRule{1}{0}{1}{1}{0}{2}{2}{1}\enskip.
\end{equation}
Combining the diagrams~\eqref{eq:rca54DualComplicated1} and~\eqref{eq:rca54DualComplicated2}
together with the simple 3-site update rules~\eqref{eq:simpleDualRules}
completes the construction of local deterministic space evolution maps.
They take the following explicit form,
\begin{equation}
  \phi(s_1,s_2,s_3,s_4,s_5,s_6,s_7)=
  \begin{cases}
    0;& s_3=s_4=s_5=0,\\
    1;& s_3=s_4=0,\, s_5=1,\\
    0;& s_3=0,\, s_4=s_5=1,\, s_7=0,\\
    1;& s_3=0,\, s_4=s_5=1,\, s_7=1,\\
    1;& s_3=1,\, s_4=s_5=0,\\
    0;& s_3=1,\, s_4=0,\, s_5=1,\\
    0;& s_1=0,\, s_3=s_4=1,\, s_5=0,\\
    1;& s_1=1,\, s_3=s_4=1,\, s_5=0.
  \end{cases}
\end{equation}
Since the update rules $s_4^{\prime}=\phi(s_1,s_2,s_3,s_4,s_5,s_6,s_7)$ do not depend
explicitly on the values of the sites $s_2$ and $s_6$, all the local
maps applied at the same step commute.

\section{Circuit representation}\label{sec:circuits}
\subsection{The dual picture}
Even though the local space evolution can be straightforwardly obtained by considering
local subconfigurations, it requires carefulness to check all the possible cases.
In this section we provide a more intuitive
\emph{circuit representation} of dynamics, which provides a simpler and more formal algebraic interpretation of local
space propagation rules.

The local $3$-site time evolution operator $U$ (cf.~\eqref{eq:defU})
acts on three consecutive sites and changes only the value of the middle site,
$U_{(s_1^{\prime},s_2^{\prime},s_3^{\prime}),(s_1,s_2,s_3)}=\delta_{s_1,s_1^{\prime}}
\delta_{\chi(s_1,s_2,s_3),s_2^{\prime}}\delta_{s_3,s_3^{\prime}}$,
which can be represented by the following graphical notation,
\begin{equation}
  \begin{tikzpicture}[baseline={([yshift=-0.5ex]current bounding box.center)},
    scale=2,rotate=90]
    \grid{3}{2}
    \prop{1}{3}{1.5}{colU}
    \node at ({6*\dx},{-\dt}) {$\scriptstyle{s_1^{\prime}}$};
    \node at ({6*\dx},{-2*\dt}) {$\scriptstyle{s_2^{\prime}}$};
    \node at ({6*\dx},{-3*\dt}) {$\scriptstyle{s_3^{\prime}}$};
    \node at (0,{-\dt}) {$\scriptstyle{s_1}$};
    \node at (0,{-2*\dt}) {$\scriptstyle{s_2}$};
    \node at (0,{-3*\dt}) {$\scriptstyle{s_3}$};
    \node at ({3*\dx},{-2*\dt}) {$\scriptstyle{U}$};
  \end{tikzpicture}.
\end{equation}
Thus, $U$ is a reversible single bit gate conditioned on the values of two
neighbouring bits. Using it, the full time evolution can be represented as a
grid with gates centered on sites $x+t\equiv 0 \pmod{2}$,
\begin{equation}\label{eq:timePropPicture}
  \begin{tikzpicture}[baseline={([yshift=-0.5ex]current bounding box.center)}]
    \begin{scope}[rotate=270,xscale=1.5,yscale=1.5]
      \node at ({16*\dx},{-9*\dt}) {${U^{\text{e}}}$};
      \node at ({14*\dx},{-9*\dt}) {${U^{\text{o}}}$};
      \node at ({12*\dx},{-9*\dt}) {${U^{\text{e}}}$};
      \node at ({10*\dx},{-9*\dt}) {${U^{\text{o}}}$};
      \node at ({8*\dx},{-9*\dt}) {${\vdots}$};
      \draw [thick,gray,->] ({18*\dx},{-9.75*\dt}) -- ({9*\dx},{-9.75*\dt}) node[midway,xshift=-5pt] {$t$};
      \draw [thick,gray,->] ({18*\dx},{-9.75*\dt}) -- ({18*\dx},{-5*\dt}) node[midway,yshift=-5pt] {$x$};
      \begin{scope}
        \clip (\dx,{-0.5*\dt}) rectangle ({17*\dx},{-8.5*\dt});
        \grid{8}{8}

        \prop{-1}{1}{1}{colU}
        \prop{1}{3}{1}{colU}
        \prop{3}{5}{1}{colU}
        \prop{5}{7}{1}{colU}
        \prop{7}{9}{1}{colU}

        \prop{0}{2}{2}{colU}
        \prop{2}{4}{2}{colU}
        \prop{4}{6}{2}{colU}
        \prop{6}{8}{2}{colU}
        \prop{8}{10}{2}{colU}

        \prop{-1}{1}{3}{colU}
        \prop{1}{3}{3}{colU}
        \prop{3}{5}{3}{colU}
        \prop{5}{7}{3}{colU}
        \prop{7}{9}{3}{colU}

        \prop{0}{2}{4}{colU}
        \prop{2}{4}{4}{colU}
        \prop{4}{6}{4}{colU}
        \prop{6}{8}{4}{colU}
        \prop{8}{10}{4}{colU}

        \prop{-1}{1}{5}{colU}
        \prop{1}{3}{5}{colU}
        \prop{3}{5}{5}{colU}
        \prop{5}{7}{5}{colU}
        \prop{7}{9}{5}{colU}

        \prop{0}{2}{6}{colU}
        \prop{2}{4}{6}{colU}
        \prop{4}{6}{6}{colU}
        \prop{6}{8}{6}{colU}
        \prop{8}{10}{6}{colU}

        \prop{-1}{1}{7}{colU}
        \prop{1}{3}{7}{colU}
        \prop{3}{5}{7}{colU}
        \prop{5}{7}{7}{colU}
        \prop{7}{9}{7}{colU}

        \prop{0}{2}{8}{colU}
        \prop{2}{4}{8}{colU}
        \prop{4}{6}{8}{colU}
        \prop{6}{8}{8}{colU}
        \prop{8}{10}{8}{colU}
      \end{scope}
    \end{scope}
  \end{tikzpicture}.
\end{equation}
For simplicity we assume periodic boundary conditions in space and time
directions. To obtain this circuit, we implicitly take
into account the commutativity of gates that share at most one site:
since the bit at the small circle does not change, neighbouring gates
can be "merged" together into a horizontal line consisting
of small and big circles. The symmetric form of~\eqref{eq:timePropPicture}
suggests that the dynamics can be mapped to a $2$-dimensional vertex model,
as described in Appendix~\ref{sec:rotatingDiagram}.
Therefore, the roles of space and time can be formally exchanged by
introducing the following local dual evolution operator~$\hat{U}$,
\begin{equation}\label{eq:defUt}
  \begin{tikzpicture}[baseline={([yshift=-0.5ex]current bounding box.center)},scale=2]
    \grid{3}{2}
    \prop{1}{3}{1.5}{colUt}
    \node at ({6*\dx},{-\dt}) {$\scriptstyle{s_3^{\prime}}$};
    \node at ({6*\dx},{-2*\dt}) {$\scriptstyle{s_2^{\prime}}$};
    \node at ({6*\dx},{-3*\dt}) {$\scriptstyle{s_1^{\prime}}$};
    \node at (0,{-\dt}) {$\scriptstyle{s_3}$};
    \node at (0,{-2*\dt}) {$\scriptstyle{s_2}$};
    \node at (0,{-3*\dt}) {$\scriptstyle{s_1}$};
    \node at ({3*\dx},{-2*\dt}) {$\scriptstyle{\hat{U}}$};
  \end{tikzpicture}\qquad
  \hat{U}_{(s_1^{\prime},s_2^{\prime},s_3^{\prime})\,(s_1,s_2,s_3)}=
  \delta_{s_1^{\prime},s_1}
  \delta_{s_3^{\prime},s_3}
  U_{(s_2,s_1,s_2^{\prime}),(s_2,s_3,s_2^{\prime})}.
\end{equation}
Then the picture~\eqref{eq:timePropPicture} can be replaced by the
following
\begin{equation}\label{eq:spacePropPicture}
  \begin{tikzpicture}[baseline={([yshift=-0.5ex]current bounding box.center)}]
    \begin{scope}[scale=1.5]
      \node at ({2*\dx},{0.25*\dt}) {${\ \hat{U}^{\text{e}}}$};
      \node at ({4*\dx},{0.25*\dt}) {${\ \hat{U}^{\text{o}}}$};
      \node at ({6*\dx},{0.25*\dt}) {${\ \hat{U}^{\text{e}}}$};
      \node at ({8*\dx},{0.25*\dt}) {${\ \hat{U}^{\text{o}}}$};
      \node at ({10*\dx},0) {${\cdots}$};
      \begin{scope}
        \clip (\dx,{-0.5*\dt}) rectangle ({17*\dx},{-8.5*\dt});
        \grid{8}{8}

        \prop{0}{2}{1}{colUt}
        \prop{2}{4}{1}{colUt}
        \prop{4}{6}{1}{colUt}
        \prop{6}{8}{1}{colUt}
        \prop{8}{10}{1}{colUt}

        \prop{-1}{1}{2}{colUt}
        \prop{1}{3}{2}{colUt}
        \prop{3}{5}{2}{colUt}
        \prop{5}{7}{2}{colUt}
        \prop{7}{9}{2}{colUt}

        \prop{0}{2}{3}{colUt}
        \prop{2}{4}{3}{colUt}
        \prop{4}{6}{3}{colUt}
        \prop{6}{8}{3}{colUt}
        \prop{8}{10}{3}{colUt}

        \prop{-1}{1}{4}{colUt}
        \prop{1}{3}{4}{colUt}
        \prop{3}{5}{4}{colUt}
        \prop{5}{7}{4}{colUt}
        \prop{7}{9}{4}{colUt}

        \prop{0}{2}{5}{colUt}
        \prop{2}{4}{5}{colUt}
        \prop{4}{6}{5}{colUt}
        \prop{6}{8}{5}{colUt}
        \prop{8}{10}{5}{colUt}

        \prop{-1}{1}{6}{colUt}
        \prop{1}{3}{6}{colUt}
        \prop{3}{5}{6}{colUt}
        \prop{5}{7}{6}{colUt}
        \prop{7}{9}{6}{colUt}

        \prop{0}{2}{7}{colUt}
        \prop{2}{4}{7}{colUt}
        \prop{4}{6}{7}{colUt}
        \prop{6}{8}{7}{colUt}
        \prop{8}{10}{7}{colUt}

        \prop{-1}{1}{8}{colUt}
        \prop{1}{3}{8}{colUt}
        \prop{3}{5}{8}{colUt}
        \prop{5}{7}{8}{colUt}
        \prop{7}{9}{8}{colUt}
      \end{scope}
    \end{scope}
  \end{tikzpicture},
\end{equation}
where $\hat{U}^{\text{e/o}}$ are the products of local operators acting on
even/odd triplets of sites,
\begin{equation}
  \hat{U}^{\text{e}}=\prod_t \hat{U}_{2t},\qquad
  \hat{U}^{\text{o}}=\prod_t \hat{U}_{2t+1}.
\end{equation}
Here, $\hat{U}_t$ denotes the local operator $\hat{U}$ acting nontrivially
on the triplet of sites $(t-1,t,t+1)$,
\begin{equation}
  \hat{U}_t\equiv \unit^{\otimes t-2} \otimes \hat{U} \otimes \unit^{\otimes 2m-t-1}.
\end{equation}

The local operators acting on all odd (or all even) triples commute, but they 
are clearly not deterministic,
\begin{equation}
  \hat{U}=\begin{bmatrix}
    1 &   &   &   &   &   &   &   \\
    & 0 &   & 1 &   &   &   &   \\
    &   & 0 &   &   &   &   &   \\
    & 1 &   & 1 &   &   &   &   \\
    &   &   &   & 0 &   & 1 &   \\
    &   &   &   &   & 1 &   &   \\
    &   &   &   & 1 &   & 1 &   \\
    &   &   &   &   &   &   & 0 \\
  \end{bmatrix}.
\end{equation}
Nonetheless, as we show in the remainder of this section, projecting to the
reduced space of allowed time-states (see the discussion
in~\ref{sec:timestates}), space propagation can be expressed as a product of
local deterministic gates with bigger support.

\subsection{Projected dual propagators}
We start by noting that the local dual operator~$\hat{U}$ projects to the
subspace of allowed $3$-site configurations and commutes with the projector
introduced in~\eqref{eq:projector},
\begin{equation}\label{eq:localProjection}
  \hat{U}=P\hat{U}=\hat{U}P.
\end{equation}
Therefore, defining $P^{\text{e/o}}$ as the products of projectors on appropriate
triples of sites,
\begin{equation}
  P^{\text{e}}=\prod_t P_{2t},\qquad
  P^{\text{o}}=\prod_t P_{2t+1},
\end{equation}
a relation similar to~\eqref{eq:localProjection} holds for
the even/odd dual propagators~$\hat{U}^{\text{e/o}}$,
\begin{equation}
  \hat{U}^{\text{e}}=P^{\text{e}}\hat{U}^{\text{e}}=\hat{U}^{\text{e}}P^{\text{e}},\qquad
  \hat{U}^{\text{o}}=P^{\text{o}}\hat{U}^{\text{o}}=\hat{U}^{\text{o}}P^{\text{o}}.\qquad
\end{equation}
This allows us to represent the space evolution on the restricted space in
terms of \emph{projected dual operators} $\tilde{U}^{\text{e/o}}$,
\begin{equation}\label{eq:commProj}
  \tilde{U}^{\text{e}}=P^{\text{o}} \hat{U}^{\text{e}} P^{\text{o}},\qquad
  \tilde{U}^{\text{o}}=P^{\text{e}} \hat{U}^{\text{o}} P^{\text{e}}.
\end{equation}
Explicitly, by projecting to the space spanned by allowed time configurations
at the beginning and at the end, the space evolution for $2m$ sites can be
equivalently expressed in terms of the projected space propagators as
\begin{equation}
  P^{\text{e}}P^{\text{o}}
  \underbrace{\hat{U}^{\text{e}}\cdots \hat{U}^{\text{o}}\hat{U}^{\text{e}}}_{2m}
  P^{\text{e}}P^{\text{o}}
  = \underbrace{\tilde{U}^{\text{e}}\tilde{U}^{\text{o}}\cdots \tilde{U}^{\text{e}}}_{2m},
\end{equation}
which follows directly from the definition of
$\tilde{U}^{\text{e/o}}$~\eqref{eq:commProj} and the commutativity of local
projectors centered at different sites. This equality can be visualised
graphically, by first introducing the following representation for the
projector $P$,
\begin{equation}\label{eq:defP}
  \begin{tikzpicture}[baseline={([yshift=-0.6ex]current bounding box.center)},scale=1.25]
    \grid{3}{2}
    \proj{1}{3}{1.5}{colP}
    \node at ({6*\dx},{-\dt}) {$\scriptstyle{s_1^{\prime}}$};
    \node at ({6*\dx},{-2*\dt}) {$\scriptstyle{s_2^{\prime}}$};
    \node at ({6*\dx},{-3*\dt}) {$\scriptstyle{s_3^{\prime}}$};
    \node at (0,{-\dt}) {$\scriptstyle{s_1}$};
    \node at (0,{-2*\dt}) {$\scriptstyle{s_2}$};
    \node at (0,{-3*\dt}) {$\scriptstyle{s_3}$};
    \node at ({3.75*\dx},{-2.375*\dt}) {$\scriptstyle{P}$};
  \end{tikzpicture},
\end{equation}
and transforming the diagram~\eqref{eq:spacePropPicture} as
\begin{equation}\label{eq:spacePropDiagram2}
  \begin{tikzpicture}[baseline={([yshift=-2.1ex]current bounding box.center)}]
    \clip (\dx,{\dt}) rectangle ({7.4*\dx},{-15.5*\dt});
    \node[label={[scale=0.7]center: $\ \hat{U}^{\text{o}}$}] at ({2*\dx},{0.25*\dt}) {};
    \node[label={[scale=0.7]center: $\ \hat{U}^{\text{e}}$}] at ({4*\dx},{0.25*\dt}) {};
    \node[label={[scale=0.7]center: $\ \hat{U}^{\text{o}}$}] at ({6*\dx},{0.25*\dt}) {};

    \begin{scope}
      \clip (\dx,{-0.5*\dt}) rectangle ({7*\dx},{-15.5*\dt});
      \grid{15}{3}{colUt}

      \prop{2}{4}{1}{colUt}
      \prop{10}{12}{1}{colUt}
      \prop{6}{8}{1}{colUt}
      \prop{14}{16}{1}{colUt}

      \prop{0}{2}{1}{colUt}
      \prop{8}{10}{1}{colUt}
      \prop{4}{6}{1}{colUt}
      \prop{12}{14}{1}{colUt}

      \prop{5}{7}{2}{colUt}
      \prop{13}{15}{2}{colUt}
      \prop{1}{3}{2}{colUt}
      \prop{9}{11}{2}{colUt}

      \prop{-1}{1}{2}{colUt}
      \prop{7}{9}{2}{colUt}
      \prop{15}{17}{2}{colUt}
      \prop{3}{5}{2}{colUt}
      \prop{11}{13}{2}{colUt}

      \prop{2}{4}{3}{colUt}
      \prop{10}{12}{3}{colUt}
      \prop{6}{8}{3}{colUt}
      \prop{14}{16}{3}{colUt}

      \prop{0}{2}{3}{colUt}
      \prop{8}{10}{3}{colUt}
      \prop{4}{6}{3}{colUt}
      \prop{12}{14}{3}{colUt}
    \end{scope}
  \end{tikzpicture}\!\rightarrow\!
  \begin{tikzpicture}[baseline={([yshift=-2.1ex]current bounding box.center)}]
    \clip (\dx,{\dt}) rectangle ({17*\dx},{-15.5*\dt});
    \node at ({3.5*\dx},{0.25*\dt}) {$\scriptstyle{\,\hat{U}^{\text{o}}}$};
    \node at ({9.5*\dx},{0.25*\dt}) {$\scriptstyle{\,\hat{U}^{\text{e}}}$};
    \node at ({15.5*\dx},{0.25*\dt}) {$\scriptstyle{\,\hat{U}^{\text{o}}}$};
    \begin{scope}
      \clip (\dx,{-0.5*\dt}) rectangle ({17*\dx},{-15.5*\dt});
      \grid{15}{8}
      \prop{2}{4}{1}{colUt}
      \prop{10}{12}{1}{colUt}
      \prop{6}{8}{1}{colUt}
      \prop{14}{16}{1}{colUt}

      \prop{0}{2}{2}{colUt}
      \prop{8}{10}{2}{colUt}
      \prop{4}{6}{2}{colUt}
      \prop{12}{14}{2}{colUt}

      \prop{2}{4}{7}{colUt}
      \prop{10}{12}{7}{colUt}
      \prop{6}{8}{7}{colUt}
      \prop{14}{16}{7}{colUt}

      \prop{0}{2}{8}{colUt}
      \prop{8}{10}{8}{colUt}
      \prop{4}{6}{8}{colUt}
      \prop{12}{14}{8}{colUt}

      \prop{5}{7}{4}{colUt}
      \prop{13}{15}{4}{colUt}
      \prop{1}{3}{4}{colUt}
      \prop{9}{11}{4}{colUt}

      \prop{-1}{1}{5}{colUt}
      \prop{7}{9}{5}{colUt}
      \prop{15}{17}{5}{colUt}
      \prop{3}{5}{5}{colUt}
      \prop{11}{13}{5}{colUt}
    \end{scope}
  \end{tikzpicture}\!\rightarrow\!
  \begin{tikzpicture}[baseline={([yshift=-2.1ex]current bounding box.center)}]
    \clip (\dx,{\dt}) rectangle ({33*\dx},{-15.5*\dt});
    \node at ({3.5*\dx},{0.25*\dt}) {$\scriptstyle{\,\hat{U}^{\text{o}}}$};
    \node at ({7.5*\dx},{0.25*\dt}) {$\scriptstyle{\,P^{\text{o}}}$};
    \node at ({13.5*\dx},{0.25*\dt}) {$\scriptstyle{\,P^{\text{e}}}$};
    \node at ({17.5*\dx},{0.25*\dt}) {$\scriptstyle{\,\hat{U}^{\text{e}}}$};
    \node at ({21.5*\dx},{0.25*\dt}) {$\scriptstyle{\,P^{\text{e}}}$};
    \node at ({27.5*\dx},{0.25*\dt}) {$\scriptstyle{\,P^{\text{o}}}$};
    \node at ({31.5*\dx},{0.25*\dt}) {$\scriptstyle{\,\hat{U}^{\text{o}}}$};
    \begin{scope}
      \clip (\dx,{-0.5*\dt}) rectangle ({33*\dx},{-15.5*\dt});
      \grid{15}{16}
      \prop{2}{4}{1}{colUt}
      \prop{10}{12}{1}{colUt}
      \prop{6}{8}{1}{colUt}
      \prop{14}{16}{1}{colUt}

      \prop{0}{2}{2}{colUt}
      \prop{8}{10}{2}{colUt}
      \prop{4}{6}{2}{colUt}
      \prop{12}{14}{2}{colUt}

      \proj{5}{7}{6}{colP}
      \proj{13}{15}{6}{colP}
      \proj{1}{3}{6}{colP}
      \proj{9}{11}{6}{colP}

      \proj{-1}{1}{7}{colP}
      \proj{7}{9}{7}{colP}
      \proj{15}{17}{7}{colP}
      \proj{3}{5}{7}{colP}
      \proj{11}{13}{7}{colP}

      \proj{2}{4}{3}{colP}
      \proj{10}{12}{3}{colP}
      \proj{6}{8}{3}{colP}
      \proj{14}{16}{3}{colP}

      \proj{0}{2}{4}{colP}
      \proj{8}{10}{4}{colP}
      \proj{4}{6}{4}{colP}
      \proj{12}{14}{4}{colP}

      \prop{2}{4}{15}{colUt}
      \prop{10}{12}{15}{colUt}
      \prop{6}{8}{15}{colUt}
      \prop{14}{16}{15}{colUt}

      \prop{0}{2}{16}{colUt}
      \prop{8}{10}{16}{colUt}
      \prop{4}{6}{16}{colUt}
      \prop{12}{14}{16}{colUt}

      \proj{2}{4}{13}{colP}
      \proj{10}{12}{13}{colP}
      \proj{6}{8}{13}{colP}
      \proj{14}{16}{13}{colP}

      \proj{0}{2}{14}{colP}
      \proj{8}{10}{14}{colP}
      \proj{4}{6}{14}{colP}
      \proj{12}{14}{14}{colP}

      \proj{5}{7}{10}{colP}
      \proj{13}{15}{10}{colP}
      \proj{1}{3}{10}{colP}
      \proj{9}{11}{10}{colP}

      \proj{-1}{1}{11}{colP}
      \proj{7}{9}{11}{colP}
      \proj{15}{17}{11}{colP}
      \proj{3}{5}{11}{colP}
      \proj{11}{13}{11}{colP}

      \prop{5}{7}{8}{colUt}
      \prop{13}{15}{8}{colUt}
      \prop{1}{3}{8}{colUt}
      \prop{9}{11}{8}{colUt}

      \prop{-1}{1}{9}{colUt}
      \prop{7}{9}{9}{colUt}
      \prop{15}{17}{9}{colUt}
      \prop{3}{5}{9}{colUt}
      \prop{11}{13}{9}{colUt}
    \end{scope}
  \end{tikzpicture}\!\rightarrow\!
  \begin{tikzpicture}[baseline={([yshift=-2.1ex]current bounding box.center)}]
    \clip (\dx,{\dt}) rectangle ({33*\dx},{-15.5*\dt});
    \draw [decorate,decoration={brace,amplitude=3pt}]({11*\dx},{-0.5*\dt}) -- ({23*\dx},{-0.5*\dt}) node[midway,xshift=3pt,yshift=8pt] {$\scriptstyle{\tilde{U}^{\text{e}}}$};
    \begin{scope}
      \clip (\dx,{-0.5*\dt}) rectangle ({33*\dx},{-15.5*\dt});
      \grid{15}{16}
      \proj{5}{7}{3}{colP}
      \proj{13}{15}{3}{colP}
      \proj{1}{3}{3}{colP}
      \proj{9}{11}{3}{colP}

      \proj{-1}{1}{4}{colP}
      \proj{7}{9}{4}{colP}
      \proj{15}{17}{4}{colP}
      \proj{3}{5}{4}{colP}
      \proj{11}{13}{4}{colP}

      \prop{2}{4}{1}{colUt}
      \prop{10}{12}{1}{colUt}
      \prop{6}{8}{1}{colUt}
      \prop{14}{16}{1}{colUt}

      \prop{0}{2}{2}{colUt}
      \prop{8}{10}{2}{colUt}
      \prop{4}{6}{2}{colUt}
      \prop{12}{14}{2}{colUt}

      \proj{5}{7}{13}{colP}
      \proj{13}{15}{13}{colP}
      \proj{1}{3}{13}{colP}
      \proj{9}{11}{13}{colP}

      \proj{-1}{1}{14}{colP}
      \proj{7}{9}{14}{colP}
      \proj{15}{17}{14}{colP}
      \proj{3}{5}{14}{colP}
      \proj{11}{13}{14}{colP}

      \prop{2}{4}{15}{colUt}
      \prop{10}{12}{15}{colUt}
      \prop{6}{8}{15}{colUt}
      \prop{14}{16}{15}{colUt}

      \prop{0}{2}{16}{colUt}
      \prop{8}{10}{16}{colUt}
      \prop{4}{6}{16}{colUt}
      \prop{12}{14}{16}{colUt}

      \proj{2}{4}{6}{colP}
      \proj{10}{12}{6}{colP}
      \proj{6}{8}{6}{colP}
      \proj{14}{16}{6}{colP}

      \proj{0}{2}{7}{colP}
      \proj{8}{10}{7}{colP}
      \proj{4}{6}{7}{colP}
      \proj{12}{14}{7}{colP}

      \prop{5}{7}{8}{colUt}
      \prop{13}{15}{8}{colUt}
      \prop{1}{3}{8}{colUt}
      \prop{9}{11}{8}{colUt}

      \prop{-1}{1}{9}{colUt}
      \prop{7}{9}{9}{colUt}
      \prop{15}{17}{9}{colUt}
      \prop{3}{5}{9}{colUt}
      \prop{11}{13}{9}{colUt}

      \proj{2}{4}{10}{colP}
      \proj{10}{12}{10}{colP}
      \proj{6}{8}{10}{colP}
      \proj{14}{16}{10}{colP}

      \proj{0}{2}{11}{colP}
      \proj{8}{10}{11}{colP}
      \proj{4}{6}{11}{colP}
      \proj{12}{14}{11}{colP}
    \end{scope}
  \end{tikzpicture},
\end{equation}
where we took into account the fact that the only combinations of noncommuting
gates are the ones with the big circle of one gate sitting on the same line as
the small circle of another one. Explicitly, in this case the following $3$ pairs
\emph{do not} commute,
\begin{equation}
  \begin{tikzpicture}[baseline={([yshift=-0.6ex]current bounding box.center)}]
    \grid{4}{2}
    \prop{1}{3}{1}{colUt}
    \prop{2}{4}{2}{colUt}
  \end{tikzpicture}
  \neq
  \begin{tikzpicture}[baseline={([yshift=-0.6ex]current bounding box.center)}]
    \grid{4}{2}
    \prop{1}{3}{2}{colUt}
    \prop{2}{4}{1}{colUt}
  \end{tikzpicture},\qquad
  \begin{tikzpicture}[baseline={([yshift=-0.6ex]current bounding box.center)}]
    \grid{4}{2}
    \prop{1}{3}{1}{colUt}
    \proj{2}{4}{2}{colP}
  \end{tikzpicture}
  \neq
  \begin{tikzpicture}[baseline={([yshift=-0.6ex]current bounding box.center)}]
    \grid{4}{2}
    \prop{1}{3}{2}{colUt}
    \proj{2}{4}{1}{colP}
  \end{tikzpicture},\qquad
  \begin{tikzpicture}[baseline={([yshift=-0.6ex]current bounding box.center)}]
    \grid{4}{2}
    \prop{2}{4}{1}{colUt}
    \proj{1}{3}{2}{colP}
  \end{tikzpicture}
  \neq
  \begin{tikzpicture}[baseline={([yshift=-0.6ex]current bounding box.center)}]
    \grid{4}{2}
    \prop{2}{4}{2}{colUt}
    \proj{1}{3}{1}{colP}
  \end{tikzpicture}.
\end{equation}

\subsection{Deterministic local 7-site gates}
The one-step space evolution operators~$\tilde{U}^{\text{e/o}}$ can be written
as products of local gates with support $7$ that are deterministic on the
reduced configuration subspace, i.e.\ it is possible to suitably define local
propagators~$\tilde{V}$ and~$\tilde{W}$ so that $\tilde{U}^{\text{o/e}}$ take
the following form,
\begin{equation}\label{eq:detSE}
  \begin{aligned}
    \tilde{U}^{\text{e}}&=
    \Big(\prod_{t}\tilde{W}_{8t+10}\Big)
    \Big(\prod_{t}\tilde{W}_{8t+6}\Big)
    \Big(\prod_{t}\tilde{V}_{8t+8}\Big)
    \Big(\prod_{t}\tilde{V}_{8t+4}\Big),\\
    \tilde{U}^{\text{o}}&=
    \Big(\prod_{t}\tilde{W}_{8t+11}\Big)
    \Big(\prod_{t}\tilde{W}_{8t+7}\Big)
    \Big(\prod_{t}\tilde{V}_{8t+9}\Big)
    \Big(\prod_{t}\tilde{V}_{8t+5}\Big),
  \end{aligned}
\end{equation}
where the subscript denotes the middle site of the subchain on which the local
evolution operators acts, i.e.\ $\tilde{W}_t$ acts nontrivially on the sites $t-3$,
$t-2$, $t-1$, $t$, $t+1$, $t+2$ and $t+3$. Graphically, this is represented by
the following diagram,
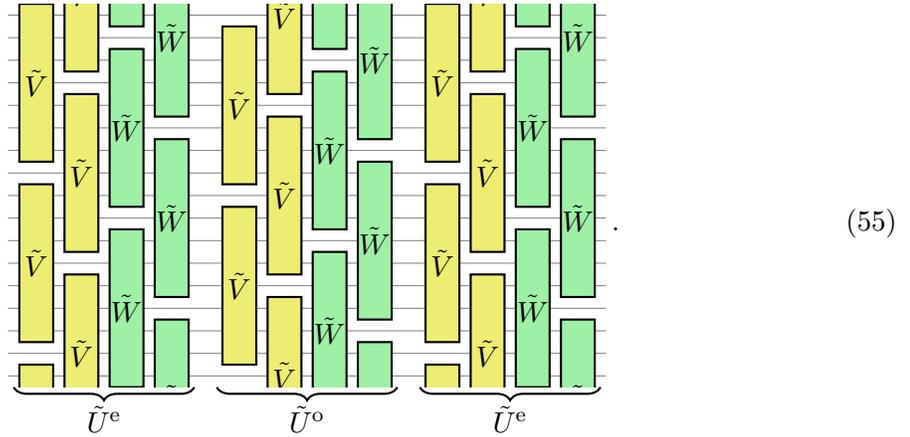
\begin{equation}\label{eq:7sitePropGraph}
  \begin{tikzpicture}[baseline={([yshift=-0.6ex]current bounding box.center)}]
    \draw [thick,decorate,decoration={brace,amplitude=5pt,mirror}]
    (\dx,{-17.5*\dt}) -- ({17*\dx},{-17.5*\dt}) node[midway,yshift=-12pt] {$\tilde{U}^{\text{e}}$};
    \draw [thick,decorate,decoration={brace,amplitude=5pt,mirror}]
    ({19*\dx},{-17.5*\dt}) -- ({35*\dx},{-17.5*\dt}) node[midway,yshift=-12pt] {$\tilde{U}^{\text{o}}$};
    \draw [thick,decorate,decoration={brace,amplitude=5pt,mirror}]
    ({37*\dx},{-17.5*\dt}) -- ({53*\dx},{-17.5*\dt}) node[midway,yshift=-12pt] {$\tilde{U}^{\text{e}}$};

    \clip (0,{-0.5*\dt}) rectangle ({54*\dx},{-17.5*\dt});
    \addgrid{17}{26}

    \rect{1}{7}{1}{colA}{\tilde{V}}
    \rect{9}{15}{1}{colA}{\tilde{V}}
    \rect{17}{23}{1}{colA}{\tilde{V}}

    \rect{-3}{3}{2}{colA}{\tilde{V}}
    \rect{5}{11}{2}{colA}{\tilde{V}}
    \rect{13}{19}{2}{colA}{\tilde{V}}

    \rect{-5}{1}{3}{colB}{\tilde{W}}
    \rect{3}{9}{3}{colB}{\tilde{W}}
    \rect{11}{17}{3}{colB}{\tilde{W}}

    \rect{-1}{5}{4}{colB}{\tilde{W}}
    \rect{7}{13}{4}{colB}{\tilde{W}}
    \rect{15}{21}{4}{colB}{\tilde{W}}

    \rect{2}{8}{5.5}{colA}{\tilde{V}}
    \rect{10}{16}{5.5}{colA}{\tilde{V}}

    \rect{-4}{2}{7.5}{colB}{\tilde{W}}
    \rect{4}{10}{7.5}{colB}{\tilde{W}}
    \rect{12}{18}{7.5}{colB}{\tilde{W}}

    \rect{-2}{4}{6.5}{colA}{\tilde{V}}
    \rect{6}{12}{6.5}{colA}{\tilde{V}}
    \rect{14}{20}{6.5}{colA}{\tilde{V}}

    \rect{0}{6}{8.5}{colB}{\tilde{W}}
    \rect{8}{14}{8.5}{colB}{\tilde{W}}
    \rect{16}{22}{8.5}{colB}{\tilde{W}}

    \rect{1}{7}{10}{colA}{\tilde{V}}
    \rect{9}{15}{10}{colA}{\tilde{V}}
    \rect{17}{23}{10}{colA}{\tilde{V}}

    \rect{-5}{1}{12}{colB}{\tilde{W}}
    \rect{3}{9}{12}{colB}{\tilde{W}}
    \rect{11}{17}{12}{colB}{\tilde{W}}

    \rect{-3}{3}{11}{colA}{\tilde{V}}
    \rect{5}{11}{11}{colA}{\tilde{V}}
    \rect{13}{19}{11}{colA}{\tilde{V}}

    \rect{-1}{5}{13}{colB}{\tilde{W}}
    \rect{7}{13}{13}{colB}{\tilde{W}}
    \rect{15}{21}{13}{colB}{\tilde{W}}
  \end{tikzpicture}.
\end{equation}
The operators $\tilde{V}$, $\tilde{W}$ can be explicitly written in terms of
the dual $3$-site operators $\hat{U}$ by introducing the following~$5$-site
projector $Q$,
\begin{equation}
  \begin{tikzpicture}[baseline={([yshift=-0.6ex]current bounding box.center)},scale=1.5]
    \grid{5}{2}
    \proj{1}{5}{1.5}{colPt}
    \node at ({6*\dx},{-\dt}) {$\scriptstyle{s_5^{\prime}}$};
    \node at ({6*\dx},{-2*\dt}) {$\scriptstyle{s_4^{\prime}}$};
    \node at ({6*\dx},{-3*\dt}) {$\scriptstyle{s_3^{\prime}}$};
    \node at ({6*\dx},{-4*\dt}) {$\scriptstyle{s_2^{\prime}}$};
    \node at ({6*\dx},{-5*\dt}) {$\scriptstyle{s_1^{\prime}}$};
    \node at (0,{-\dt}) {$\scriptstyle{s_5}$};
    \node at (0,{-2*\dt}) {$\scriptstyle{s_4}$};
    \node at (0,{-3*\dt}) {$\scriptstyle{s_3}$};
    \node at (0,{-4*\dt}) {$\scriptstyle{s_2}$};
    \node at (0,{-5*\dt}) {$\scriptstyle{s_1}$};
    \node at ({3.75*\dx},{-3.375*\dt}) {$\scriptstyle{Q}$};
  \end{tikzpicture},\quad
  \begin{aligned}
    Q_{(s_1^{\prime},s_2^{\prime},s_3^{\prime},s_4^{\prime},s_5^{\prime}),(s_1,s_2,s_3,s_4,s_5)}&=
    \delta_{s_1,s_1^{\prime}}
    \delta_{s_2,s_2^{\prime}}
    \delta_{s_3,s_3^{\prime}}
    \delta_{s_4,s_4^{\prime}}
    \delta_{s_5,s_5^{\prime}}\\
    &\mkern-36mu
    \cdot \big(1-\delta_{s_2,0}\delta_{s_3,1}\delta_{s_1+s_4,1}\big)
    \big(1-\delta_{s_4,0}\delta_{s_3,1}\delta_{s_2+s_5,1}\big).
  \end{aligned}
\end{equation}
Then the $7$-site gates can be expressed as a~$3$-site projected
operator~$\hat{U}$, sandwiched between two $P$ projectors on one and two $Q$
projectors on the other side,
\begin{equation}\label{eq:defDetGates}
  \begin{tikzpicture}[baseline={([yshift=-0.6ex]current bounding box.center)}]
    \addgrid{7}{2}
    \rect{1}{7}{1}{colA}{\tilde{V}}
  \end{tikzpicture}\equiv
  \begin{tikzpicture}[baseline={([yshift=-0.6ex]current bounding box.center)}]
    \grid{7}{5}
    \proj{2}{4}{1}{colP}
    \proj{4}{6}{2}{colP}
    \prop{3}{5}{3}{colUt}
    \proj{1}{5}{4}{colPt}
    \proj{3}{7}{5}{colPt}
  \end{tikzpicture}, \qquad
  \begin{tikzpicture}[baseline={([yshift=-0.6ex]current bounding box.center)}]
    \addgrid{7}{2}
    \rect{1}{7}{1}{colB}{\tilde{W}}
  \end{tikzpicture}\equiv
  \begin{tikzpicture}[baseline={([yshift=-0.6ex]current bounding box.center)}]
    \grid{7}{5}
    \proj{1}{5}{1}{colPt}
    \proj{3}{7}{2}{colPt}
    \prop{3}{5}{3}{colUt}
    \proj{2}{4}{4}{colP}
    \proj{4}{6}{5}{colP}
  \end{tikzpicture},
\end{equation}
or equivalently
\begin{equation}
  \tilde{V}_{t}=Q_{t+1}Q_{t-1}\tilde{U}_{t}P_{t+1}P_{t-1},\qquad
  \tilde{W}_{t}=P_{t+1}P_{t-1}\tilde{U}_{t}Q_{t+1}Q_{t-1}.
\end{equation}
These gates are deterministic on the restricted space of allowed
time-configurations, since the following holds,
\begin{equation}
  \tilde{V}_{t}\tilde{V}_{t}^T=
  \tilde{W}^T_{t}\tilde{W}_{t}=Q_{t-1}P_{t}Q_{t+1},\qquad
  \tilde{V}^T_{t}\tilde{V}_{t}=
  \tilde{W}_{t}\tilde{W}_{t}^T=P_{t-1}P_{t}P_{t+1},
\end{equation}
where right-hand-sides are diagonal projection matrices with matrix elements
that can only be $0$ or $1$. Therefore, to see that the space evolution is
local and deterministic, we only have to show that the
diagrams~\eqref{eq:7sitePropGraph} and~\eqref{eq:spacePropDiagram2} are
equivalent. The proof is provided in Appendix~\ref{sec:equivalence}.

\section{Equilibrium time states}\label{sec:eqTS}
Similar ideas can be employed to find equilibrium time-states, i.e.\ the
probability distributions of time-configurations under the assumption of the
underlying system being in equilibrium (or stationary state). This provides an alternative derivation
of the results of Ref.~\cite{vMPA2019} that does not explicitly rely on the
quasi-particle interpretation of dynamics. To simplify the discussion we
first consider the infinite temperature state in~\ref{subsec:MEstateTS}, where
we show how to use the properties of the local propagator~\eqref{eq:defU} and
its dual~\eqref{eq:defUt} to represent the time-state as a layer of observables
squeezed between two vertically oriented MPSs with Schmidt rank~$2$.
In~\ref{subsec:GstateTS} we generalize this to the class of equilibrium states
introduced in section~\ref{sec:MacroStates}.
In~\ref{subsec:MPSrepresentationTS} we then reformulate the result in terms of a
single MPS with Schmidt rank~$4$ and thus reproduce the main results
of~\cite{vMPA2019}.

\subsection{Maximum entropy state}\label{subsec:MEstateTS}
In the case of maximum entropy (or infinite temperature)
stationary state~$\vec{p}_{\infty}=2^{-2n}(\vec{\omega}^T)^{\otimes 2n}$, the vectorial form
of the finite-size multi-time correlation function~\eqref{eq:multiTimeCorrelations}
exhibits a simple diagrammatic representation,
\begin{equation}\label{eq:multiCorrCircuit}
  C^{(2n)}_{a_1,a_2,a_3,\ldots,a_{2m}}(\vec{p}_{\infty})=2^{-2n}
  \begin{tikzpicture}[baseline={([yshift=-0.6ex]current bounding box.center)},scale=1.5,rotate=90]
    \draw [decoration={brace},decorate] ({18*\dx},{-0.5*\dt}) -- ({18*\dx},{-16.5*\dt}) node [midway,above] {$2n$};
    \grid{16}{8}
    \foreach \u in {1,...,16}{
      \draw[thick,gray,fill=gray] ({17*\dx},{-\u*\dt}) circle (1.25pt);
      \draw[thick,gray,fill=gray] ({\dx},{-\u*\dt}) circle (1.25pt);
    }
    \draw[very thick] ({3*\dx},{-0.5*\dt}) arc (0:150:{0.5*\r});
    \draw[very thick] ({5*\dx},{-0.5*\dt}) arc (0:150:{0.5*\r});
    \draw[very thick] ({7*\dx},{-0.5*\dt}) arc (0:150:{0.5*\r});
    \draw[very thick] ({9*\dx},{-0.5*\dt}) arc (0:150:{0.5*\r});
    \draw[very thick] ({11*\dx},{-0.5*\dt}) arc (0:150:{0.5*\r});
    \draw[very thick] ({13*\dx},{-0.5*\dt}) arc (0:150:{0.5*\r});
    \draw[very thick] ({15*\dx},{-0.5*\dt}) arc (0:150:{0.5*\r});
    \draw[very thick] ({3*\dx},{-16.5*\dt}) arc (360:210:{0.5*\r});
    \draw[very thick] ({5*\dx},{-16.5*\dt}) arc (360:210:{0.5*\r});
    \draw[very thick] ({7*\dx},{-16.5*\dt}) arc (360:210:{0.5*\r});
    \draw[very thick] ({9*\dx},{-16.5*\dt}) arc (360:210:{0.5*\r});
    \draw[very thick] ({11*\dx},{-16.5*\dt}) arc (360:210:{0.5*\r});
    \draw[very thick] ({13*\dx},{-16.5*\dt}) arc (360:210:{0.5*\r});
    \draw[very thick] ({15*\dx},{-16.5*\dt}) arc (360:210:{0.5*\r});
    \begin{scope}
      \clip ({-1.5*\dx},{-0.5*\dt}) rectangle ({18*\dx},{-16.5*\dt});
      \foreach \u in {-1,1,3,5,...,15}{
        \prop{\u}{(\u+2)}{1.5}{colU}
        \prop{\u}{(\u+2)}{3.5}{colU}
        \prop{\u}{(\u+2)}{5.5}{colU}
        \prop{\u}{(\u+2)}{7.5}{colU}
      }
      \foreach \u in {0,2,4,...,16}{
        \prop{\u}{(\u+2)}{2.5}{colU}
        \prop{\u}{(\u+2)}{4.5}{colU}
        \prop{\u}{(\u+2)}{6.5}{colU}
      }
    \end{scope}
    \obs{8}{0.92}{0.5*\dx}{colObs}
    \obs{9}{1.9}{0.5*\dx}{colObs}
    \obs{8}{2.9}{0.5*\dx}{colObs}
    \obs{9}{3.9}{0.5*\dx}{colObs}
    \obs{8}{4.9}{0.5*\dx}{colObs}
    \obs{9}{5.9}{0.5*\dx}{colObs}
    \obs{8}{6.9}{0.5*\dx}{colObs}
    \obs{9}{8}{0.5*\dx}{colObs}
  \end{tikzpicture},
\end{equation}
where the red squares represent (in general different) one-site observables and
the grey circles denote one-site row (and column) vectors~$\vec{\omega}$
(and~$\vec{\omega}^T$).  The local time-evolution operator~$U$ is
deterministic, which implies that it maps the three-site maximum entropy state
into itself,
\begin{equation}\label{eq:deterministicGates}
  \begin{aligned}
    U\left(\vec{\omega}\otimes\vec{\omega}\otimes\vec{\omega}\right)^T
    &=\left(\vec{\omega}\otimes\vec{\omega}\otimes\vec{\omega}\right)^T,& \qquad
    \left(\vec{\omega}\otimes\vec{\omega}\otimes\vec{\omega}\right)U
    &=\vec{\omega}\otimes\vec{\omega}\otimes\vec{\omega},\\
    \begin{tikzpicture}[baseline={([yshift=-0.6ex]current bounding box.center)},scale=2,rotate=90]
      \grid{3}{2}
      \prop{1}{3}{1.5}{colU}
      \draw[thick,gray,fill=gray] ({\dx},{-\dt}) circle (1.25pt);
      \draw[thick,gray,fill=gray] ({\dx},{-2*\dt}) circle (1.25pt);
      \draw[thick,gray,fill=gray] ({\dx},{-3*\dt}) circle (1.25pt);
    \end{tikzpicture}&\equiv
    \begin{tikzpicture}[baseline={([yshift=-0.6ex]current bounding box.center)},scale=2,rotate=90]
      \grid{3}{2}
      \draw[thick,gray,fill=gray] ({\dx},{-\dt}) circle (1.25pt);
      \draw[thick,gray,fill=gray] ({\dx},{-2*\dt}) circle (1.25pt);
      \draw[thick,gray,fill=gray] ({\dx},{-3*\dt}) circle (1.25pt);
    \end{tikzpicture}, & 
    \begin{tikzpicture}[baseline={([yshift=-0.6ex]current bounding box.center)},scale=2,rotate=90]
      \grid{3}{2}
      \prop{1}{3}{1.5}{colU}
      \draw[thick,gray,fill=gray] ({5*\dx},{-\dt}) circle (1.25pt);
      \draw[thick,gray,fill=gray] ({5*\dx},{-2*\dt}) circle (1.25pt);
      \draw[thick,gray,fill=gray] ({5*\dx},{-3*\dt}) circle (1.25pt);
    \end{tikzpicture}&\equiv
    \begin{tikzpicture}[baseline={([yshift=-0.6ex]current bounding box.center)},scale=2,rotate=90]
      \grid{3}{2}
      \draw[thick,gray,fill=gray] ({5*\dx},{-\dt}) circle (1.25pt);
      \draw[thick,gray,fill=gray] ({5*\dx},{-2*\dt}) circle (1.25pt);
      \draw[thick,gray,fill=gray] ({5*\dx},{-3*\dt}) circle (1.25pt);
    \end{tikzpicture}.
  \end{aligned}
\end{equation}
This immediately allows us to simplify the diagrammatic expression by removing the gates
from the top and bottom to obtain a light-cone structure,
\begin{equation} \label{eq:InfTempCorrs}
  C_{a_1,a_2,\ldots,a_{2m}}(\vec{p}_{\infty})=
  2^{-2m}
  \begin{tikzpicture}[baseline={([yshift=-0.6ex]current bounding box.center)},scale=1.75,rotate=90]
    \draw[thick,gray,fill=gray] ({9*\dx},{-7*\dt}) circle (1.25pt);
    \draw[thick,gray,fill=gray] ({5*\dx},{-7*\dt}) circle (1.25pt);
    \draw[gray,very thin] ({5*\dx},{-7*\dt}) -- ({9*\dx},{-7*\dt});
    \draw[thick,gray,fill=gray] ({3*\dx},{-8*\dt}) circle (1.25pt);
    \draw[thick,gray,fill=gray] ({11*\dx},{-8*\dt}) circle (1.25pt);
    \draw[gray,very thin] ({3*\dx},{-8*\dt}) -- ({11*\dx},{-8*\dt});
    \draw[thick,gray,fill=gray] ({\dx},{-9*\dt}) circle (1.25pt);
    \draw[thick,gray,fill=gray] ({13*\dx},{-9*\dt}) circle (1.25pt);
    \draw[gray,very thin] ({\dx},{-9*\dt}) -- ({13*\dx},{-9*\dt});
    \draw[thick,gray,fill=gray] ({\dx},{-10*\dt}) circle (1.25pt);
    \draw[thick,gray,fill=gray] ({16*\dx},{-10*\dt}) circle (1.25pt);
    \draw[gray,very thin] ({\dx},{-10*\dt}) -- ({16*\dx},{-10*\dt});
    \draw[thick,gray,fill=gray] ({\dx},{-11*\dt}) circle (1.25pt);
    \draw[thick,gray,fill=gray] ({16*\dx},{-11*\dt}) circle (1.25pt);
    \draw[gray,very thin] ({\dx},{-11*\dt}) -- ({16*\dx},{-11*\dt});
    \draw[thick,gray,fill=gray] ({3*\dx},{-12*\dt}) circle (1.25pt);
    \draw[thick,gray,fill=gray] ({16*\dx},{-12*\dt}) circle (1.25pt);
    \draw[gray,very thin] ({3*\dx},{-12*\dt}) -- ({16*\dx},{-12*\dt});
    \draw[thick,gray,fill=gray] ({5*\dx},{-13*\dt}) circle (1.25pt);
    \draw[thick,gray,fill=gray] ({13*\dx},{-13*\dt}) circle (1.25pt);
    \draw[gray,very thin] ({5*\dx},{-13*\dt}) -- ({13*\dx},{-13*\dt});
    \draw[thick,gray,fill=gray] ({7*\dx},{-14*\dt}) circle (1.25pt);
    \draw[thick,gray,fill=gray] ({11*\dx},{-14*\dt}) circle (1.25pt);
    \draw[gray,very thin] ({7*\dx},{-14*\dt}) -- ({11*\dx},{-14*\dt});

    \prop{9}{11}{1.5}{colU}
    \prop{8}{10}{2.5}{colU}
    \prop{10}{12}{2.5}{colU}
    \prop{7}{9}{3.5}{colU}
    \prop{9}{11}{3.5}{colU}
    \prop{11}{13}{3.5}{colU}
    \prop{8}{10}{4.5}{colU}
    \prop{10}{12}{4.5}{colU}
    \prop{12}{14}{4.5}{colU}
    \prop{9}{11}{5.5}{colU}
    \prop{11}{13}{5.5}{colU}
    \prop{10}{12}{6.5}{colU}

    \obs{10}{0.92}{0.5*\dx}{colObs}
    \obs{11}{1.9}{0.5*\dx}{colObs}
    \obs{10}{2.9}{0.5*\dx}{colObs}
    \obs{11}{3.9}{0.5*\dx}{colObs}
    \obs{10}{4.9}{0.5*\dx}{colObs}
    \obs{11}{5.9}{0.5*\dx}{colObs}
    \obs{10}{6.9}{0.5*\dx}{colObs}
    \obs{11}{7.5}{0.5*\dx}{colObs}
  \end{tikzpicture}
  \equiv 2^{-2m}
  \begin{tikzpicture}[baseline={([yshift=-0.6ex]current bounding box.center)},scale=1.75]
    \draw[gray,very thin] ({6*\dx},{-\dt}) -- ({10*\dx},{-\dt});
    \draw[gray,very thin] ({4*\dx},{-2*\dt}) -- ({12*\dx},{-2*\dt});
    \draw[gray,very thin] ({2*\dx},{-3*\dt}) -- ({14*\dx},{-3*\dt});
    \draw[gray,very thin] ({0*\dx},{-4*\dt}) -- ({14*\dx},{-4*\dt});
    \draw[gray,very thin] ({0*\dx},{-5*\dt}) -- ({14*\dx},{-5*\dt});
    \draw[gray,very thin] ({0*\dx},{-6*\dt}) -- ({12*\dx},{-6*\dt});
    \draw[gray,very thin] ({2*\dx},{-7*\dt}) -- ({10*\dx},{-7*\dt});
    \draw[gray,very thin] ({4*\dx},{-8*\dt}) -- ({8*\dx},{-8*\dt});

    \draw[thick,gray,fill=gray] ({6*\dx},{-\dt}) circle(1.25pt);
    \draw[thick,gray,fill=gray] ({10*\dx},{-\dt}) circle(1.25pt);
    \draw[thick,gray,fill=gray] ({4*\dx},{-2*\dt}) circle(1.25pt);
    \draw[thick,gray,fill=gray] ({12*\dx},{-2*\dt}) circle(1.25pt);
    \draw[thick,gray,fill=gray] ({2*\dx},{-3*\dt}) circle(1.25pt);
    \draw[thick,gray,fill=gray] ({14*\dx},{-3*\dt}) circle(1.25pt);
    \draw[thick,gray,fill=gray] ({0*\dx},{-4*\dt}) circle(1.25pt);
    \draw[thick,gray,fill=gray] ({14*\dx},{-4*\dt}) circle(1.25pt);
    \draw[thick,gray,fill=gray] ({0*\dx},{-5*\dt}) circle(1.25pt);
    \draw[thick,gray,fill=gray] ({14*\dx},{-5*\dt}) circle(1.25pt);
    \draw[thick,gray,fill=gray] ({0*\dx},{-6*\dt}) circle(1.25pt);
    \draw[thick,gray,fill=gray] ({12*\dx},{-6*\dt}) circle(1.25pt);
    \draw[thick,gray,fill=gray] ({2*\dx},{-7*\dt}) circle(1.25pt);
    \draw[thick,gray,fill=gray] ({10*\dx},{-7*\dt}) circle(1.25pt);
    \draw[thick,gray,fill=gray] ({4*\dx},{-8*\dt}) circle(1.25pt);
    \draw[thick,gray,fill=gray] ({8*\dx},{-8*\dt}) circle(1.25pt);

    \obs{1}{3.5}{0.5*\dx}{colObs}
    \obs{2}{3.4}{0.5*\dx}{colObs}
    \obs{3}{3.6}{0.5*\dx}{colObs}
    \obs{4}{3.4}{0.5*\dx}{colObs}
    \obs{5}{3.6}{0.5*\dx}{colObs}
    \obs{6}{3.4}{0.5*\dx}{colObs}
    \obs{7}{3.6}{0.5*\dx}{colObs}
    \obs{8}{3.5}{0.5*\dx}{colObs}

    \prop{4}{6}{1}{colUt}
    \prop{3}{5}{2}{colUt}
    \prop{5}{7}{2}{colUt}
    \prop{2}{4}{3}{colUt}
    \prop{4}{6}{3}{colUt}
    \prop{6}{8}{3}{colUt}
    \prop{1}{3}{4}{colUt}
    \prop{3}{5}{4}{colUt}
    \prop{5}{7}{4}{colUt}
    \prop{2}{4}{5}{colUt}
    \prop{4}{6}{5}{colUt}
    \prop{3}{5}{6}{colUt}
  \end{tikzpicture}.
\end{equation}
Note that the normalization factor is different with respect
to~\eqref{eq:multiCorrCircuit}, because of the normalization of
vectors~$\vec{\omega}$, namely $\vec{\omega}\vec{\omega}^T=2$.  The right hand
diagram follows from the definition of (in general, non-deterministic) dual
gates (\ref{eq:defUt}), and the fact that the observables can be understood as
diagonal operators (see Appendix~\ref{sec:rotatingDiagram} for more details).

Up to now we made no assumption on the structure of dual evolution; the right-hand
side of Eq.~\eqref{eq:InfTempCorrs} follows from the deterministic nature
of time evolution and the formal definition of the dual gate~$\hat{U}$~\eqref{eq:defUt}.
It holds for any deterministic $3$-site propagator that nontrivially acts only on the middle
site.~\footnote{To be more precise, the exact requirement is the validity
  of~\eqref{eq:deterministicGates}, which is satisfied by any bistochastic
matrix.} In our case the dual propagation is deterministic as well, therefore,
in analogy to dual unitary circuits~\cite{bertini2019exact}, we expect the diagram
to further simplify. However, since the definition of local deterministic gates is
rather involved (see Eq.~\eqref{eq:defDetGates}), the deterministic property cannot
be directly used to reduce the diagram~\eqref{eq:InfTempCorrs}. Instead, we take advantage
of the following two diagrammatic relations fulfilled by~$\hat{U}$,
\begin{equation}\label{eq:infTrels}
  \begin{tikzpicture}[baseline={([yshift=-0.6ex]current bounding box.center)},scale=1.5]
    \grid{5}{3.5}
    \prop{2}{4}{1.5}{colUt}
    \prop{1}{3}{2.5}{colUt}
    \prop{3}{5}{2.5}{colUt}
    \proj{2}{4}{3.5}{colP}
    \draw[thick,gray,fill=gray] ({1*\dx},{-1*\dt}) circle (1.25pt);
    \draw[thick,gray,fill=gray] ({1*\dx},{-2*\dt}) circle (1.25pt);
    \draw[thick,gray,fill=gray] ({1*\dx},{-3*\dt}) circle (1.25pt);
    \draw[thick,gray,fill=gray] ({1*\dx},{-4*\dt}) circle (1.25pt);
    \draw[thick,gray,fill=gray] ({1*\dx},{-5*\dt}) circle (1.25pt);
  \end{tikzpicture}\equiv
  \begin{tikzpicture}[baseline={([yshift=-0.6ex]current bounding box.center)},scale=1.5]
    \grid{5}{2.5}
    \prop{1}{3}{1.5}{colUt}
    \prop{3}{5}{1.5}{colUt}
    \proj{2}{4}{2.5}{colP}
    \draw[thick,gray,fill=gray] ({1*\dx},{-1*\dt}) circle (1.25pt);
    \draw[thick,gray,fill=gray] ({1*\dx},{-2*\dt}) circle (1.25pt);
    \draw[thick,gray,fill=gray] ({1*\dx},{-3*\dt}) circle (1.25pt);
    \draw[thick,gray,fill=gray] ({1*\dx},{-4*\dt}) circle (1.25pt);
    \draw[thick,gray,fill=gray] ({1*\dx},{-5*\dt}) circle (1.25pt);
  \end{tikzpicture},\qquad
  \begin{tikzpicture}[baseline={([yshift=-0.6ex]current bounding box.center)},scale=1.5]
    \grid{6}{3.5}
    \prop{1}{3}{1.5}{colUt}
    \prop{3}{5}{1.5}{colUt}
    \prop{2}{4}{2.5}{colUt}
    \prop{4}{6}{2.5}{colUt}
    \proj{3}{5}{3.5}{colP}
    \draw[thick,gray,fill=gray] ({1*\dx},{-1*\dt}) circle (1.25pt);
    \draw[thick,gray,fill=gray] ({1*\dx},{-2*\dt}) circle (1.25pt);
    \draw[thick,gray,fill=gray] ({1*\dx},{-3*\dt}) circle (1.25pt);
    \draw[thick,gray,fill=gray] ({1*\dx},{-4*\dt}) circle (1.25pt);
    \draw[thick,gray,fill=gray] ({1*\dx},{-5*\dt}) circle (1.25pt);
    \draw[thick,gray,fill=gray] ({1*\dx},{-6*\dt}) circle (1.25pt);
  \end{tikzpicture}\equiv
  \begin{tikzpicture}[baseline={([yshift=-0.6ex]current bounding box.center)},scale=1.5]
    \grid{6}{3.5}
    \prop{1}{3}{1.5}{colUt}
    \prop{2}{4}{2.5}{colUt}
    \prop{4}{6}{2.5}{colUt}
    \proj{3}{5}{3.5}{colP}
    \draw[thick,gray,fill=gray] ({1*\dx},{-1*\dt}) circle (1.25pt);
    \draw[thick,gray,fill=gray] ({1*\dx},{-2*\dt}) circle (1.25pt);
    \draw[thick,gray,fill=gray] ({1*\dx},{-3*\dt}) circle (1.25pt);
    \draw[thick,gray,fill=gray] ({1*\dx},{-4*\dt}) circle (1.25pt);
    \draw[thick,gray,fill=gray] ({1*\dx},{-5*\dt}) circle (1.25pt);
    \draw[thick,gray,fill=gray] ({1*\dx},{-6*\dt}) circle (1.25pt);
  \end{tikzpicture}.
\end{equation}
We stress that even though these diagrams are conceptually similar to those used to
prove the deterministic property of space evolution
(cf.~\eqref{eq:nontrivialDiagrams}), the precise relation between the two is
not clear at present.

Because of~$\hat{U}^T=\hat{U}$, also the left-right reversed diagrams hold.
Additionally, since the observables are diagonal, they all commute with the
three-site projector~$P$. This, together with the
relations~\eqref{eq:infTrels}, allows us to transform the correlation function
$C_{a_1,a_2,\ldots,a_{2m}}(\vec{p}_{\infty})$ into a diagram with only the two
inner-most layers of dual gates left,
\begin{equation}
  C_{a_1,a_2,a_3,\ldots,a_{2m}}(\vec{p}_{\infty}) = 2^{-2m}
  \begin{tikzpicture}[baseline={([yshift=-0.6ex]current bounding box.center)},scale=1.5]
    \draw[gray,very thin] ({4*\dx},{0*\dt}) -- ({10*\dx},{0*\dt});
    \draw[gray,very thin] ({2*\dx},{-1*\dt}) -- ({10*\dx},{-1*\dt});
    \draw[gray,very thin] ({2*\dx},{-2*\dt}) -- ({10*\dx},{-2*\dt});
    \draw[gray,very thin] ({2*\dx},{-3*\dt}) -- ({10*\dx},{-3*\dt});
    \draw[gray,very thin] ({2*\dx},{-4*\dt}) -- ({10*\dx},{-4*\dt});
    \draw[gray,very thin] ({2*\dx},{-5*\dt}) -- ({10*\dx},{-5*\dt});
    \draw[gray,very thin] ({2*\dx},{-6*\dt}) -- ({10*\dx},{-6*\dt});
    \draw[gray,very thin] ({2*\dx},{-9*\dt}) -- ({10*\dx},{-9*\dt});
    \draw[gray,very thin] ({2*\dx},{-10*\dt}) -- ({10*\dx},{-10*\dt});
    \draw[gray,very thin] ({2*\dx},{-11*\dt}) -- ({8*\dx},{-11*\dt});
    \draw[thick,gray,fill=gray] ({4*\dx},{0*\dt}) circle(1.25pt);
    \draw[thick,gray,fill=gray] ({2*\dx},{-1*\dt}) circle(1.25pt);
    \draw[thick,gray,fill=gray] ({2*\dx},{-2*\dt}) circle(1.25pt);
    \draw[thick,gray,fill=gray] ({2*\dx},{-3*\dt}) circle(1.25pt);
    \draw[thick,gray,fill=gray] ({2*\dx},{-4*\dt}) circle(1.25pt);
    \draw[thick,gray,fill=gray] ({2*\dx},{-5*\dt}) circle(1.25pt);
    \draw[thick,gray,fill=gray] ({2*\dx},{-6*\dt}) circle(1.25pt);
    \draw[thick,gray,fill=gray] ({2*\dx},{-9*\dt}) circle(1.25pt);
    \draw[thick,gray,fill=gray] ({2*\dx},{-10*\dt}) circle(1.25pt);
    \draw[thick,gray,fill=gray] ({2*\dx},{-11*\dt}) circle(1.25pt);
    \draw[thick,gray,fill=gray] ({10*\dx},{0*\dt}) circle(1.25pt);
    \draw[thick,gray,fill=gray] ({10*\dx},{-1*\dt}) circle(1.25pt);
    \draw[thick,gray,fill=gray] ({10*\dx},{-2*\dt}) circle(1.25pt);
    \draw[thick,gray,fill=gray] ({10*\dx},{-3*\dt}) circle(1.25pt);
    \draw[thick,gray,fill=gray] ({10*\dx},{-4*\dt}) circle(1.25pt);
    \draw[thick,gray,fill=gray] ({10*\dx},{-5*\dt}) circle(1.25pt);
    \draw[thick,gray,fill=gray] ({10*\dx},{-6*\dt}) circle(1.25pt);
    \draw[thick,gray,fill=gray] ({10*\dx},{-9*\dt}) circle(1.25pt);
    \draw[thick,gray,fill=gray] ({10*\dx},{-10*\dt}) circle(1.25pt);
    \draw[thick,gray,fill=gray] ({8*\dx},{-11*\dt}) circle(1.25pt);
    \node[rotate=90] at ({4*\dx},{-7.5*\dt}){\ldots};
    \node[rotate=90] at ({8*\dx},{-7.5*\dt}){\ldots};
    \prop{1}{3}{2}{colUt}
    \prop{3}{5}{2}{colUt}
    \prop{0}{2}{4}{colUt}
    \prop{2}{4}{4}{colUt}
    \prop{4}{6}{4}{colUt}
    \begin{scope}
      \clip ({10*\dx},{0*\dt}) rectangle ({3*\dx},{-6.5*\dt});
      \prop{5}{7}{2}{colUt}
      \prop{6}{8}{4}{colUt}
    \end{scope}
    \begin{scope}
      \clip ({10*\dx},{-11*\dt}) rectangle ({3*\dx},{-8.5*\dt});
      \prop{7}{9}{2}{colUt}
      \prop{8}{10}{4}{colUt}
    \end{scope}
    \prop{9}{11}{2}{colUt}
    \obs{0}{3}{0.5*\dx}{colObs}
    \obs{1}{3}{0.5*\dx}{colObs}
    \obs{2}{3}{0.5*\dx}{colObs}
    \obs{3}{3}{0.5*\dx}{colObs}
    \obs{4}{3}{0.5*\dx}{colObs}
    \obs{5}{3}{0.5*\dx}{colObs}
    \obs{6}{3}{0.5*\dx}{colObs}
    \obs{9}{3}{0.5*\dx}{colObs}
    \obs{10}{3}{0.5*\dx}{colObs}
    \obs{11}{3}{0.5*\dx}{colObs}
  \end{tikzpicture}.
\end{equation}
This can be put in a more convenient form by introducing left and right edge
matrix product states, $\bbra{L}\vec{A}_1 \vec{B}_2^{\prime}\vec{A}_3\cdots
\vec{A}_{2m-1}\kket{R}$ and $\bbra{L}\vec{A}_2 \vec{B}_3\vec{A}_4\cdots
\vec{A}_{2m}\kket{R}$, to replace the two remaining dual gate layers.
The auxiliary space is $2$-dimensional, with the following boundary vectors
\begin{equation}\label{eq:vertMPAbv}
  \begin{tikzpicture}[baseline={([yshift=-0.6ex]current bounding box.center)},scale=2]
    \draw[thick] (0,{-0.5*\dt}) -- (0,{0*\dt});
    \vbvec{0.5}{0}
  \end{tikzpicture}\equiv \bbra{L}=\begin{bmatrix}1 &1\end{bmatrix},\qquad
  \begin{tikzpicture}[baseline={([yshift=-0.6ex]current bounding box.center)},scale=2]
    \draw[thick] (0,{-0.5*\dt}) -- (0,{0*\dt});
    \vbvec{0}{0}
  \end{tikzpicture}\equiv \kket{R}=\begin{bmatrix}1\\1\end{bmatrix}.
\end{equation}
The matrices~$A_s$ are diagonal with one nonzero entry,
\begin{equation}\label{eq:vertMPAmA}
  \begin{tikzpicture}[baseline={([yshift=-0.6ex]current bounding box.center)},scale=2]
    \draw[very thin, gray] (0,{-0.5*\dt}) -- ({\dx},{-0.5*\dt});
    \draw[thick] (0,{-1*\dt}) -- (0,{0*\dt});
    \vA{0.5}{0}{colvMPA}
  \end{tikzpicture}\equiv \vec{A}\equiv
  \begin{tikzpicture}[baseline={([yshift=-0.6ex]current bounding box.center)},scale=2]
    \draw[very thin, gray] (0,{-0.5*\dt}) -- ({-\dx},{-0.5*\dt});
    \draw[thick] (0,{-1*\dt}) -- (0,{0*\dt});
    \vA{0.5}{0}{colvMPA}
  \end{tikzpicture},\qquad
  A_0=\begin{bmatrix}1 & 0 \\ 0 & 0\end{bmatrix},\qquad
  A_1=\begin{bmatrix}0 & 0 \\ 0 & 1\end{bmatrix},
\end{equation}
while the matrix elements of $B_s$, $B_s^{\prime}$, diagrammatically
represented by the squares
\begin{equation}
  \begin{tikzpicture}[baseline={([yshift=-0.6ex]current bounding box.center)},scale=2]
    \draw[very thin, gray] (0,{-0.5*\dt}) -- ({-\dx},{-0.5*\dt});
    \draw[thick] (0,{-1*\dt}) -- (0,{0*\dt});
    \vB{0.5}{0}{colvMPAp}
  \end{tikzpicture}\equiv \vec{B},\qquad
  \begin{tikzpicture}[baseline={([yshift=-0.6ex]current bounding box.center)},scale=2]
    \draw[very thin, gray] (0,{-0.5*\dt}) -- ({\dx},{-0.5*\dt});
    \draw[thick] (0,{-1*\dt}) -- (0,{0*\dt});
    \vB{0.5}{0}{colvMPA}
  \end{tikzpicture}\equiv \vec{B}^{\prime},\qquad
  B_0=B^{\prime}_0=\begin{bmatrix}1&1\\1&1\end{bmatrix},\qquad
  B_1=B^{\prime}_1=\begin{bmatrix}0&2\\2&0\end{bmatrix},
\end{equation}
are determined by requiring the following relations,
\begin{equation}
  \begin{tikzpicture}[baseline={([yshift=-0.6ex]current bounding box.center)},scale=1.5]
    \grid{3}{2}
    \draw[thick,gray,fill=gray] ({1*\dx},{-1*\dt}) circle (1.25pt);
    \draw[thick,gray,fill=gray] ({1*\dx},{-2*\dt}) circle (1.25pt);
    \draw[thick,gray,fill=gray] ({1*\dx},{-3*\dt}) circle (1.25pt);
    \prop{1}{3}{1.5}{colUt}
  \end{tikzpicture}\equiv
  \begin{tikzpicture}[baseline={([yshift=-0.6ex]current bounding box.center)},scale=1.5]
    \grid{3}{1}
    \draw[thick] ({\dx},{-0.5*\dt}) -- ({\dx},{-3.5*\dt});
    \vbvec{0.5}{0.5}
    \vA{1}{0.5}{colvMPA}
    \vB{2}{0.5}{colvMPA}
    \vA{3}{0.5}{colvMPA}
    \vbvec{3.5}{0.5}
  \end{tikzpicture},\qquad
  \begin{tikzpicture}[baseline={([yshift=-0.6ex]current bounding box.center)},scale=1.5]
    \grid{3}{2}
    \draw[thick,gray,fill=gray] ({5*\dx},{-1*\dt}) circle (1.25pt);
    \draw[thick,gray,fill=gray] ({5*\dx},{-2*\dt}) circle (1.25pt);
    \draw[thick,gray,fill=gray] ({5*\dx},{-3*\dt}) circle (1.25pt);
    \prop{1}{3}{1.5}{colUt}
  \end{tikzpicture}\equiv
  \begin{tikzpicture}[baseline={([yshift=-0.6ex]current bounding box.center)},scale=1.5]
    \grid{3}{1}
    \draw[thick] ({3*\dx},{-0.5*\dt}) -- ({3*\dx},{-3.5*\dt});
    \vbvec{0.5}{1.5}
    \vA{1}{1.5}{colvMPAp}
    \vB{2}{1.5}{colvMPAp}
    \vA{3}{1.5}{colvMPAp}
    \vbvec{3.5}{1.5}
  \end{tikzpicture},\qquad
  \begin{tikzpicture}[baseline={([yshift=0ex]current bounding box.center)},scale=1.5]
    \grid{3}{2}
    \draw[thick,gray,fill=gray] ({1*\dx},{-1*\dt}) circle (1.25pt);
    \draw[thick,gray,fill=gray] ({1*\dx},{-2*\dt}) circle (1.25pt);
    \draw[thick] ({\dx},{-2.5*\dt}) -- ({\dx},{-3.5*\dt});
    \vbvec{2.5}{0.5}
    \vA{3}{0.5}{colvMPA}
    \prop{1}{3}{1.5}{colUt}
  \end{tikzpicture}\equiv
  \begin{tikzpicture}[baseline={([yshift=-0.6ex]current bounding box.center)},scale=1.5]
    \grid{3}{1}
    \draw[thick] ({\dx},{-0.5*\dt}) -- ({\dx},{-3.5*\dt});
    \vbvec{0.5}{0.5}
    \vA{1}{0.5}{colvMPA}
    \vB{2}{0.5}{colvMPA}
    \vA{3}{0.5}{colvMPA}
  \end{tikzpicture},\qquad
  \begin{tikzpicture}[baseline={([yshift=0ex]current bounding box.center)},scale=1.5]
    \grid{3}{2}
    \draw[thick,gray,fill=gray] ({5*\dx},{-1*\dt}) circle (1.25pt);
    \draw[thick,gray,fill=gray] ({5*\dx},{-2*\dt}) circle (1.25pt);
    \draw[thick] ({5*\dx},{-2.5*\dt}) -- ({5*\dx},{-3.5*\dt});
    \vbvec{2.5}{2.5}
    \vA{3}{2.5}{colvMPAp}
    \prop{1}{3}{1.5}{colUt}
  \end{tikzpicture}\equiv
  \begin{tikzpicture}[baseline={([yshift=-0.6ex]current bounding box.center)},scale=1.5]
    \grid{3}{1}
    \draw[thick] ({3*\dx},{-0.5*\dt}) -- ({3*\dx},{-3.5*\dt});
    \vbvec{0.5}{1.5}
    \vA{1}{1.5}{colvMPAp}
    \vB{2}{1.5}{colvMPAp}
    \vA{3}{1.5}{colvMPAp}
  \end{tikzpicture}.
\end{equation}
Note that the first pair of diagrams follows from the second one
due to~$\bbra{L}A_0\kket{R}=\bbra{L}A_1\kket{R}=1$.
The correlation function can finally be rewritten as
\begin{equation}\label{eq:InfTempFinalCircuit}
  C_{a_1,a_2,a_3,\ldots,a_{2m}}(\vec{p}_{\infty}) = 2^{-2m}
  \begin{tikzpicture}[baseline={([yshift=-0.6ex]current bounding box.center)},scale=1.6]
    \draw[gray,very thin] ({4*\dx},{0*\dt}) -- ({8*\dx},{0*\dt});
    \draw[gray,very thin] ({4*\dx},{-1*\dt}) -- ({8*\dx},{-1*\dt});
    \draw[gray,very thin] ({4*\dx},{-2*\dt}) -- ({8*\dx},{-2*\dt});
    \draw[gray,very thin] ({4*\dx},{-5*\dt}) -- ({8*\dx},{-5*\dt});
    \draw[gray,very thin] ({4*\dx},{-6*\dt}) -- ({8*\dx},{-6*\dt});
    \draw[thick,gray,fill=gray] ({4*\dx},{0*\dt}) circle(1.25pt);
    \draw[thick,gray,fill=gray] ({8*\dx},{-6*\dt}) circle(1.25pt);
    \node[rotate=90] at ({6*\dx},{-3.5*\dt}){\ldots};
    \draw[thick] ({4*\dx},{-0.5*\dt}) -- ({4*\dx},{-2.5*\dt});
    \draw[thick] ({4*\dx},{-4.5*\dt}) -- ({4*\dx},{-6.5*\dt});
    \draw[thick] ({8*\dx},{0.5*\dt}) -- ({8*\dx},{-2.5*\dt});
    \draw[thick] ({8*\dx},{-4.5*\dt}) -- ({8*\dx},{-5.5*\dt});
    \vbvec{0.5}{2}
    \vA{1}{2}{colvMPA}
    \vB{2}{2}{colvMPA}
    \vB{5}{2}{colvMPA}
    \vA{6}{2}{colvMPA}
    \vbvec{6.5}{2}
    \vbvec{-0.5}{4}
    \vA{0}{4}{colvMPAp}
    \vB{1}{4}{colvMPAp}
    \vA{2}{4}{colvMPAp}
    \vA{5}{4}{colvMPAp}
    \vbvec{5.5}{4}
    \obs{0}{3}{0.5*\dx}{colObs}
    \obs{1}{3}{0.5*\dx}{colObs}
    \obs{2}{3}{0.5*\dx}{colObs}
    \obs{5}{3}{0.5*\dx}{colObs}
    \obs{6}{3}{0.5*\dx}{colObs}
  \end{tikzpicture}.
\end{equation}
The matrix-product-state (MPS) form of the time-state follows directly from
here. However, before showing it explicitly, we first generalize the result to
the class of equilibrium states~$\vec{p}$, introduced in
section~\ref{sec:MacroStates}.

\subsection{Multi-time correlations for generic equilibrium states}\label{subsec:GstateTS}
To conveniently express multi-time correlations for the class of
states~$\vec{p}$ that can be expressed in a matrix product
form~\eqref{eq:eqStates}, we introduce the diagrammatic notation for the MPS,
\begin{equation}
  \vec{W}\equiv
  \begin{tikzpicture}[baseline={([yshift=-0.6ex]current bounding box.center)},scale=2,rotate=90]
    \clip ({0*\dx},{-0.5*\dt}) rectangle ({2*\dx},{-1.5*\dt});
    \draw[double,thick] ({\dx},{-0.5*\dt}) -- ({\dx},{-1.5*\dt});
    \draw[thin,gray] ({\dx},{-1*\dt}) -- ({2*\dx},{-1*\dt});
    \mW{1}{0.5}{colMPA}
  \end{tikzpicture},\qquad
  \vec{W}^{\prime}\equiv
  \begin{tikzpicture}[baseline={([yshift=-0.6ex]current bounding box.center)},scale=2,rotate=90]
    \clip ({0*\dx},{-0.5*\dt}) rectangle ({2*\dx},{-1.5*\dt});
    \draw[double,thick] ({\dx},{-0.5*\dt}) -- ({\dx},{-1.5*\dt});
    \draw[thin,gray] ({\dx},{-1*\dt}) -- ({2*\dx},{-1*\dt});
    \mV{1}{0.5}{colMPA}
  \end{tikzpicture},\qquad
  S\equiv
  \begin{tikzpicture}[baseline={([yshift=-0.6ex]current bounding box.center)},scale=2,rotate=90]
    \draw[double,thick] ({\dx},{-0.5*\dt}) -- ({\dx},{-1.5*\dt});
    \mS{1}{0.5}{colMPA}
  \end{tikzpicture},
\end{equation}
which allows us to diagrammatically express finite-size correlation
function as,
\begin{equation}
  C^{(2n)}_{a_1,a_2,a_3,\ldots,a_{2m}}(\vec{p})=
  \frac{1}{Z_{2n}}
  \begin{tikzpicture}[baseline={([yshift=-0.6ex]current bounding box.center)},scale=1.5,rotate=90]
    \grid{16}{8}
    \foreach \u in {1,...,16}{
      \draw[thick,gray,fill=gray] ({17*\dx},{-\u*\dt}) circle (1.25pt);
    }
    \draw[thick,double] ({\dx},{-0.5*\dt}) -- ({\dx},{-16.5*\dt});
    \draw[thick,double] ({\dx},{-0.5*\dt}) arc (0:150:{0.5*\r});
    \draw[thick,double] ({\dx},{-16.5*\dt}) arc (360:210:{0.5*\r});
    \draw[very thick] ({3*\dx},{-0.5*\dt}) arc (0:150:{0.5*\r});
    \draw[very thick] ({5*\dx},{-0.5*\dt}) arc (0:150:{0.5*\r});
    \draw[very thick] ({7*\dx},{-0.5*\dt}) arc (0:150:{0.5*\r});
    \draw[very thick] ({9*\dx},{-0.5*\dt}) arc (0:150:{0.5*\r});
    \draw[very thick] ({11*\dx},{-0.5*\dt}) arc (0:150:{0.5*\r});
    \draw[very thick] ({13*\dx},{-0.5*\dt}) arc (0:150:{0.5*\r});
    \draw[very thick] ({15*\dx},{-0.5*\dt}) arc (0:150:{0.5*\r});
    \draw[very thick] ({3*\dx},{-16.5*\dt}) arc (360:210:{0.5*\r});
    \draw[very thick] ({5*\dx},{-16.5*\dt}) arc (360:210:{0.5*\r});
    \draw[very thick] ({7*\dx},{-16.5*\dt}) arc (360:210:{0.5*\r});
    \draw[very thick] ({9*\dx},{-16.5*\dt}) arc (360:210:{0.5*\r});
    \draw[very thick] ({11*\dx},{-16.5*\dt}) arc (360:210:{0.5*\r});
    \draw[very thick] ({13*\dx},{-16.5*\dt}) arc (360:210:{0.5*\r});
    \draw[very thick] ({15*\dx},{-16.5*\dt}) arc (360:210:{0.5*\r});
    \foreach \u in {1,3,...,15}{
      \mW{\u}{0.5}{colMPA}
    }
    \foreach \u in {2,4,...,16}{
      \mV{\u}{0.5}{colMPA}
    }
    \begin{scope}
      \clip ({-1.5*\dx},{-0.5*\dt}) rectangle ({18*\dx},{-16.5*\dt});
      \foreach \u in {-1,1,3,5,...,15}{
        \prop{\u}{(\u+2)}{1.5}{colU}
        \prop{\u}{(\u+2)}{3.5}{colU}
        \prop{\u}{(\u+2)}{5.5}{colU}
        \prop{\u}{(\u+2)}{7.5}{colU}
      }
      \foreach \u in {0,2,4,...,16}{
        \prop{\u}{(\u+2)}{2.5}{colU}
        \prop{\u}{(\u+2)}{4.5}{colU}
        \prop{\u}{(\u+2)}{6.5}{colU}
      }
    \end{scope}
    \obs{8}{0.92}{0.5*\dx}{colObs}
    \obs{9}{1.9}{0.5*\dx}{colObs}
    \obs{8}{2.9}{0.5*\dx}{colObs}
    \obs{9}{3.9}{0.5*\dx}{colObs}
    \obs{8}{4.9}{0.5*\dx}{colObs}
    \obs{9}{5.9}{0.5*\dx}{colObs}
    \obs{8}{6.9}{0.5*\dx}{colObs}
    \obs{9}{8}{0.5*\dx}{colObs}
  \end{tikzpicture}.
\end{equation}
Similarly to the case of the maximum entropy state, the deterministic time evolution
implies that the gates outside of the light-cone can be removed. To prove this,
in addition to~\eqref{eq:deterministicGates}, we use the following three-site
algebraic relations fulfilled by the state~$\vec{p}$, 
\begin{equation}\label{eq:cubicAlgebraDiag}
  \begin{aligned}
    \begin{tikzpicture}[baseline={([yshift=-0.6ex]current bounding box.center)},scale=2,rotate=90]
      \clip (0,{-0.5*\dt}) rectangle ({5.5*\dx},{-4*\dt});
      \grid{3}{2}
      \prop{1}{3}{1.5}{colU}
      \draw[double,thick] ({\dx},{-0.5*\dt}) -- ({\dx},{-4*\dt});
      \mW{1}{0.5}{colMPA}
      \mV{2}{0.5}{colMPA}
      \mW{3}{0.5}{colMPA}
      \mS{3.5}{0.5}{colMPA}
    \end{tikzpicture}
    &\equiv
    \begin{tikzpicture}[baseline={([yshift=-0.6ex]current bounding box.center)},scale=2,rotate=90]
      \clip (0,{-0.5*\dt}) rectangle ({5.5*\dx},{-4*\dt});
      \grid{3}{2}
      \draw[double,thick] ({\dx},{-0.5*\dt}) -- ({\dx},{-3.5*\dt});
      \mW{1}{0.5}{colMPA}
      \mS{1.5}{0.5}{colMPA}
      \mW{2}{0.5}{colMPA}
      \mV{3}{0.5}{colMPA}
    \end{tikzpicture},&\qquad
    \begin{tikzpicture}[baseline={([yshift=-0.6ex]current bounding box.center)},scale=2,rotate=90]
      \clip (0,{-0.5*\dt}) rectangle ({5.5*\dx},{-4*\dt});
      \grid{3}{2}
      \prop{1}{3}{1.5}{colU}
      \draw[double,thick] ({\dx},{-0.5*\dt}) -- ({\dx},{-3.5*\dt});
      \mV{1}{0.5}{colMPA}
      \mS{1.5}{0.5}{colMPA}
      \mV{2}{0.5}{colMPA}
      \mW{3}{0.5}{colMPA}
    \end{tikzpicture}
    &\equiv
    \begin{tikzpicture}[baseline={([yshift=-0.6ex]current bounding box.center)},scale=2,rotate=90]
      \clip (0,{-0.5*\dt}) rectangle ({5.5*\dx},{-4*\dt});
      \grid{3}{2}
      \draw[double,thick] ({\dx},{-0.5*\dt}) -- ({\dx},{-4*\dt});
      \mV{1}{0.5}{colMPA}
      \mW{2}{0.5}{colMPA}
      \mV{3}{0.5}{colMPA}
      \mS{3.5}{0.5}{colMPA}
    \end{tikzpicture},
    \\
    \begin{tikzpicture}[baseline={([yshift=-0.6ex]current bounding box.center)},scale=2,rotate=90]
      \clip (0,{-0.5*\dt}) rectangle ({5.5*\dx},{-4*\dt});
      \grid{3}{2}
      \prop{1}{3}{1.5}{colU}
      \draw[double,thick] ({\dx},{-0.5*\dt}) -- ({\dx},{-3.5*\dt});
      \mW{1}{0.5}{colMPA}
      \mV{2}{0.5}{colMPA}
      \mW{3}{0.5}{colMPA}
    \end{tikzpicture}
    &\equiv
    \begin{tikzpicture}[baseline={([yshift=-0.6ex]current bounding box.center)},scale=2,rotate=90]
      \clip (0,{-0.5*\dt}) rectangle ({5.5*\dx},{-4*\dt});
      \grid{3}{2}
      \draw[double,thick] ({\dx},{-0.5*\dt}) -- ({\dx},{-4*\dt});
      \mW{1}{0.5}{colMPA}
      \mS{1.5}{0.5}{colMPA}
      \mW{2}{0.5}{colMPA}
      \mV{3}{0.5}{colMPA}
      \mS{3.5}{0.5}{colMPA}
    \end{tikzpicture}, &
    \begin{tikzpicture}[baseline={([yshift=-0.6ex]current bounding box.center)},scale=2,rotate=90]
      \clip (0,{-0.5*\dt}) rectangle ({5.5*\dx},{-4*\dt});
      \grid{3}{2}
      \prop{1}{3}{1.5}{colU}
      \draw[double,thick] ({\dx},{-0.5*\dt}) -- ({\dx},{-4*\dt});
      \mV{1}{0.5}{colMPA}
      \mS{1.5}{0.5}{colMPA}
      \mV{2}{0.5}{colMPA}
      \mW{3}{0.5}{colMPA}
      \mS{3.5}{0.5}{colMPA}
    \end{tikzpicture}
    &\equiv
    \begin{tikzpicture}[baseline={([yshift=-0.6ex]current bounding box.center)},scale=2,rotate=90]
      \clip (0,{-0.5*\dt}) rectangle ({5.5*\dx},{-4*\dt});
      \grid{3}{2}
      \draw[double,thick] ({\dx},{-0.5*\dt}) -- ({\dx},{-3.5*\dt});
      \mV{1}{0.5}{colMPA}
      \mW{2}{0.5}{colMPA}
      \mV{3}{0.5}{colMPA}
    \end{tikzpicture}.
  \end{aligned}
\end{equation}
The first relation is a diagrammatic analogue of equation~\eqref{eq:cubicRel1},
while the others follow directly from it by using~$U=U^{-1}$, and~$S=S^{-1}$,
as well as the simple mapping between~$\vec{W}$ and~$\vec{W}^{\prime}$.
Additionally, we define~$\bra{l}$ and $\ket{r}$ as the left and right
eigenvector of the matrix~$(W_0^{\prime}+W^{\prime}_1)(W_0+W_1)$ that
correspond to the leading eigenvalue~$\lambda$, respectively
\begin{equation}
  \begin{tikzpicture}[baseline={([yshift=-2.2ex]current bounding box.center)},scale=2,rotate=90]
    \clip ({-0.5*\dx},{-0.4*\dt}) rectangle ({2.5*\dx},{-2.5*\dt});
    \draw[thick,gray,fill=gray] ({2*\dx},{-1*\dt}) circle (1.25pt);
    \draw[gray,very thin] ({2*\dx},{-1*\dt}) -- ({0},{-1*\dt});
    \draw[thick,gray,fill=gray] ({2*\dx},{-2*\dt}) circle (1.25pt);
    \draw[gray,very thin] ({2*\dx},{-2*\dt}) -- ({0},{-2*\dt});

    \draw[double,thick] (0,{-0.5*\dt}) -- (0,{-2.5*\dt});
    \mbvec{0.5}{0}
    \mV{1}{0}{colMPA}
    \mW{2}{0}{colMPA}
  \end{tikzpicture}\equiv
  \lambda\,
  \begin{tikzpicture}[baseline={([yshift=-2.2ex]current bounding box.center)},scale=2,rotate=90]
    \clip ({-0.5*\dx},{-0.4*\dt}) rectangle ({2.5*\dx},{-\dt});
    \draw[double,thick] (0,{-0.5*\dt}) -- (0,{-\dt});
    \mbvec{0.5}{0}
  \end{tikzpicture},\qquad
  \begin{tikzpicture}[baseline={([yshift=-2.2ex]current bounding box.center)},scale=2,rotate=90]
    \clip ({-0.5*\dx},{-0.5*\dt}) rectangle ({2.5*\dx},{-2.6*\dt});
    \draw[thick,gray,fill=gray] ({2*\dx},{-1*\dt}) circle (1.25pt);
    \draw[gray,very thin] ({2*\dx},{-1*\dt}) -- ({0},{-1*\dt});
    \draw[thick,gray,fill=gray] ({2*\dx},{-2*\dt}) circle (1.25pt);
    \draw[gray,very thin] ({2*\dx},{-2*\dt}) -- ({0},{-2*\dt});

    \draw[double,thick] (0,{-0.5*\dt}) -- (0,{-2.5*\dt});
    \mV{1}{0}{colMPA}
    \mW{2}{0}{colMPA}
    \mbvec{2.5}{0}
  \end{tikzpicture}\equiv
  \lambda\,
  \begin{tikzpicture}[baseline={([yshift=-2.2 ex]current bounding box.center)},scale=2,rotate=90]
    \clip ({-0.5*\dx},{-0.5*\dt}) rectangle ({2.5*\dx},{-1.1*\dt});
    \draw[double,thick] (0,{-0.5*\dt}) -- (0,{-\dt});
    \mbvec{1}{0}
  \end{tikzpicture}.
\end{equation}
Explicitly,~$\lambda$ is the leading root of the following cubic equation
\begin{equation}
  \lambda^3-(1+3\xi\omega)\lambda^2-(\xi+\omega+\xi\omega(1-3\xi\omega))\lambda
  -\xi\omega(1-\xi\omega)^2=0,
\end{equation}
and the leading eigenvectors~$\ket{r}$, $\bra{l}$ can be parametrized
with~$\lambda$, $\xi$ and $\omega$ as
\begin{equation}
  \begin{aligned}
    \ket{r}&=\begin{bmatrix}
      \xi(\lambda-\xi\omega+\omega)\quad&
      (\lambda-\xi\omega)^2-(\lambda+\omega)\quad&
      \xi(\lambda-\xi\omega+\xi)
    \end{bmatrix},\\
    \bra{l}&=\begin{bmatrix}
      (\lambda-\xi\omega)^2-\xi\omega\quad&
      \xi(\lambda-\xi\omega+\omega)\quad&
      \omega(\lambda-\xi\omega+\xi)
    \end{bmatrix}^T.
  \end{aligned}
\end{equation}
The relations~\eqref{eq:cubicAlgebraDiag} immediately imply that the
multi-time correlation function can be expressed in a form analogous
to~\eqref{eq:InfTempCorrs},
\begin{equation}\label{eq:DiagramCorr}
  C_{a_1,a_2,\ldots,a_{2m}}(\vec{p})=
  \frac{\lambda^{-m}}{\braket{l}{r}}
  \begin{tikzpicture}[baseline={([yshift=-0.6ex]current bounding box.center)},scale=1.75,rotate=90]
    \draw[thick,gray,fill=gray] ({9*\dx},{-7*\dt}) circle (1.25pt);
    \draw[gray,very thin] ({5*\dx},{-7*\dt}) -- ({9*\dx},{-7*\dt});
    \draw[thick,gray,fill=gray] ({11*\dx},{-8*\dt}) circle (1.25pt);
    \draw[gray,very thin] ({3*\dx},{-8*\dt}) -- ({11*\dx},{-8*\dt});
    \draw[thick,gray,fill=gray] ({13*\dx},{-9*\dt}) circle (1.25pt);
    \draw[gray,very thin] ({\dx},{-9*\dt}) -- ({13*\dx},{-9*\dt});
    \draw[thick,gray,fill=gray] ({16*\dx},{-10*\dt}) circle (1.25pt);
    \draw[gray,very thin] ({\dx},{-10*\dt}) -- ({16*\dx},{-10*\dt});
    \draw[thick,gray,fill=gray] ({16*\dx},{-11*\dt}) circle (1.25pt);
    \draw[gray,very thin] ({\dx},{-11*\dt}) -- ({16*\dx},{-11*\dt});
    \draw[thick,gray,fill=gray] ({16*\dx},{-12*\dt}) circle (1.25pt);
    \draw[gray,very thin] ({3*\dx},{-12*\dt}) -- ({16*\dx},{-12*\dt});
    \draw[thick,gray,fill=gray] ({13*\dx},{-13*\dt}) circle (1.25pt);
    \draw[gray,very thin] ({5*\dx},{-13*\dt}) -- ({13*\dx},{-13*\dt});
    \draw[thick,gray,fill=gray] ({11*\dx},{-14*\dt}) circle (1.25pt);
    \draw[gray,very thin] ({7*\dx},{-14*\dt}) -- ({11*\dx},{-14*\dt});

    \draw[double,thick] ({6*\dx},{-6.5*\dt}) -- ({\dx},{-9*\dt}) -- ({\dx},{-11*\dt}) -- ({8*\dx},{-14.5*\dt});

    \mbvec{6.5}{3}
    \mV{7}{2.5}{colMPA}
    \mS{7.5}{2}{colMPA}
    \mV{8}{1.5}{colMPA}
    \mS{8.5}{1}{colMPA}
    \mV{9}{0.5}{colMPA}
    \mS{9.5}{0.5}{colMPA}
    \mV{10}{0.5}{colMPA}
    \mW{11}{0.5}{colMPA}
    \mS{11.5}{1}{colMPA}
    \mW{12}{1.5}{colMPA}
    \mS{12.5}{2}{colMPA}
    \mW{13}{2.5}{colMPA}
    \mS{13.5}{3}{colMPA}
    \mW{14}{3.5}{colMPA}
    \mbvec{14.5}{4}

    \prop{9}{11}{1.5}{colU}
    \prop{8}{10}{2.5}{colU}
    \prop{10}{12}{2.5}{colU}
    \prop{7}{9}{3.5}{colU}
    \prop{9}{11}{3.5}{colU}
    \prop{11}{13}{3.5}{colU}
    \prop{8}{10}{4.5}{colU}
    \prop{10}{12}{4.5}{colU}
    \prop{12}{14}{4.5}{colU}
    \prop{9}{11}{5.5}{colU}
    \prop{11}{13}{5.5}{colU}
    \prop{10}{12}{6.5}{colU}

    \obs{10}{0.92}{0.5*\dx}{colObs}
    \obs{11}{1.9}{0.5*\dx}{colObs}
    \obs{10}{2.9}{0.5*\dx}{colObs}
    \obs{11}{3.9}{0.5*\dx}{colObs}
    \obs{10}{4.9}{0.5*\dx}{colObs}
    \obs{11}{5.9}{0.5*\dx}{colObs}
    \obs{10}{6.9}{0.5*\dx}{colObs}
    \obs{11}{7.5}{0.5*\dx}{colObs}
  \end{tikzpicture}=
  \frac{\lambda^{-m}}{\braket{l}{r}}\,
  \begin{tikzpicture}[baseline={([yshift=-0.6ex]current bounding box.center)},scale=1.75]
    \draw[gray,very thin] ({6*\dx},{-\dt}) -- ({10*\dx},{-\dt});
    \draw[gray,very thin] ({4*\dx},{-2*\dt}) -- ({12*\dx},{-2*\dt});
    \draw[gray,very thin] ({2*\dx},{-3*\dt}) -- ({14*\dx},{-3*\dt});
    \draw[gray,very thin] ({0*\dx},{-4*\dt}) -- ({14*\dx},{-4*\dt});
    \draw[gray,very thin] ({0*\dx},{-5*\dt}) -- ({14*\dx},{-5*\dt});
    \draw[gray,very thin] ({0*\dx},{-6*\dt}) -- ({12*\dx},{-6*\dt});
    \draw[gray,very thin] ({2*\dx},{-7*\dt}) -- ({10*\dx},{-7*\dt});
    \draw[gray,very thin] ({4*\dx},{-8*\dt}) -- ({8*\dx},{-8*\dt});

    \draw[thick,gray,fill=gray] ({6*\dx},{-\dt}) circle(1.25pt);
    \draw[thick,gray,fill=gray] ({4*\dx},{-2*\dt}) circle(1.25pt);
    \draw[thick,gray,fill=gray] ({2*\dx},{-3*\dt}) circle(1.25pt);
    \draw[thick,gray,fill=gray] ({0*\dx},{-4*\dt}) circle(1.25pt);
    \draw[thick,gray,fill=gray] ({10*\dx},{-\dt}) circle(1.25pt);
    \draw[thick,gray,fill=gray] ({12*\dx},{-2*\dt}) circle(1.25pt);
    \draw[thick,gray,fill=gray] ({14*\dx},{-3*\dt}) circle(1.25pt);
    \draw[thick,gray,fill=gray] ({8*\dx},{-8*\dt}) circle(1.25pt);

    \obs{1}{3.5}{0.5*\dx}{colObs}
    \obs{2}{3.4}{0.5*\dx}{colObs}
    \obs{3}{3.6}{0.5*\dx}{colObs}
    \obs{4}{3.4}{0.5*\dx}{colObs}
    \obs{5}{3.6}{0.5*\dx}{colObs}
    \obs{6}{3.4}{0.5*\dx}{colObs}
    \obs{7}{3.6}{0.5*\dx}{colObs}
    \obs{8}{3.5}{0.5*\dx}{colObs}

    \prop{4}{6}{1}{colUt}
    \prop{3}{5}{2}{colUt}
    \prop{5}{7}{2}{colUt}
    \prop{2}{4}{3}{colUt}
    \prop{4}{6}{3}{colUt}
    \prop{6}{8}{3}{colUt}
    \prop{1}{3}{4}{colUt}
    \prop{3}{5}{4}{colUt}
    \prop{5}{7}{4}{colUt}
    \prop{2}{4}{5}{colUt}
    \prop{4}{6}{5}{colUt}
    \prop{3}{5}{6}{colUt}
    \draw[double,thick] ({-\dx},{-5*\dt}) -- (0,{-5*\dt}) -- (0,{-6*\dt}) -- ({4*\dx},{-8*\dt});  
    \draw[double,rounded corners, thick] ({4*\dx},{-8*\dt}) -- ({4*\dx},{-8.5*\dt})
    -- ({10*\dx},{-8.5*\dt}) -- ({10*\dx},{-7*\dt});
    \draw[double,thick] ({10*\dx},{-7*\dt}) -- ({14*\dx},{-5*\dt}) -- ({14*\dx},{-4*\dt}) -- ({15*\dx},{-4*\dt});
    \bvec{5}{-0.5}
    \V{5}{0}{colMPA}
    \mS{5.5}{0}{colMPA}
    \V{6}{0}{colMPA}
    \mS{6.5}{0.5}{colMPA}
    \V{7}{1}{colMPA}
    \mS{7.5}{1.5}{colMPA}
    \V{8}{2}{colMPA}
    \W{7}{5}{colMPA}
    \mS{6.5}{5.5}{colMPA}
    \W{6}{6}{colMPA}
    \mS{5.5}{6.5}{colMPA}
    \W{5}{7}{colMPA}
    \mS{4.5}{7}{colMPA}
    \W{4}{7}{colMPA}
    \bvec{4}{7.5}
  \end{tikzpicture}.
\end{equation}

One of the key ingredients in simplifying the circuit into a form analogous
to~\eqref{eq:InfTempFinalCircuit} is a special factorization property
of the equilibrium state~\vec{p}, introduced in App.~B~of~\cite{vMPA2019}.
For any configuration of three consecutive edge sites~$(s_1,s_2,s_3)$ the
leftmost (rightmost) matrix can be absorbed into the left (right) boundary
vector and replaced with a configuration-dependent prefactor. Namely, it is
possible to define tensors of coefficients~$\alpha_{s_1 s_2 s_3}$, $\alpha^{\prime}_{s_1 s_2
s_3}$, $\beta_{s_1 s_2 s_3}$ and $\beta^{\prime}_{s_1 s_2 s_3}$ so that the
following holds,
\begin{equation}\label{eq:factCond}
  \begin{aligned}
    W_{s_1} S W_{s_2} S W_{s_3} \ket{r}
    &= \alpha_{s_1 s_2 s_3} W_{s_1} S W_{s_2} \ket{r},& \ 
    W^{\prime}_{s_1} W_{s_2} S W_{s_3} \ket{r}
    &= \beta_{s_1 s_2 s_3}  W^{\prime}_{s_1} W_{s_2}\ket{r},\\
    \bra{l}W^{\prime}_{s_1} S W^{\prime}_{s_2} S W^{\prime}_{s_3}
    &= \alpha^{\prime}_{s_1 s_2 s_3} \bra{l} W^{\prime}_{s_2} S W^{\prime}_{s_3}, &
    \bra{l}W^{\prime}_{s_1} S W^{\prime}_{s_2} W_{s_3} 
    &= \beta^{\prime}_{s_1 s_2 s_3} \bra{l} W^{\prime}_{s_2} W_{s_3}.
  \end{aligned}
\end{equation}
As a consequence one is able to define vertically oriented left and right
MPSs that replace layers of dual gates.
These are analogous to the left and right edge states
introduced for the maximum entropy case. The boundary vectors~$\bbra{L}$
and $\kket{R}$, as well as the matrices~$A_s$ are defined in
equations~\eqref{eq:vertMPAbv} and~\eqref{eq:vertMPAmA}. The matrix elements
of~$B^{\prime}_s$ are determined by the following relations,
\begin{equation}\label{eq:genLeftMPS}
  \begin{tikzpicture}[baseline={([yshift=-0.6ex]current bounding box.center)},scale=1.5]
    \draw[gray,very thin] ({\dx},{-1*\dt}) -- ({6.5*\dx},{-1*\dt});
    \draw[gray,very thin] ({\dx},{-2*\dt}) -- ({6.5*\dx},{-2*\dt});
    \draw[gray,very thin] ({\dx},{-3*\dt}) -- ({6.5*\dx},{-3*\dt});
    \draw[gray,very thin] ({\dx},{-4*\dt}) -- ({6.5*\dx},{-4*\dt});
    \draw[gray,very thin] ({3*\dx},{-5*\dt}) -- ({6.5*\dx},{-5*\dt});
    \prop{2}{4}{1.5}{colUt}
    \proj{1}{3}{2.25}{colP}
    \proj{3}{5}{2.75}{colP}
    \draw[thick,gray,fill=gray] ({\dx},{-1*\dt}) circle(1.25pt);
    \draw[thick,gray,fill=gray] ({\dx},{-2*\dt}) circle(1.25pt);
    \draw[double,thick] (0,{-3*\dt}) -- ({\dx},{-3*\dt}) -- ({\dx},{-4*\dt}) -- ({4*\dx},{-5.5*\dt});
    \bvec{3}{0}
    \V{3}{0.5}{colMPA}
    \mS{3.5}{0.5}{colMPA}
    \V{4}{0.5}{colMPA}
    \mS{4.5}{1}{colMPA}
    \V{5}{1.5}{colMPA}
  \end{tikzpicture}
  \equiv
  \begin{tikzpicture}[baseline={([yshift=-0.6ex]current bounding box.center)},scale=1.5]
    \draw[gray,very thin] ({\dx},{-1*\dt}) -- ({6.5*\dx},{-1*\dt});
    \draw[gray,very thin] ({\dx},{-2*\dt}) -- ({6.5*\dx},{-2*\dt});
    \draw[gray,very thin] ({\dx},{-3*\dt}) -- ({6.5*\dx},{-3*\dt});
    \draw[gray,very thin] ({\dx},{-4*\dt}) -- ({6.5*\dx},{-4*\dt});
    \draw[gray,very thin] ({4*\dx},{-5*\dt}) -- ({6.5*\dx},{-5*\dt});
    \draw[gray,very thin] ({2*\dx},{-5*\dt}) -- ({2*\dx},{-4*\dt});
    \proj{1}{3}{2.25}{colP}
    \proj{3}{5}{2.75}{colP}
    \draw[thick,gray,fill=gray] ({\dx},{-1*\dt}) circle(1.25pt);
    \draw[thick] ({\dx},{-1.5*\dt}) -- ({\dx},{-4.5*\dt});
    \vbvec{1.5}{0.5}
    \vA{2}{0.5}{colvMPA}
    \vB{3}{0.5}{colvMPA}
    \vA{4}{0.5}{colvMPA}
    \vbvec{4.5}{0.5}
    \draw[double,thick] ({1*\dx},{-5*\dt}) -- ({4*\dx},{-5*\dt}) -- ({5*\dx},{-5.5*\dt});
    \bvec{5}{0.5}{colMPA}
    \V{5}{1}{colMPA}
    \mS{5}{1.5}{colMPA}
    \V{5}{2}{colMPA}
  \end{tikzpicture},\qquad
  \begin{tikzpicture}[baseline={([yshift=-0.6ex]current bounding box.center)},scale=1.5]
    \draw[gray,very thin] (0,{-1*\dt}) -- ({5*\dx},{-1*\dt});
    \draw[gray,very thin] (0,{-2*\dt}) -- ({5*\dx},{-2*\dt});
    \draw[gray,very thin] (0,{-3*\dt}) -- ({5*\dx},{-3*\dt});
    \draw[gray,very thin] (0,{-4*\dt}) -- ({5*\dx},{-4*\dt});
    \draw[gray,very thin] (0,{-5*\dt}) -- ({5*\dx},{-5*\dt});
    \draw[thick] (0,{-0.5*\dt}) -- (0,{-1.4*\dt});
    \draw[thick] (0,{-1.6*\dt}) -- (0,{-5.5*\dt});
    \vA{1}{0}{colvMPA}
    \vbvec{1.4}{0}
    \vbvec{1.6}{0}
    \vA{2}{0}{colvMPA}
    \vB{3}{0}{colvMPA}
    \vA{4}{0}{colvMPA}
    \vB{5}{0}{colvMPA}
    \prop{1}{3}{1}{colUt}
    \prop{3}{5}{1}{colUt}
    \proj{2}{4}{2}{colP}
  \end{tikzpicture}\equiv
  \begin{tikzpicture}[baseline={([yshift=-0.6ex]current bounding box.center)},scale=1.5]
    \draw[gray,very thin] (0,{-1*\dt}) -- ({5*\dx},{-1*\dt});
    \draw[gray,very thin] (0,{-2*\dt}) -- ({5*\dx},{-2*\dt});
    \draw[gray,very thin] (0,{-3*\dt}) -- ({5*\dx},{-3*\dt});
    \draw[gray,very thin] (0,{-4*\dt}) -- ({5*\dx},{-4*\dt});
    \draw[gray,very thin] (0,{-5*\dt}) -- ({5*\dx},{-5*\dt});
    \draw[thick] (0,{-0.5*\dt}) -- (0,{-3.4*\dt});
    \draw[thick] (0,{-3.6*\dt}) -- (0,{-5.5*\dt});
    \vA{1}{0}{colvMPA}
    \vB{2}{0}{colvMPA}
    \vA{3}{0}{colvMPA}
    \vbvec{3.4}{0}
    \vbvec{3.6}{0}
    \vA{4}{0}{colvMPA}
    \vB{5}{0}{colvMPA}
    \prop{3}{5}{1}{colUt}
    \proj{2}{4}{2}{colP}
  \end{tikzpicture},\qquad
  \begin{tikzpicture}[baseline={([yshift=0.6ex]current bounding box.center)},scale=1.5]
    \draw[gray,very thin] (0,{-1*\dt}) -- ({8.5*\dx},{-1*\dt});
    \draw[gray,very thin] (0,{-2*\dt}) -- ({8.5*\dx},{-2*\dt});
    \draw[gray,very thin] (0,{-3*\dt}) -- ({8.5*\dx},{-3*\dt});
    \draw[gray,very thin] (0,{-4*\dt}) -- ({8.5*\dx},{-4*\dt});
    \draw[gray,very thin] ({3*\dx},{-5*\dt})  -- ({8.5*\dx},{-5*\dt});
    \draw[gray,very thin] ({1*\dx},{-5*\dt}) -- ({1*\dx},{-4*\dt});
    \draw[gray,very thin] ({5*\dx},{-6*\dt})  -- ({8.5*\dx},{-6*\dt});
    \prop{1}{3}{2.5}{colUt}
    \prop{3}{5}{2.5}{colUt}
    \proj{2}{4}{3.25}{colP}
    \proj{4}{6}{3.75}{colP}
    \draw[thick] ({0},{-0.5*\dt}) -- ({0},{-1.4*\dt});
    \draw[thick] ({0},{-1.6*\dt}) -- ({0},{-4.5*\dt});
    \vA{1}{0}{colvMPA}
    \vbvec{1.4}{0}
    \vbvec{1.6}{0}
    \vA{2}{0}{colvMPA}
    \vB{3}{0}{colvMPA}
    \vA{4}{0}{colvMPA}
    \vbvec{4.5}{0}
    \draw[double,thick] (0,{-5*\dt}) -- ({3*\dx},{-5*\dt}) -- ({6*\dx},{-6.5*\dt});
    \bvec{5}{0}{colMPA}
    \V{5}{0.5}{colMPA}
    \mS{5}{1}{colMPA}
    \V{5}{1.5}{colMPA}
    \mS{5.5}{2}{colMPA}
    \V{6}{2.5}{colMPA}
  \end{tikzpicture}\equiv
  \begin{tikzpicture}[baseline={([yshift=0.6ex]current bounding box.center)},scale=1.5]
    \draw[gray,very thin] (0,{-1*\dt}) -- ({5.5*\dx},{-1*\dt});
    \draw[gray,very thin] (0,{-2*\dt}) -- ({5.5*\dx},{-2*\dt});
    \draw[gray,very thin] (0,{-3*\dt}) -- ({5.5*\dx},{-3*\dt});
    \draw[gray,very thin] (0,{-4*\dt}) -- ({5.5*\dx},{-4*\dt});
    \draw[gray,very thin] (0,{-5*\dt})  -- ({5.5*\dx},{-5*\dt});
    \draw[gray,very thin] ({1*\dx},{-6*\dt}) -- ({1*\dx},{-5*\dt});
    \draw[gray,very thin] ({3*\dx},{-6*\dt})  -- ({5.5*\dx},{-6*\dt});
    \proj{2}{4}{1.75}{colP}
    \proj{4}{6}{2.25}{colP}
    \draw[thick] ({0},{-0.5*\dt}) -- ({0},{-5.5*\dt});
    \vA{1}{0}{colvMPA}
    \vB{2}{0}{colvMPA}
    \vA{3}{0}{colvMPA}
    \vB{4}{0}{colvMPA}
    \vA{5}{0}{colvMPA}
    \vbvec{5.5}{0}
    \draw[double,thick] (0,{-6*\dt}) -- ({3*\dx},{-6*\dt}) -- ({4*\dx},{-6.5*\dt});
    \bvec{6}{0}{colMPA}
    \V{6}{0.5}{colMPA}
    \mS{6}{1}{colMPA}
    \V{6}{1.5}{colMPA}
  \end{tikzpicture},
\end{equation}
while the matrices~$B_s$ fulfill the analogous identities for the right edge,
\begin{equation}\label{eq:genRightMPS}
  \begin{tikzpicture}[baseline={([yshift=-0.6ex]current bounding box.center)},scale=1.5]
    \draw[gray,very thin] ({-\dx},{-1*\dt}) -- ({-6.5*\dx},{-1*\dt});
    \draw[gray,very thin] ({-\dx},{-2*\dt}) -- ({-6.5*\dx},{-2*\dt});
    \draw[gray,very thin] ({-\dx},{-3*\dt}) -- ({-6.5*\dx},{-3*\dt});
    \draw[gray,very thin] ({-\dx},{-4*\dt}) -- ({-6.5*\dx},{-4*\dt});
    \draw[gray,very thin] ({-3*\dx},{-5*\dt}) -- ({-6.5*\dx},{-5*\dt});
    \prop{2}{4}{-1.5}{colUt}
    \proj{1}{3}{-2.25}{colP}
    \proj{3}{5}{-2.75}{colP}
    \draw[thick,gray,fill=gray] ({-\dx},{-1*\dt}) circle(1.25pt);
    \draw[thick,gray,fill=gray] ({-\dx},{-2*\dt}) circle(1.25pt);
    \draw[double,thick] (0,{-3*\dt}) -- ({-\dx},{-3*\dt}) -- ({-\dx},{-4*\dt}) -- ({-4*\dx},{-5.5*\dt});
    \bvec{3}{-0}
    \W{3}{-0.5}{colMPA}
    \mS{3.5}{-0.5}{colMPA}
    \W{4}{-0.5}{colMPA}
    \mS{4.5}{-1}{colMPA}
    \W{5}{-1.5}{colMPA}
  \end{tikzpicture}
  \equiv
  \begin{tikzpicture}[baseline={([yshift=-0.6ex]current bounding box.center)},scale=1.5]
    \draw[gray,very thin] ({-\dx},{-1*\dt}) -- ({-6.5*\dx},{-1*\dt});
    \draw[gray,very thin] ({-\dx},{-2*\dt}) -- ({-6.5*\dx},{-2*\dt});
    \draw[gray,very thin] ({-\dx},{-3*\dt}) -- ({-6.5*\dx},{-3*\dt});
    \draw[gray,very thin] ({-\dx},{-4*\dt}) -- ({-6.5*\dx},{-4*\dt});
    \draw[gray,very thin] ({-4*\dx},{-5*\dt}) -- ({-6.5*\dx},{-5*\dt});
    \draw[gray,very thin] ({-2*\dx},{-5*\dt}) -- ({-2*\dx},{-4*\dt});
    \proj{1}{3}{-2.25}{colP}
    \proj{3}{5}{-2.75}{colP}
    \draw[thick,gray,fill=gray] ({-\dx},{-1*\dt}) circle(1.25pt);
    \draw[thick] ({-\dx},{-1.5*\dt}) -- ({-\dx},{-4.5*\dt});
    \vbvec{1.5}{-0.5}
    \vA{2}{-0.5}{colvMPAp}
    \vB{3}{-0.5}{colvMPAp}
    \vA{4}{-0.5}{colvMPAp}
    \vbvec{4.5}{-0.5}
    \draw[double,thick] ({-1*\dx},{-5*\dt}) -- ({-4*\dx},{-5*\dt}) -- ({-5*\dx},{-5.5*\dt});
    \bvec{5}{-0.5}{colMPA}
    \W{5}{-1}{colMPA}
    \mS{5}{-1.5}{colMPA}
    \W{5}{-2}{colMPA}
  \end{tikzpicture},\qquad
  \begin{tikzpicture}[baseline={([yshift=-0.6ex]current bounding box.center)},scale=1.5]
    \draw[gray,very thin] (0,{-1*\dt}) -- ({-5*\dx},{-1*\dt});
    \draw[gray,very thin] (0,{-2*\dt}) -- ({-5*\dx},{-2*\dt});
    \draw[gray,very thin] (0,{-3*\dt}) -- ({-5*\dx},{-3*\dt});
    \draw[gray,very thin] (0,{-4*\dt}) -- ({-5*\dx},{-4*\dt});
    \draw[gray,very thin] (0,{-5*\dt}) -- ({-5*\dx},{-5*\dt});
    \draw[thick] (0,{-0.5*\dt}) -- (0,{-1.4*\dt});
    \draw[thick] (0,{-1.6*\dt}) -- (0,{-5.5*\dt});
    \vA{1}{0}{colvMPAp}
    \vbvec{1.4}{0}
    \vbvec{1.6}{0}
    \vA{2}{0}{colvMPAp}
    \vB{3}{0}{colvMPAp}
    \vA{4}{0}{colvMPAp}
    \vB{5}{0}{colvMPAp}
    \prop{1}{3}{-1}{colUt}
    \prop{3}{5}{-1}{colUt}
    \proj{2}{4}{-2}{colP}
  \end{tikzpicture}\equiv
  \begin{tikzpicture}[baseline={([yshift=-0.6ex]current bounding box.center)},scale=1.5]
    \draw[gray,very thin] (0,{-1*\dt}) -- ({-5*\dx},{-1*\dt});
    \draw[gray,very thin] (0,{-2*\dt}) -- ({-5*\dx},{-2*\dt});
    \draw[gray,very thin] (0,{-3*\dt}) -- ({-5*\dx},{-3*\dt});
    \draw[gray,very thin] (0,{-4*\dt}) -- ({-5*\dx},{-4*\dt});
    \draw[gray,very thin] (0,{-5*\dt}) -- ({-5*\dx},{-5*\dt});
    \draw[thick] (0,{-0.5*\dt}) -- (0,{-3.4*\dt});
    \draw[thick] (0,{-3.6*\dt}) -- (0,{-5.5*\dt});
    \vA{1}{0}{colvMPAp}
    \vB{2}{0}{colvMPAp}
    \vA{3}{0}{colvMPAp}
    \vbvec{3.4}{0}
    \vbvec{3.6}{0}
    \vA{4}{0}{colvMPAp}
    \vB{5}{0}{colvMPAp}
    \prop{3}{5}{-1}{colUt}
    \proj{2}{4}{-2}{colP}
  \end{tikzpicture},\qquad
  \begin{tikzpicture}[baseline={([yshift=0.6ex]current bounding box.center)},scale=1.5]
    \draw[gray,very thin] (0,{-1*\dt}) -- ({-8.5*\dx},{-1*\dt});
    \draw[gray,very thin] (0,{-2*\dt}) -- ({-8.5*\dx},{-2*\dt});
    \draw[gray,very thin] (0,{-3*\dt}) -- ({-8.5*\dx},{-3*\dt});
    \draw[gray,very thin] (0,{-4*\dt}) -- ({-8.5*\dx},{-4*\dt});
    \draw[gray,very thin] ({-3*\dx},{-5*\dt})  -- ({-8.5*\dx},{-5*\dt});
    \draw[gray,very thin] ({-1*\dx},{-5*\dt}) -- ({-1*\dx},{-4*\dt});
    \draw[gray,very thin] ({-5*\dx},{-6*\dt})  -- ({-8.5*\dx},{-6*\dt});
    \prop{1}{3}{-2.5}{colUt}
    \prop{3}{5}{-2.5}{colUt}
    \proj{2}{4}{-3.25}{colP}
    \proj{4}{6}{-3.75}{colP}
    \draw[thick] ({0},{-0.5*\dt}) -- ({0},{-1.4*\dt});
    \draw[thick] ({0},{-1.6*\dt}) -- ({0},{-4.5*\dt});
    \vA{1}{0}{colvMPAp}
    \vbvec{1.4}{0}
    \vbvec{1.6}{0}
    \vA{2}{0}{colvMPAp}
    \vB{3}{0}{colvMPAp}
    \vA{4}{0}{colvMPAp}
    \vbvec{4.5}{0}
    \draw[double,thick] (0,{-5*\dt}) -- ({-3*\dx},{-5*\dt}) -- ({-6*\dx},{-6.5*\dt});
    \bvec{5}{0}{colMPA}
    \W{5}{-0.5}{colMPA}
    \mS{5}{-1}{colMPA}
    \W{5}{-1.5}{colMPA}
    \mS{5.5}{-2}{colMPA}
    \W{6}{-2.5}{colMPA}
  \end{tikzpicture}\equiv
  \begin{tikzpicture}[baseline={([yshift=0.6ex]current bounding box.center)},scale=1.5]
    \draw[gray,very thin] (0,{-1*\dt}) -- ({-5.5*\dx},{-1*\dt});
    \draw[gray,very thin] (0,{-2*\dt}) -- ({-5.5*\dx},{-2*\dt});
    \draw[gray,very thin] (0,{-3*\dt}) -- ({-5.5*\dx},{-3*\dt});
    \draw[gray,very thin] (0,{-4*\dt}) -- ({-5.5*\dx},{-4*\dt});
    \draw[gray,very thin] (0,{-5*\dt})  -- ({-5.5*\dx},{-5*\dt});
    \draw[gray,very thin] ({-1*\dx},{-6*\dt}) -- ({-1*\dx},{-5*\dt});
    \draw[gray,very thin] ({-3*\dx},{-6*\dt})  -- ({-5.5*\dx},{-6*\dt});
    \proj{2}{4}{-1.75}{colP}
    \proj{4}{6}{-2.25}{colP}
    \draw[thick] ({0},{-0.5*\dt}) -- ({0},{-5.5*\dt});
    \vA{1}{0}{colvMPAp}
    \vB{2}{0}{colvMPAp}
    \vA{3}{0}{colvMPAp}
    \vB{4}{0}{colvMPAp}
    \vA{5}{0}{colvMPAp}
    \vbvec{5.5}{0}
    \draw[double,thick] (0,{-6*\dt}) -- ({-3*\dx},{-6*\dt}) -- ({-4*\dx},{-6.5*\dt});
    \bvec{6}{0}{colMPA}
    \W{6}{-0.5}{colMPA}
    \mS{6}{-1}{colMPA}
    \W{6}{-1.5}{colMPA}
  \end{tikzpicture}.
\end{equation}
The solution to these relations can be explicitly expressed in terms of tensors
$\alpha_{s_1 s_2 s_3}$, $\alpha^{\prime}_{s_1 s_2 s_3}$ as
\begin{equation}
  \begin{aligned}
    B_0&=\begin{bmatrix}
      \alpha_{000}& \alpha_{001}\\
      \alpha_{000}& \alpha_{001}
    \end{bmatrix}, & \qquad
    B_1&=\begin{bmatrix}
      0&\alpha_{000}+\alpha_{001}\\
      \alpha_{000}+\alpha_{001}&0
    \end{bmatrix},\\
    B_0^{\prime}&=\begin{bmatrix}
      \alpha^{\prime}_{000}& \alpha^{\prime}_{100}\\
      \alpha^{\prime}_{000}& \alpha^{\prime}_{100}
    \end{bmatrix}, &
    B_1^{\prime}&=\begin{bmatrix}
      0&\alpha^{\prime}_{000}+\alpha^{\prime}_{100}\\
      \alpha^{\prime}_{000}+\alpha^{\prime}_{100}&0
    \end{bmatrix}.
  \end{aligned}
\end{equation}
Additional details and the explicit values of the coefficient tensors are provided in
Appendix~\ref{sec:factorSol}.

Using the relations~\eqref{eq:genLeftMPS} and~\eqref{eq:genRightMPS}, together with
the observation that $\bbra{L}A_s\kket{R}=1$ for any $s$, namely
\begin{equation}
  \begin{tikzpicture}[baseline={([yshift=-0.6ex]current bounding box.center)},scale=1.5]
    \draw[very thin, gray] (0,{-0.5*\dt}) -- ({-\dx},{-0.5*\dt});
    \draw[thick] (0,{-1*\dt}) -- (0,{0*\dt});
    \vbvec{1}{0}
    \vA{0.5}{0}{colvMPAp}
    \vbvec{0}{0}
  \end{tikzpicture}\equiv
  \begin{tikzpicture}[baseline={([yshift=-0.6ex]current bounding box.center)},scale=1.5]
    \draw[very thin, gray] (0,{-0.5*\dt}) -- ({-\dx},{-0.5*\dt});
    \draw[thick,gray,fill=gray] (0,{-0.5*\dt}) circle(1.25pt);
  \end{tikzpicture},\qquad
  \begin{tikzpicture}[baseline={([yshift=-0.6ex]current bounding box.center)},scale=1.5]
    \draw[very thin, gray] (0,{-0.5*\dt}) -- ({\dx},{-0.5*\dt});
    \draw[thick] (0,{-1*\dt}) -- (0,{0*\dt});
    \vbvec{1}{0}
    \vA{0.5}{0}{colvMPAp}
    \vbvec{0}{0}
  \end{tikzpicture}\equiv
  \begin{tikzpicture}[baseline={([yshift=-0.6ex]current bounding box.center)},scale=1.5]
    \draw[very thin, gray] (0,{-0.5*\dt}) -- ({\dx},{-0.5*\dt});
    \draw[thick,gray,fill=gray] (0,{-0.5*\dt}) circle(1.25pt);
  \end{tikzpicture},
\end{equation}
the layers of dual gates in the diagram~\eqref{eq:DiagramCorr} can be
removed one after another, until we are left with the innermost two layers
squeezed between two vertical matrix product states,
\begin{equation}
  C_{a_1,a_2,a_3,\ldots,a_{2m}}(\vec{p})=\frac{\lambda^{-m}}{\braket{l}{r}}
  \begin{tikzpicture}[baseline={([yshift=-0.6ex]current bounding box.center)},scale=1.5]
    \draw[gray,very thin] ({4*\dx},{0*\dt}) -- ({12*\dx},{0*\dt});
    \draw[gray,very thin] ({0*\dx},{-1*\dt}) -- ({12*\dx},{-1*\dt});
    \draw[gray,very thin] ({0*\dx},{-2*\dt}) -- ({12*\dx},{-2*\dt});
    \draw[gray,very thin] ({0*\dx},{-3*\dt}) -- ({12*\dx},{-3*\dt});
    \draw[gray,very thin] ({0*\dx},{-4*\dt}) -- ({12*\dx},{-4*\dt});
    \draw[gray,very thin] ({0*\dx},{-5*\dt}) -- ({12*\dx},{-5*\dt});
    \draw[gray,very thin] ({0*\dx},{-6*\dt}) -- ({12*\dx},{-6*\dt});
    \draw[gray,very thin] ({0*\dx},{-9*\dt}) -- ({12*\dx},{-9*\dt});
    \draw[gray,very thin] ({0*\dx},{-10*\dt}) -- ({9*\dx},{-10*\dt});
    \draw[gray,very thin] ({3*\dx},{-11*\dt}) -- ({8*\dx},{-11*\dt});
    \draw[gray,very thin] ({1*\dx},{-11*\dt}) -- ({1*\dx},{-10*\dt});

    \draw[thick,gray,fill=gray] ({0*\dx},{-1*\dt}) circle(1.25pt);
    \draw[thick,gray,fill=gray] ({4*\dx},{0*\dt}) circle(1.25pt);
    \draw[thick,gray,fill=gray] ({8*\dx},{-11*\dt}) circle(1.25pt);
    \draw[thick,gray,fill=gray] ({12*\dx},{0*\dt}) circle(1.25pt);

    \draw[double,thick,rounded corners] ({3*\dx},{-11*\dt}) -- ({3*\dx},{-11.5*\dt})
    -- ({9*\dx},{-11.5*\dt}) -- ({9*\dx},{-10*\dt});
    \draw[double,thick] (0,{-11*\dt}) -- ({3*\dx},{-11*\dt});
    \bvec{11}{0}
    \V{11}{0.5}{colMPA}
    \mS{11}{1}{colMPA}
    \V{11}{1.5}{colMPA}

    \draw[double,thick] ({12*\dx},{-10*\dt}) -- ({9*\dx},{-10*\dt});
    \bvec{10}{6}
    \W{10}{5.5}{colMPA}
    \mS{10}{5}{colMPA}
    \W{10}{4.5}{colMPA}

    \draw[thick] (0,{-10.5*\dt}) -- (0,{-8.5*\dt});
    \draw[thick] (0,{-6.5*\dt}) -- (0,{-1.5*\dt});

    \draw[thick] ({12*\dx},{-9.5*\dt}) -- ({12*\dx},{-8.5*\dt});
    \draw[thick] ({12*\dx},{-6.5*\dt}) -- ({12*\dx},{-0.5*\dt});

    \node[rotate=90] at (0,{-7.5*\dt}){\ldots};
    \node[rotate=90] at ({12*\dx},{-7.5*\dt}){\ldots};
    \vbvec{9.5}{6}
    \vA{9}{6}{colvMPAp}
    \vB{6}{6}{colvMPAp}
    \vA{5}{6}{colvMPAp}
    \vB{4}{6}{colvMPAp}
    \vA{3}{6}{colvMPAp}
    \vB{2}{6}{colvMPAp}
    \vA{1}{6}{colvMPAp}
    \vbvec{0.5}{6}

    \vbvec{10.5}{0}
    \vA{10}{0}{colvMPA}
    \vB{9}{0}{colvMPA}

    \vA{6}{0}{colvMPA}
    \vB{5}{0}{colvMPA}
    \vA{4}{0}{colvMPA}
    \vB{3}{0}{colvMPA}
    \vA{2}{0}{colvMPA}
    \vbvec{1.5}{0}
    \node[rotate=90] at ({4*\dx},{-7.5*\dt}){\ldots};
    \node[rotate=90] at ({6*\dx},{-7.5*\dt}){\ldots};
    \node[rotate=90] at ({8*\dx},{-7.5*\dt}){\ldots};
    \begin{scope}
      \clip ({10*\dx},{\dt}) rectangle ({3*\dx},{-6.5*\dt});
      \prop{1}{3}{2}{colUt}
      \prop{3}{5}{2}{colUt}
      \prop{5}{7}{2}{colUt}
      \prop{0}{2}{4}{colUt}
      \prop{2}{4}{4}{colUt}
      \prop{4}{6}{4}{colUt}
      \prop{6}{8}{4}{colUt}
    \end{scope}
    \begin{scope}
      \clip ({10*\dx},{-11.5*\dt}) rectangle ({3*\dx},{-8.5*\dt});
      \prop{7}{9}{2}{colUt}
      \prop{9}{11}{2}{colUt}
      \prop{6}{8}{4}{colUt}
      \prop{8}{10}{4}{colUt}
    \end{scope}
    \obs{0}{3}{0.5*\dx}{colObs}
    \obs{1}{3}{0.5*\dx}{colObs}
    \obs{2}{3}{0.5*\dx}{colObs}
    \obs{3}{3}{0.5*\dx}{colObs}
    \obs{4}{3}{0.5*\dx}{colObs}
    \obs{5}{3}{0.5*\dx}{colObs}
    \obs{6}{3}{0.5*\dx}{colObs}
    \obs{9}{3}{0.5*\dx}{colObs}
    \obs{10}{3}{0.5*\dx}{colObs}
    \obs{11}{3}{0.5*\dx}{colObs}
  \end{tikzpicture}.
\end{equation}
To remove the last two layers, we note that the observables commute with all
the projectors since they are diagonal in the same basis and structure
of~$\vec{B}$, $\vec{B}^{\prime}$ implies that left (right) vertical states are
invariant under projectors centered at odd (even) sites,
\begin{equation}
  \begin{tikzpicture}[baseline={([yshift=-0.6ex]current bounding box.center)},scale=1.5]
    \draw[gray,very thin] (0,{-1*\dt}) -- ({4*\dx},{-1*\dt});
    \draw[gray,very thin] (0,{-2*\dt}) -- ({4*\dx},{-2*\dt});
    \draw[gray,very thin] (0,{-3*\dt}) -- ({4*\dx},{-3*\dt});
    \proj{1}{3}{0.5}{colP}
    \obs{1}{1.5}{0.5*\dx}{colObs}
    \obs{2}{1.5}{0.5*\dx}{colObs}
    \obs{3}{1.5}{0.5*\dx}{colObs}
  \end{tikzpicture}\equiv
  \begin{tikzpicture}[baseline={([yshift=-0.6ex]current bounding box.center)},scale=1.5]
    \draw[gray,very thin] (0,{-1*\dt}) -- ({4*\dx},{-1*\dt});
    \draw[gray,very thin] (0,{-2*\dt}) -- ({4*\dx},{-2*\dt});
    \draw[gray,very thin] (0,{-3*\dt}) -- ({4*\dx},{-3*\dt});
    \proj{1}{3}{1.5}{colP}
    \obs{1}{0.5}{0.5*\dx}{colObs}
    \obs{2}{0.5}{0.5*\dx}{colObs}
    \obs{3}{0.5}{0.5*\dx}{colObs}
  \end{tikzpicture},\qquad
  \begin{tikzpicture}[baseline={([yshift=-0.6ex]current bounding box.center)},scale=1.5]
    \draw[gray,very thin] (0,{-1*\dt}) -- ({1*\dx},{-1*\dt});
    \draw[gray,very thin] (0,{-2*\dt}) -- ({1*\dx},{-2*\dt});
    \draw[gray,very thin] (0,{-3*\dt}) -- ({1*\dx},{-3*\dt});
    \draw[thick] (0,{-0.5*\dt}) -- (0,{-3.5*\dt});
    \vA{1}{0}{colvMPA}
    \vB{2}{0}{colvMPA}
    \vA{3}{0}{colvMPA}
  \end{tikzpicture}\equiv
  \begin{tikzpicture}[baseline={([yshift=-0.6ex]current bounding box.center)},scale=1.5]
    \draw[gray,very thin] (0,{-1*\dt}) -- ({3*\dx},{-1*\dt});
    \draw[gray,very thin] (0,{-2*\dt}) -- ({3*\dx},{-2*\dt});
    \draw[gray,very thin] (0,{-3*\dt}) -- ({3*\dx},{-3*\dt});
    \draw[thick] (0,{-0.5*\dt}) -- (0,{-3.5*\dt});
    \vA{1}{0}{colvMPA}
    \vB{2}{0}{colvMPA}
    \vA{3}{0}{colvMPA}
    \proj{1}{3}{1}{colP}
  \end{tikzpicture},\qquad
  \begin{tikzpicture}[baseline={([yshift=-0.6ex]current bounding box.center)},scale=1.5]
    \draw[gray,very thin] (0,{-1*\dt}) -- ({-1*\dx},{-1*\dt});
    \draw[gray,very thin] (0,{-2*\dt}) -- ({-1*\dx},{-2*\dt});
    \draw[gray,very thin] (0,{-3*\dt}) -- ({-1*\dx},{-3*\dt});
    \draw[thick] (0,{-0.5*\dt}) -- (0,{-3.5*\dt});
    \vA{1}{0}{colvMPAp}
    \vB{2}{0}{colvMPAp}
    \vA{3}{0}{colvMPAp}
  \end{tikzpicture}\equiv
  \begin{tikzpicture}[baseline={([yshift=-0.6ex]current bounding box.center)},scale=1.5]
    \draw[gray,very thin] (0,{-1*\dt}) -- ({-3*\dx},{-1*\dt});
    \draw[gray,very thin] (0,{-2*\dt}) -- ({-3*\dx},{-2*\dt});
    \draw[gray,very thin] (0,{-3*\dt}) -- ({-3*\dx},{-3*\dt});
    \draw[thick] (0,{-0.5*\dt}) -- (0,{-3.5*\dt});
    \vA{1}{0}{colvMPAp}
    \vB{2}{0}{colvMPAp}
    \vA{3}{0}{colvMPAp}
    \proj{1}{3}{-1}{colP}
  \end{tikzpicture}.
\end{equation}
Additionally we have
\begin{equation}
  \begin{tikzpicture}[baseline={([yshift=-0.6ex]current bounding box.center)},scale=1.5]
    \draw[gray,very thin] (0,{-1*\dt}) -- ({8*\dx},{-1*\dt});
    \draw[gray,very thin] (0,{-2*\dt}) -- ({8*\dx},{-2*\dt});
    \draw[gray,very thin] (0,{-3*\dt}) -- ({8*\dx},{-3*\dt});
    \draw[gray,very thin] (0,{-4*\dt}) -- ({8*\dx},{-4*\dt});
    \draw[gray,very thin] ({3*\dx},{-5*\dt})  -- ({8*\dx},{-5*\dt});
    \draw[gray,very thin] ({1*\dx},{-5*\dt}) -- ({1*\dx},{-4*\dt});
    \draw[gray,very thin] ({6*\dx},{-5.75*\dt}) -- ({6*\dx},{-5*\dt-0.15*\dt});
    \draw[gray,very thin] ({6*\dx},{-5*\dt-0.15*\dt}) arc (-90:90:0.15*\dt);
    \draw[gray,very thin] ({6*\dx},{-5*\dt+0.15*\dt}) -- ({6*\dx},{-4*\dt});
    \prop{1}{3}{2.25}{colUt}
    \prop{3}{5}{2.25}{colUt}
    \proj{2}{4}{3.5}{colP}
    \draw[thick] ({0},{-0.5*\dt}) -- ({0},{-1.4*\dt});
    \draw[thick] ({0},{-1.6*\dt}) -- ({0},{-4.5*\dt});
    \vA{1}{0}{colvMPA}
    \vbvec{1.4}{0}
    \vbvec{1.6}{0}
    \vA{2}{0}{colvMPA}
    \vB{3}{0}{colvMPA}
    \vA{4}{0}{colvMPA}
    \vbvec{4.5}{0}
    \draw[double,thick] (0,{-5*\dt}) -- ({3*\dx},{-5*\dt}) -- ({6*\dx},{-5.75*\dt}) -- ({7*\dx},{-5.75*\dt});
    \bvec{5}{0}{colMPA}
    \V{5}{0.5}{colMPA}
    \mS{5}{1}{colMPA}
    \V{5}{1.5}{colMPA}
    \W{5.75}{3}{colMPA}
  \end{tikzpicture}\equiv
  \begin{tikzpicture}[baseline={([yshift=-0.6ex]current bounding box.center)},scale=1.5]
    \draw[gray,very thin] (0,{-1*\dt}) -- ({4*\dx},{-1*\dt});
    \draw[gray,very thin] (0,{-2*\dt}) -- ({4*\dx},{-2*\dt});
    \draw[gray,very thin] (0,{-3*\dt}) -- ({4*\dx},{-3*\dt});
    \draw[gray,very thin] (0,{-4*\dt}) -- ({4*\dx},{-4*\dt});
    \draw[gray,very thin] (0,{-5*\dt})  -- ({4*\dx},{-5*\dt});
    \draw[gray,very thin] ({1*\dx},{-6*\dt}) -- ({1*\dx},{-5*\dt});
    \draw[gray,very thin] ({2*\dx},{-6*\dt}) -- ({2*\dx},{-5*\dt-0.15*\dt});
    \draw[gray,very thin] ({2*\dx},{-5*\dt-0.15*\dt}) arc (-90:90:0.15*\dt);
    \draw[gray,very thin] ({2*\dx},{-5*\dt+0.15*\dt}) -- ({2*\dx},{-4*\dt});
    \proj{2}{4}{1.5}{colP}
    \draw[thick] ({0},{-0.5*\dt}) -- ({0},{-5.5*\dt});
    \vA{1}{0}{colvMPA}
    \vB{2}{0}{colvMPA}
    \vA{3}{0}{colvMPA}
    \vB{4}{0}{colvMPA}
    \vA{5}{0}{colvMPA}
    \vbvec{5.5}{0}
    \draw[double,thick] (0,{-6*\dt}) -- ({3*\dx},{-6*\dt});
    \bvec{6}{0}{colMPA}
    \V{6}{0.5}{colMPA}
    \W{6}{1}{colMPA}
  \end{tikzpicture},\qquad
  \begin{tikzpicture}[baseline={([yshift=0.6ex]current bounding box.center)},scale=1.5]
    \draw[gray,very thin] (0,{-1*\dt}) -- ({-8.5*\dx},{-1*\dt});
    \draw[gray,very thin] (0,{-2*\dt}) -- ({-8.5*\dx},{-2*\dt});
    \draw[gray,very thin] (0,{-3*\dt}) -- ({-8.5*\dx},{-3*\dt});
    \draw[gray,very thin] (0,{-4*\dt}) -- ({-8.5*\dx},{-4*\dt});
    \draw[gray,very thin] ({-3*\dx},{-5*\dt})  -- ({-8.5*\dx},{-5*\dt});
    \draw[gray,very thin] ({-1*\dx},{-5*\dt}) -- ({-1*\dx},{-4*\dt});
    \draw[gray,very thin] ({-5*\dx},{-6*\dt})  -- ({-8.5*\dx},{-6*\dt});
    \prop{1}{3}{-2.5}{colUt}
    \prop{3}{5}{-2.5}{colUt}
    \proj{2}{4}{-3.25}{colP}
    \proj{4}{6}{-3.75}{colP}
    \draw[thick] ({0},{-0.5*\dt}) -- ({0},{-1.4*\dt});
    \draw[thick] ({0},{-1.6*\dt}) -- ({0},{-4.5*\dt});
    \vA{1}{0}{colvMPAp}
    \vbvec{1.4}{0}
    \vbvec{1.6}{0}
    \vA{2}{0}{colvMPAp}
    \vB{3}{0}{colvMPAp}
    \vA{4}{0}{colvMPAp}
    \vbvec{4.5}{0}
    \draw[double,thick] (0,{-5*\dt}) -- ({-3*\dx},{-5*\dt}) -- ({-6*\dx},{-6.5*\dt});
    \bvec{5}{0}{colMPA}
    \W{5}{-0.5}{colMPA}
    \mS{5}{-1}{colMPA}
    \W{5}{-1.5}{colMPA}
    \V{6}{-2.5}{colMPA}
  \end{tikzpicture}\equiv
  \begin{tikzpicture}[baseline={([yshift=0.6ex]current bounding box.center)},scale=1.5]
    \draw[gray,very thin] (0,{-1*\dt}) -- ({-5.5*\dx},{-1*\dt});
    \draw[gray,very thin] (0,{-2*\dt}) -- ({-5.5*\dx},{-2*\dt});
    \draw[gray,very thin] (0,{-3*\dt}) -- ({-5.5*\dx},{-3*\dt});
    \draw[gray,very thin] (0,{-4*\dt}) -- ({-5.5*\dx},{-4*\dt});
    \draw[gray,very thin] (0,{-5*\dt})  -- ({-5.5*\dx},{-5*\dt});
    \draw[gray,very thin] ({-1*\dx},{-6*\dt}) -- ({-1*\dx},{-5*\dt});
    \draw[gray,very thin] ({-3*\dx},{-6*\dt})  -- ({-5.5*\dx},{-6*\dt});
    \proj{2}{4}{-1.75}{colP}
    \proj{4}{6}{-2.25}{colP}
    \draw[thick] ({0},{-0.5*\dt}) -- ({0},{-5.5*\dt});
    \vA{1}{0}{colvMPAp}
    \vB{2}{0}{colvMPAp}
    \vA{3}{0}{colvMPAp}
    \vB{4}{0}{colvMPAp}
    \vA{5}{0}{colvMPAp}
    \vbvec{5.5}{0}
    \draw[double,thick] (0,{-6*\dt}) -- ({-3*\dx},{-6*\dt}) -- ({-4*\dx},{-6.5*\dt});
    \bvec{6}{0}{colMPA}
    \W{6}{-0.5}{colMPA}
    \V{6}{-1.5}{colMPA}
  \end{tikzpicture}.
\end{equation}
These relations are  analogous to the right-most diagrams
from~\eqref{eq:genLeftMPS} and~\eqref{eq:genRightMPS}, and imply that the
multi-time correlation function can be finally written as follows
\begin{equation}\label{eq:multiCorrsDiagram}
  C_{a_1,a_2,a_3,\ldots,a_{2m}}(\vec{p})=\frac{\lambda^{-m}}{\braket{l}{r}}
  \begin{tikzpicture}[baseline={([yshift=-0.6ex]current bounding box.center)},scale=1.5]
    \draw[gray,very thin] ({1.5*\dx},{0*\dt}) -- ({5*\dx},{0*\dt});
    \draw[gray,very thin] ({0*\dx},{-1*\dt}) -- ({5*\dx},{-1*\dt});
    \draw[gray,very thin] ({0*\dx},{-2*\dt}) -- ({5*\dx},{-2*\dt});
    \draw[gray,very thin] ({0*\dx},{-3*\dt}) -- ({5*\dx},{-3*\dt});
    \draw[gray,very thin] ({0*\dx},{-6*\dt}) -- ({5*\dx},{-6*\dt});
    \draw[gray,very thin] ({0*\dx},{-7*\dt}) -- ({5*\dx},{-7*\dt});
    \draw[gray,very thin] ({0*\dx},{-8*\dt}) -- ({3.5*\dx},{-8*\dt});
    \draw[gray,very thin] ({1*\dx},{-9*\dt}) -- ({1*\dx},{-8*\dt});
    \draw[gray,very thin] ({4*\dx},{-9*\dt}) -- ({4*\dx},{-7*\dt});
    \draw[thick,gray,fill=gray] ({1.5*\dx},{0*\dt}) circle(1.25pt);
    \draw[thick,gray,fill=gray] ({3.5*\dx},{-8*\dt}) circle(1.25pt);
    \draw[double,thick,rounded corners] (0,{-9*\dt}) -- ({5*\dx},{-9*\dt});
    \bvec{9}{0}
    \V{9}{0.5}{colMPA}
    \W{9}{2}{colMPA}
    \bvec{9}{2.5}

    \draw[thick] (0,{-8.5*\dt}) -- (0,{-5.5*\dt});
    \draw[thick] (0,{-3.5*\dt}) -- (0,{-0.5*\dt});

    \draw[thick] ({5*\dx},{-7.5*\dt}) -- ({5*\dx},{-5.5*\dt});
    \draw[thick] ({5*\dx},{-3.5*\dt}) -- ({5*\dx},{0.5*\dt});

    \node[rotate=90] at ({0*\dx},{-4.5*\dt}){\ldots};
    \node[rotate=90] at ({2.5*\dx},{-4.5*\dt}){\ldots};
    \node[rotate=90] at ({5*\dx},{-4.5*\dt}){\ldots};
    \vbvec{7.5}{2.5}
    \vA{7}{2.5}{colvMPAp}
    \vB{6}{2.5}{colvMPAp}
    \vB{3}{2.5}{colvMPAp}
    \vA{2}{2.5}{colvMPAp}
    \vB{1}{2.5}{colvMPAp}
    \vA{0}{2.5}{colvMPAp}
    \vbvec{-0.5}{2.5}
    \vbvec{8.5}{0}
    \vA{8}{0}{colvMPA}
    \vB{7}{0}{colvMPA}
    \vA{6}{0}{colvMPA}
    \vA{3}{0}{colvMPA}
    \vB{2}{0}{colvMPA}
    \vA{1}{0}{colvMPA}
    \vbvec{0.5}{0}
    \obs{0}{1.25}{0.5*\dx}{colObs}
    \obs{1}{1.25}{0.5*\dx}{colObs}
    \obs{2}{1.25}{0.5*\dx}{colObs}
    \obs{3}{1.25}{0.5*\dx}{colObs}
    \obs{6}{1.25}{0.5*\dx}{colObs}
    \obs{7}{1.25}{0.5*\dx}{colObs}
    \obs{8}{1.25}{0.5*\dx}{colObs}
  \end{tikzpicture}.
\end{equation}
In components Eq.~\eqref{eq:multiCorrsDiagram} reads as
\begin{equation}\label{eq:multiCorrsFinal}
  C_{a_1,\ldots ,a_{2m}}
  =\smashoperator{\sum_{s_1,s_2,s_3,\ldots,s_{2m}}}\frac{\bra{l}W^{\prime}_{s_1}W_{s_2}\ket{r}}{\lambda^k\braket{l}{r}}
  \bbra{L}A_{s_1} B^{\prime}_{s_2}\cdots A_{s_{2m-1}}\kket{R}
  \prod_{j=1}^{2m} a_j(s_j)
  \bbra{L}A_{s_2} B_{s_3}\cdots A_{s_{2m}}\kket{R}.
\end{equation}

\subsection{Matrix product representation of the time state}\label{subsec:MPSrepresentationTS}
\label{subsec:MPSmultitimeCorrs}
The above results can be expressed in terms of a time state, as defined in
Section~\ref{sec:timestates}. The equilibrium time
state~$\vec{q}\in\mathbb{R}^{2^{2m}}$ corresponding to the equilibrium
state~$\vec{p}$ uniquely fixes multi-time correlation functions, which by
definition implies
\begin{equation}
  C_{a_1,a_2,a_3,\ldots,a_{2m}}=\smashoperator{\sum_{s_1,s_2,s_3,\ldots,s_{2m}}}
  q_{s_1 s_2 s_3\ldots s_{2m}}
  \prod_{j=1}^{2m}a_{j}(s_j).
\end{equation}
We can then read the probabilities of time-configurations $q_{s_1 s_2\ldots s_{2m}}$
directly from~\eqref{eq:multiCorrsFinal} as
\begin{equation}
  q_{s_1 s_2 s_3\ldots s_{2m}}=
  \frac{\bra{l}W^{\prime}_{s_1}W_{s_2}\ket{r}}{\lambda^k\braket{l}{r}}
  \bbra{L}A_{s_1} B^{\prime}_{s_2}A_{s_3}\cdots A_{s_{2m-1}}\kket{R}
  \bbra{L}A_{s_2} B_{s_3}A_{s_4} \cdots A_{s_{2m}}\kket{R}.
\end{equation}
From here, an MPS representation is obtained by introducing matrices~$\tilde{A}_s$,
$\tilde{A}^{\prime}_s$ that act on the $4$-dimensional auxiliary space as
\begin{equation}\label{eq:defTildeA}
  \tilde{A}_s=A_s\otimes B_s,\qquad \tilde{A}^{\prime}_s=B^{\prime}_s\otimes A_s,
\end{equation}
and defining boundary vectors $\nkket{\tilde{R}}$, $\nbbra{\tilde{L}}$ as the solutions
to the following relations,
\begin{equation}\label{eq:defTildeBoundary}
  \begin{aligned}
    \nbbra{\tilde{L}}\tilde{A}_{s_1} \tilde{A}^{\prime}_{s_2}&= 
    \frac{\bra{l}W^{\prime}_{s_1}W_{s_2}\ket{r}}{\braket{l}{r}}
    \Big(\bbra{L}A_{s_1}B^{\prime}_{s_2}\Big) \otimes \Big(\bbra{L}A_{s_2}\Big),\\
    \tilde{A}_{s_1}\tilde{A}^{\prime}_{s_2} \nkket{\tilde{R}}&=
    \Big(A_{s_1}\kket{R}\Big)\otimes \Big(B_{s_1}A_{s_2}\kket{R}\Big).
  \end{aligned}
\end{equation}
The time-state can thus be written in the matrix product form as
\begin{equation}\label{eq:MPStimestate}
  \vec{q}=\frac{1}{\lambda^k} \nbbra{\tilde{L}}
  \tilde{\vec{A}}_1
  \tilde{\vec{A}}^{\prime}_2
  \tilde{\vec{A}}_3\cdots
  \tilde{\vec{A}}^{\prime}_{2m}
  \nkket{\tilde{R}}.
\end{equation}
As shown in Appendix~\ref{sec:tsMPSequiv}, this form of the time-state is
equivalent to the MPS introduced in~\cite{vMPA2019}.

\section{Conclusion}\label{sec:concl}
In this paper we have studied the properties of space evolution in Rule 54 reversible
cellular automaton. We have shown that space translation of time configurations
(i.e.\ configurations at the same position in the time direction) can be formulated as
a reversible cellular automaton. In other words, the spatial dynamics
can be expressed in terms of local deterministic maps with finite support. We have provided two 
different interpretations of local space evolution; as $7$-site local deterministic
maps, or equivalently, as a composition of non-deterministic $3$-site gates and
$3$-site projectors onto the subspace of allowed configurations.

The result is interesting from two different points of view. On one hand, due
to the existence of time states, the space dynamics of RCA54 can be studied as a
novel solvable deterministic interacting model, where quasi-particles move with
fixed velocities~$\pm 1$ and undergo pairwise scattering. The main difference
with respect to the usual (temporal) dynamics in RCA54 is the nature of two-body
interaction, which speeds the particles up instead of slowing them down,
i.e.\ it is repulsive rather than attractive.
Arguably the more interesting perspective is to use the properties of the space
dynamics to express nontrivial dynamical physical quantities. We have
demonstrated this approach can be fruitful by finding an alternative derivation
of the MPS form of \emph{equilibrium time-states}, i.e.\ probability distributions
that uniquely determine multi-time correlation functions at the same position.

The paper opens several interesting open questions. For instance, what is limit
of this approach? It would be interesting to see whether the circuit picture
can provide a new perspective on two-point spatio-temporal correlation
functions~\cite{TMPA2018} or time evolution of density matrices in the quantum
version of the model~\cite{alba2019RCA54}. This would provide a new perspective
that does not explicitly rely on the quasi-particle interpretation of the
dynamics, and is hence more robust and easier to generalize. Furthermore, one
would like to understand whether RCA54 is an isolated example or it belongs to
the bigger class of {\em dual reversible} cellular automata with a different but finite support of
local evolution maps in space and time directions. Such a class would provide a
generalization of dual unitary models~\cite{bertini2019exact} which could
support richer transport properties, while many results would still be obtained
exactly.

\section*{Acknowledgements}
KK thanks Bruno Bertini for insightful discussions and valuable comments on the
manuscript.

\paragraph{Funding Information} This work has been supported by the European Research Council under the
Advanced Grant No. 694544 -- OMNES, and by the Slovenian Research Agency (ARRS)
under the Programme P1-0402.  

\begin{appendix}
\section{Equivalence between the deterministic and nondeterministic dual gates}\label{sec:equivalence}
The validity of~\eqref{eq:detSE} can be demonstrated graphically. First we
recall that $P$ and $Q$ are projectors, i.e.\
\begin{equation}
  \begin{tikzpicture}[baseline={([yshift=-0.6ex]current bounding box.center)}]
    \grid{3}{2}
    \proj{1}{3}{1}{colP}
    \proj{1}{3}{2}{colP}
  \end{tikzpicture}
  =
  \begin{tikzpicture}[baseline={([yshift=-0.6ex]current bounding box.center)}]
    \grid{3}{1}
    \proj{1}{3}{1}{colP}
  \end{tikzpicture},\qquad
  \begin{tikzpicture}[baseline={([yshift=-0.6ex]current bounding box.center)}]
    \grid{5}{2}
    \proj{1}{5}{1}{colPt}
    \proj{1}{5}{2}{colPt}
  \end{tikzpicture}
  =
  \begin{tikzpicture}[baseline={([yshift=-0.6ex]current bounding box.center)}]
    \grid{5}{1}
    \proj{1}{5}{1}{colPt}
  \end{tikzpicture}.
\end{equation}
Furthermore, all the local projectors commute and $\hat{U}$ commutes with all
the projectors with which it shares at most one site.
Explicitly, this implies the following diagrams,
\begin{equation}\label{eq:commutativity}
  \begin{tikzpicture}[baseline={([yshift=-0.6ex]current bounding box.center)}]
    \grid{7}{2}
    \proj{1}{5}{1}{colPt}
    \prop{5}{7}{2}{colUt}
  \end{tikzpicture}=
  \begin{tikzpicture}[baseline={([yshift=-0.6ex]current bounding box.center)}]
    \grid{7}{2}
    \proj{1}{5}{2}{colPt}
    \prop{5}{7}{1}{colUt}
  \end{tikzpicture},\qquad
  \begin{tikzpicture}[baseline={([yshift=-0.6ex]current bounding box.center)}]
    \grid{7}{2}
    \proj{3}{7}{1}{colPt}
    \prop{1}{3}{2}{colUt}
  \end{tikzpicture}=
  \begin{tikzpicture}[baseline={([yshift=-0.6ex]current bounding box.center)}]
    \grid{7}{2}
    \proj{3}{7}{2}{colPt}
    \prop{1}{3}{1}{colUt}
  \end{tikzpicture},\qquad
  \begin{tikzpicture}[baseline={([yshift=-0.6ex]current bounding box.center)}]
    \grid{5}{2}
    \proj{1}{3}{1}{colP}
    \prop{3}{5}{2}{colUt}
  \end{tikzpicture}=
  \begin{tikzpicture}[baseline={([yshift=-0.6ex]current bounding box.center)}]
    \grid{5}{2}
    \proj{1}{3}{2}{colP}
    \prop{3}{5}{1}{colUt}
  \end{tikzpicture},\qquad
  \begin{tikzpicture}[baseline={([yshift=-0.6ex]current bounding box.center)}]
    \grid{5}{2}
    \prop{1}{3}{1}{colUt}
    \proj{3}{5}{2}{colP}
  \end{tikzpicture}=
  \begin{tikzpicture}[baseline={([yshift=-0.6ex]current bounding box.center)}]
    \grid{5}{2}
    \prop{1}{3}{2}{colUt}
    \proj{3}{5}{1}{colP}
  \end{tikzpicture}.
\end{equation}
The last two properties needed for the proof are less trivial,
but straightforward to check. Their diagrammatic form reads as
\begin{equation}\label{eq:nontrivialDiagrams}
  \begin{tikzpicture}[baseline={([yshift=-0.6ex]current bounding box.center)}]
    \grid{5}{5}
    \proj{2}{4}{1}{colP}
    \prop{1}{3}{2}{colUt}
    \proj{1}{5}{3}{colPt}
    \prop{3}{5}{4}{colUt}
    \proj{2}{4}{5}{colP}
  \end{tikzpicture}
  =
  \begin{tikzpicture}[baseline={([yshift=-0.6ex]current bounding box.center)}]
    \grid{5}{4}
    \proj{2}{4}{1}{colP}
    \prop{1}{3}{2}{colUt}
    \prop{3}{5}{3}{colUt}
    \proj{2}{4}{4}{colP}
  \end{tikzpicture},\qquad
  \begin{tikzpicture}[baseline={([yshift=-0.6ex]current bounding box.center)}]
    \grid{5}{5}
    \proj{2}{4}{5}{colP}
    \prop{1}{3}{4}{colUt}
    \proj{1}{5}{3}{colPt}
    \prop{3}{5}{2}{colUt}
    \proj{2}{4}{1}{colP}
  \end{tikzpicture}=
  \begin{tikzpicture}[baseline={([yshift=-0.6ex]current bounding box.center)}]
    \grid{5}{4}
    \proj{2}{4}{4}{colP}
    \prop{1}{3}{3}{colUt}
    \prop{3}{5}{2}{colUt}
    \proj{2}{4}{1}{colP}
  \end{tikzpicture}.
\end{equation}
Using these equalities, we can now easily show the equivalence
between~\eqref{eq:7sitePropGraph} and~\eqref{eq:spacePropDiagram2}. First we
express the propagator~$\tilde{U}^{\text{e}}$ from~\eqref{eq:7sitePropGraph}
in terms of gates~$P$, $Q$ and~$\hat{U}$, 
\begin{equation}
  \begin{tikzpicture}[baseline={([yshift=-0.6ex]current bounding box.center)}]
    \clip (\dx,{-0.5*\dt}) rectangle ({41*\dx},{-15.5*\dt});
    \grid{15}{20}

    \proj{2}{4}{1}{colP}
    \proj{10}{12}{1}{colP}
    \proj{4}{6}{2}{colP}
    \proj{12}{14}{2}{colP}
    \prop{3}{5}{3}{colUt}
    \prop{11}{13}{3}{colUt}
    \proj{1}{5}{4}{colPt}
    \proj{9}{13}{4}{colPt}
    \proj{3}{7}{5}{colPt}
    \proj{11}{15}{5}{colPt}

    \proj{6}{8}{6}{colP}
    \proj{14}{16}{6}{colP}
    \proj{0}{2}{7}{colP}
    \proj{8}{10}{7}{colP}
    \prop{-1}{1}{8}{colUt}
    \prop{7}{9}{8}{colUt}
    \prop{15}{17}{8}{colUt}
    \proj{-3}{1}{10}{colPt}
    \proj{5}{9}{10}{colPt}
    \proj{13}{17}{10}{colPt}
    \proj{7}{11}{9}{colPt}
    \proj{15}{19}{9}{colPt}
    \proj{-1}{3}{9}{colPt}

    \proj{3}{7}{12}{colPt}
    \proj{11}{15}{12}{colPt}
    \proj{5}{9}{11}{colPt}
    \proj{13}{17}{11}{colPt}
    \prop{5}{7}{13}{colUt}
    \prop{13}{15}{13}{colUt}
    \proj{4}{6}{14}{colP}
    \proj{12}{14}{14}{colP}
    \proj{6}{8}{15}{colP}
    \proj{14}{16}{15}{colP}

    \proj{-1}{3}{16}{colPt}
    \proj{7}{11}{16}{colPt}
    \proj{15}{19}{16}{colPt}
    \proj{1}{5}{17}{colPt}
    \proj{9}{13}{17}{colPt}
    \prop{1}{3}{18}{colUt}
    \prop{9}{11}{18}{colUt}
    \proj{0}{2}{19}{colP}
    \proj{8}{10}{19}{colP}
    \proj{2}{4}{20}{colP}
    \proj{10}{12}{20}{colP}
  \end{tikzpicture}\rightarrow
  \begin{tikzpicture}[baseline={([yshift=-0.6ex]current bounding box.center)}]
    \clip (\dx,{-0.5*\dt}) rectangle ({35*\dx},{-15.5*\dt});
    \grid{15}{17}

    \proj{2}{4}{1}{colP}
    \proj{10}{12}{1}{colP}
    \proj{4}{6}{2}{colP}
    \proj{12}{14}{2}{colP}
    \prop{3}{5}{3}{colUt}
    \prop{11}{13}{3}{colUt}
    \proj{1}{5}{4}{colPt}
    \proj{9}{13}{4}{colPt}
    \proj{3}{7}{5}{colPt}
    \proj{11}{15}{5}{colPt}

    \proj{6}{8}{7}{colP}
    \proj{14}{16}{7}{colP}
    \proj{0}{2}{6}{colP}
    \proj{8}{10}{6}{colP}
    \prop{-1}{1}{8}{colUt}
    \prop{7}{9}{8}{colUt}
    \prop{15}{17}{8}{colUt}

    \proj{5}{9}{9}{colPt}
    \proj{13}{17}{9}{colPt}
    \prop{5}{7}{10}{colUt}
    \prop{13}{15}{10}{colUt}
    \proj{4}{6}{12}{colP}
    \proj{12}{14}{12}{colP}
    \proj{6}{8}{11}{colP}
    \proj{14}{16}{11}{colP}

    \proj{-1}{3}{13}{colPt}
    \proj{7}{11}{13}{colPt}
    \proj{15}{19}{13}{colPt}
    \proj{1}{5}{14}{colPt}
    \proj{9}{13}{14}{colPt}
    \prop{1}{3}{15}{colUt}
    \prop{9}{11}{15}{colUt}
    \proj{0}{2}{16}{colP}
    \proj{8}{10}{16}{colP}
    \proj{2}{4}{17}{colP}
    \proj{10}{12}{17}{colP}
  \end{tikzpicture},
\end{equation}
where we simplified the diagram by using the fact~$Q^2=Q$ and commutation
relations between the projectors and the gates~\eqref{eq:commutativity}. Using
the second relation of~\eqref{eq:nontrivialDiagrams} and moving around some
of the commuting gates, we obtain the following,
\begin{equation}
  \begin{tikzpicture}[baseline={([yshift=-0.6ex]current bounding box.center)}]
    \clip (\dx,{-0.5*\dt}) rectangle ({33*\dx},{-15.5*\dt});
    \grid{15}{16}

    \proj{2}{4}{1}{colP}
    \proj{10}{12}{1}{colP}
    \proj{4}{6}{2}{colP}
    \proj{12}{14}{2}{colP}
    \prop{3}{5}{3}{colUt}
    \prop{11}{13}{3}{colUt}
    \proj{1}{5}{4}{colPt}
    \proj{9}{13}{4}{colPt}
    \proj{3}{7}{5}{colPt}
    \proj{11}{15}{5}{colPt}

    \proj{6}{8}{7}{colP}
    \proj{14}{16}{7}{colP}
    \proj{0}{2}{6}{colP}
    \proj{8}{10}{6}{colP}
    \prop{-1}{1}{8}{colUt}
    \prop{7}{9}{8}{colUt}
    \prop{15}{17}{8}{colUt}

    \prop{5}{7}{9}{colUt}
    \prop{13}{15}{9}{colUt}
    \proj{4}{6}{11}{colP}
    \proj{12}{14}{11}{colP}
    \proj{6}{8}{10}{colP}
    \proj{14}{16}{10}{colP}

    \proj{-1}{3}{12}{colPt}
    \proj{7}{11}{12}{colPt}
    \proj{15}{19}{12}{colPt}
    \proj{1}{5}{13}{colPt}
    \proj{9}{13}{13}{colPt}
    \prop{1}{3}{14}{colUt}
    \prop{9}{11}{14}{colUt}
    \proj{0}{2}{15}{colP}
    \proj{8}{10}{15}{colP}
    \proj{2}{4}{16}{colP}
    \proj{10}{12}{16}{colP}
  \end{tikzpicture}\rightarrow
  \begin{tikzpicture}[baseline={([yshift=-0.6ex]current bounding box.center)}]
    \clip (\dx,{-0.5*\dt}) rectangle ({27*\dx},{-15.5*\dt});
    \grid{15}{13}

    \proj{2}{4}{1}{colP}
    \proj{10}{12}{1}{colP}
    \proj{4}{6}{2}{colP}
    \proj{12}{14}{2}{colP}
    \prop{3}{5}{3}{colUt}
    \prop{11}{13}{3}{colUt}
    \proj{3}{7}{4}{colPt}
    \proj{11}{15}{4}{colPt}

    \proj{6}{8}{1}{colP}
    \proj{14}{16}{1}{colP}

    \prop{5}{7}{5}{colUt}
    \prop{13}{15}{5}{colUt}
    \proj{4}{6}{6}{colP}
    \proj{12}{14}{6}{colP}

    \proj{1}{5}{7}{colPt}
    \proj{9}{13}{7}{colPt}

    \proj{0}{2}{8}{colP}
    \proj{8}{10}{8}{colP}
    \prop{-1}{1}{9}{colUt}
    \prop{7}{9}{9}{colUt}
    \prop{15}{17}{9}{colUt}

    \proj{6}{8}{13}{colP}
    \proj{14}{16}{13}{colP}

    \proj{-1}{3}{10}{colPt}
    \proj{7}{11}{10}{colPt}
    \proj{15}{19}{10}{colPt}
    \prop{1}{3}{11}{colUt}
    \prop{9}{11}{11}{colUt}
    \proj{0}{2}{12}{colP}
    \proj{8}{10}{12}{colP}
    \proj{2}{4}{13}{colP}
    \proj{10}{12}{13}{colP}
  \end{tikzpicture}.
\end{equation}
Now we use the first equality of~\eqref{eq:nontrivialDiagrams} and reposition the commuting
gates so that we can again apply the second equality of~\eqref{eq:nontrivialDiagrams},
\begin{equation}
  \begin{tikzpicture}[baseline={([yshift=-0.6ex]current bounding box.center)}]
    \clip (\dx,{-0.5*\dt}) rectangle ({23*\dx},{-15.5*\dt});
    \grid{15}{11}

    \proj{2}{4}{1}{colP}
    \proj{10}{12}{1}{colP}
    \proj{4}{6}{2}{colP}
    \proj{12}{14}{2}{colP}
    \prop{3}{5}{3}{colUt}
    \prop{11}{13}{3}{colUt}

    \proj{6}{8}{1}{colP}
    \proj{14}{16}{1}{colP}

    \prop{5}{7}{4}{colUt}
    \prop{13}{15}{4}{colUt}
    \proj{4}{6}{5}{colP}
    \proj{12}{14}{5}{colP}

    \proj{1}{5}{6}{colPt}
    \proj{9}{13}{6}{colPt}

    \proj{0}{2}{7}{colP}
    \proj{8}{10}{7}{colP}
    \prop{-1}{1}{8}{colUt}
    \prop{7}{9}{8}{colUt}
    \prop{15}{17}{8}{colUt}

    \proj{6}{8}{11}{colP}
    \proj{14}{16}{11}{colP}

    \prop{1}{3}{9}{colUt}
    \prop{9}{11}{9}{colUt}
    \proj{0}{2}{10}{colP}
    \proj{8}{10}{10}{colP}
    \proj{2}{4}{11}{colP}
    \proj{10}{12}{11}{colP}
  \end{tikzpicture}
  \rightarrow
  \begin{tikzpicture}[baseline={([yshift=-0.6ex]current bounding box.center)}]
    \clip (\dx,{-0.5*\dt}) rectangle ({23*\dx},{-15.5*\dt});
    \grid{15}{11}

    \proj{2}{4}{4}{colP}
    \proj{10}{12}{4}{colP}
    \proj{4}{6}{2}{colP}
    \proj{12}{14}{2}{colP}
    \prop{3}{5}{5}{colUt}
    \prop{11}{13}{5}{colUt}

    \proj{6}{8}{1}{colP}
    \proj{14}{16}{1}{colP}

    \prop{5}{7}{3}{colUt}
    \prop{13}{15}{3}{colUt}
    \proj{4}{6}{10}{colP}
    \proj{12}{14}{10}{colP}

    \proj{1}{5}{6}{colPt}
    \proj{9}{13}{6}{colPt}

    \proj{0}{2}{2}{colP}
    \proj{8}{10}{2}{colP}
    \prop{-1}{1}{9}{colUt}
    \prop{7}{9}{9}{colUt}
    \prop{15}{17}{9}{colUt}

    \proj{6}{8}{11}{colP}
    \proj{14}{16}{11}{colP}

    \prop{1}{3}{7}{colUt}
    \prop{9}{11}{7}{colUt}
    \proj{0}{2}{10}{colP}
    \proj{8}{10}{10}{colP}
    \proj{2}{4}{8}{colP}
    \proj{10}{12}{8}{colP}
  \end{tikzpicture}
  \rightarrow
  \begin{tikzpicture}[baseline={([yshift=-0.6ex]current bounding box.center)}]
    \clip (\dx,{-0.5*\dt}) rectangle ({21*\dx},{-15.5*\dt});
    \grid{15}{10}

    \proj{2}{4}{4}{colP}
    \proj{10}{12}{4}{colP}
    \proj{4}{6}{2}{colP}
    \proj{12}{14}{2}{colP}
    \prop{3}{5}{5}{colUt}
    \prop{11}{13}{5}{colUt}

    \proj{6}{8}{1}{colP}
    \proj{14}{16}{1}{colP}

    \prop{5}{7}{3}{colUt}
    \prop{13}{15}{3}{colUt}
    \proj{4}{6}{9}{colP}
    \proj{12}{14}{9}{colP}

    \proj{0}{2}{2}{colP}
    \proj{8}{10}{2}{colP}
    \prop{-1}{1}{8}{colUt}
    \prop{7}{9}{8}{colUt}
    \prop{15}{17}{8}{colUt}

    \proj{6}{8}{10}{colP}
    \proj{14}{16}{10}{colP}

    \prop{1}{3}{6}{colUt}
    \prop{9}{11}{6}{colUt}
    \proj{0}{2}{9}{colP}
    \proj{8}{10}{9}{colP}
    \proj{2}{4}{7}{colP}
    \proj{10}{12}{7}{colP}
  \end{tikzpicture}\rightarrow
  \begin{tikzpicture}[baseline={([yshift=-0.6ex]current bounding box.center)}]
    \clip (\dx,{-0.5*\dt}) rectangle ({13*\dx},{-15.5*\dt});
    \grid{15}{6}

    \proj{2}{4}{1}{colP}
    \proj{10}{12}{1}{colP}
    \proj{6}{8}{1}{colP}
    \proj{14}{16}{1}{colP}

    \proj{0}{2}{2}{colP}
    \proj{8}{10}{2}{colP}
    \proj{4}{6}{2}{colP}
    \proj{12}{14}{2}{colP}

    \prop{5}{7}{3}{colUt}
    \prop{13}{15}{3}{colUt}
    \prop{1}{3}{3}{colUt}
    \prop{9}{11}{3}{colUt}

    \prop{-1}{1}{4}{colUt}
    \prop{7}{9}{4}{colUt}
    \prop{15}{17}{4}{colUt}
    \prop{3}{5}{4}{colUt}
    \prop{11}{13}{4}{colUt}

    \proj{2}{4}{5}{colP}
    \proj{10}{12}{5}{colP}
    \proj{6}{8}{5}{colP}
    \proj{14}{16}{5}{colP}

    \proj{0}{2}{6}{colP}
    \proj{8}{10}{6}{colP}
    \proj{4}{6}{6}{colP}
    \proj{12}{14}{6}{colP}
  \end{tikzpicture}.
\end{equation}
In the final step we again rearranged the operators to
obtain~$\tilde{U}^{\text{e}}$ as defined in~\eqref{eq:spacePropDiagram2}.
The same reasoning applies to~$\tilde{U}^{\text{o}}$,
therefore~\eqref{eq:7sitePropGraph} is equivalent
to~\eqref{eq:spacePropDiagram2} and the dual propagation of RCA54 can be
expressed in terms of local, deterministic $7$-site gates on the reduced
configuration space.

\section{Dual circuit representation of correlation functions}\label{sec:rotatingDiagram}
A convenient way to show the equivalence between the diagrams in
Eq.~\eqref{eq:InfTempCorrs} is to interpret the circuits as a $2$-dimensional
vertex model. Each line segment is either in the state~$s=0$ or~$s=1$,
and the weights of vertices with large circles are given by the $3$-site
propagator~$U$ as,
\begin{equation}
  \begin{tikzpicture}[baseline={([yshift=-0.6ex]current bounding box.center)},
    scale=1.75]
    \node[label=180:$s_1$,inner sep=0] (s1)  at ({-\dt},0) {};
    \node[label=270:$s_2$,inner sep=0] (s2)  at (0,{-\dt}) {};
    \node[label=0:$s_3$,inner sep=0] (s3)  at ({\dt},0) {};
    \node[label=90:$s_4$,inner sep=0] (s4)  at (0,{\dt}) {};
    \draw[thick] (s1) -- (s3);
    \draw[thick] (s2) -- (s4);
    \draw[thick,fill=colUt] (0,0) circle (3pt);
  \end{tikzpicture}\equiv
  U_{(s_1,s_4,s_3),(s_1,s_2,s_3)}=\hat{U}_{(s_2,s_1,s_4),(s_2,s_3,s_4)}.
\end{equation}
The small circles force all the incoming lines to be in the same state,
\begin{equation}
  \begin{tikzpicture}[baseline={([yshift=-0.6ex]current bounding box.center)},
    scale=1.75]
    \def\sqrtThree{1.73205}
    \node[label=180:$s_1$,inner sep=0] (s1)  at ({-\dt},0) {};
    \node[label=240:$s_2$,inner sep=0] (s2)  at ({-\dt/2.},{-\dt*\sqrtThree/2.}) {};
    \node[label=300:$s_3$,inner sep=0] (s3)  at ({\dt/2.},{-\dt*\sqrtThree/2.}) {};
    \node[label=90:$s_k$,inner sep=0] (s4)  at (0,{\dt}) {};
    \node[label={[rotate=-75]{\bf \Large{$\cdots$}}},inner sep=0] (mid) at ({\dt/4.},{\dt*(1/2.-\sqrtThree/4)}) {};
    \draw[thick] (s1) -- (0,0);
    \draw[thick] (s2) -- (0,0);
    \draw[thick] (s3) -- (0,0);
    \draw[thick] (s4) -- (0,0);
    \draw[thick,fill=colUt] (0,0) circle (1.25pt);
  \end{tikzpicture}
  \equiv
  \delta_{s_1,s_2}\delta_{s_2,s_3}\cdots \delta_{s_{k-1},s_k},
\end{equation}
where the weight is defined for any number~$k\ge2$ of intersecting lines.
In particular, for~$k=2$ the diagram can be transformed into a straight line,
\begin{equation}
  \begin{tikzpicture}[baseline={([yshift=-0.6ex]current bounding box.center)},
    scale=1.75]
    \draw[thick] ({2*\dt},0) -- (0,0);
    \draw[thick,fill=colUt] ({\dt},0) circle (1.25pt);
  \end{tikzpicture}=
  \begin{tikzpicture}[baseline={([yshift=-0.6ex]current bounding box.center)},
    scale=1.75]
    \draw[thick] ({2*\dt},0) -- (0,0);
  \end{tikzpicture}.
\end{equation}
In this context, the one-site maximum entropy state~$\vec{\omega}$ corresponds
to the sum of the line segment in states~$0$ and~$1$. This 
implies that we can always attach or remove lines connected to the maximum
entropy state from the small circle, as long as at the end at least one such
lines remains,
\begin{equation}
  \begin{tikzpicture}[baseline={([yshift=-0.6ex]current bounding box.center)},
    scale=1.75]
    \def\sqrtTwo{1.41421}
    \def\sqrtThree{1.73205}
    \node[inner sep=0] (s0)  at ({-\dt*\sqrtThree/2.},{\dt/2.}) {};
    \node[inner sep=0] (s0p)  at ({-\dt*\sqrtThree/2.},{-\dt/2.}) {};
    \node[inner sep=0] (s1)  at ({\dt*\sqrtThree/2.},{\dt/2.}) {};
    \node[inner sep=0] (s2)  at ({\dt},0) {};
    \node[inner sep=0] (s3)  at ({\dt*\sqrtThree/2.},{-\dt/2.}) {};
    \draw[thick] (s0) -- (0,0);
    \draw[thick] (s0p) -- (0,0);
    \draw[thick] (s1) -- (0,0);
    \draw[thick] (s2) -- (0,0);
    \draw[thick] (s3) -- (0,0);
    \draw[thick,fill=colUt] (0,0) circle (1.25pt);
    \draw[thick,gray,fill=gray] ({-\dt*\sqrtThree/2.},{\dt/2.}) circle (1.25pt);
    \draw[thick,gray,fill=gray] ({-\dt*\sqrtThree/2.},{-\dt/2.}) circle (1.25pt);
  \end{tikzpicture}=
  \begin{tikzpicture}[baseline={([yshift=-0.6ex]current bounding box.center)},
    scale=1.75]
    \def\sqrtTwo{1.41421}
    \def\sqrtThree{1.73205}
    \node[inner sep=0] (s0)  at ({-\dt},0) {};
    \node[inner sep=0] (s1)  at ({\dt*\sqrtThree/2.},{\dt/2.}) {};
    \node[inner sep=0] (s2)  at ({\dt},0) {};
    \node[inner sep=0] (s3)  at ({\dt*\sqrtThree/2.},{-\dt/2.}) {};
    \draw[thick] (s0) -- (0,0);
    \draw[thick] (s1) -- (0,0);
    \draw[thick] (s2) -- (0,0);
    \draw[thick] (s3) -- (0,0);
    \draw[thick,fill=colUt] (0,0) circle (1.25pt);
    \draw[thick,gray,fill=gray] ({-\dt},0) circle (1.25pt);
  \end{tikzpicture}.
\end{equation}
Using these relations, the equivalence of the diagrams from~\eqref{eq:InfTempCorrs}
can be recast as
\begin{equation}
  \begin{tikzpicture}[baseline={([yshift=-0.6ex]current bounding box.center)},
    scale=1.75]
    \def\sqrtTwo{1.41421}
    \draw[thick] ({-3*\dt},{0*\dt}) -- ({-4*\dt},{0*\dt});
    \draw[thick] ({-2*\dt},{1*\dt}) -- ({-(2+1./\sqrtTwo)*\dt},{(1+1./\sqrtTwo)*\dt});
    \draw[thick] ({-1*\dt},{2*\dt}) -- ({-(1+1./\sqrtTwo)*\dt},{(2+1./\sqrtTwo)*\dt});
    \draw[thick] ({0*\dt},{3*\dt}) -- ({-(1./\sqrtTwo)*\dt},{(3+1./\sqrtTwo)*\dt});
    \draw[thick] ({1*\dt},{4*\dt}) -- ({1*\dt},{5*\dt});
    \draw[thick] ({2*\dt},{3*\dt}) -- ({(2.+1./\sqrtTwo)*\dt},{(3+1./\sqrtTwo)*\dt});
    \draw[thick] ({3*\dt},{2*\dt}) -- ({(3.+1./\sqrtTwo)*\dt},{(2+1./\sqrtTwo)*\dt});
    \draw[thick] ({4*\dt},{1*\dt}) -- ({5*\dt},{1*\dt});
    \draw[thick] ({3*\dt},{0*\dt}) -- ({(3+1./\sqrtTwo)*\dt},{-1./\sqrtTwo*\dt});
    \draw[thick] ({2*\dt},{-1*\dt}) -- ({(2+1./\sqrtTwo)*\dt},{-(1+1./\sqrtTwo)*\dt});
    \draw[thick] ({1*\dt},{-2*\dt}) -- ({(1+1./\sqrtTwo)*\dt},{-(2+1./\sqrtTwo)*\dt});
    \draw[thick] ({0*\dt},{-3*\dt}) -- ({0*\dt},{-4*\dt});
    \draw[thick] ({-1*\dt},{-2*\dt}) -- ({-(1+1./\sqrtTwo)*\dt},{-(2+1./\sqrtTwo)*\dt});
    \draw[thick] ({-2*\dt},{-1*\dt}) -- ({-(2+1./\sqrtTwo)*\dt},{-(1+1./\sqrtTwo)*\dt});
    \draw[thick,gray,fill=gray] ({-4*\dt},{0*\dt}) circle (1.25pt);
    \draw[thick,gray,fill=gray] ({-(2+1./\sqrtTwo)*\dt},{(1+1./\sqrtTwo)*\dt})circle (1.25pt);
    \draw[thick,gray,fill=gray] ({-(1+1./\sqrtTwo)*\dt},{(2+1./\sqrtTwo)*\dt})circle (1.25pt);
    \draw[thick,gray,fill=gray] ({-(1./\sqrtTwo)*\dt},{(3+1./\sqrtTwo)*\dt})circle (1.25pt);
    \draw[thick,gray,fill=gray] ({\dt},{5*\dt})circle (1.25pt);
    \draw[thick,gray,fill=gray] ({(2+1./\sqrtTwo)*\dt},{(3+1./\sqrtTwo)*\dt})circle (1.25pt);
    \draw[thick,gray,fill=gray] ({(3+1./\sqrtTwo)*\dt},{(2+1./\sqrtTwo)*\dt})circle (1.25pt);
    \draw[thick,gray,fill=gray] ({5*\dt},{1*\dt}) circle (1.25pt);
    \draw[thick,gray,fill=gray] ({(3+1./\sqrtTwo)*\dt},{(-1./\sqrtTwo)*\dt}) circle (1.25pt);
    \draw[thick,gray,fill=gray] ({(2+1./\sqrtTwo)*\dt},{-(1+1./\sqrtTwo)*\dt})circle (1.25pt);
    \draw[thick,gray,fill=gray] ({(1+1./\sqrtTwo)*\dt},{-(2+1./\sqrtTwo)*\dt})circle (1.25pt);
    \draw[thick,gray,fill=gray] (0,{-4*\dt}) circle (1.25pt);
    \draw[thick,gray,fill=gray] ({-(1+1./\sqrtTwo)*\dt},{-(2+1./\sqrtTwo)*\dt})circle(1.25pt);
    \draw[thick,gray,fill=gray] ({-(2+1./\sqrtTwo)*\dt},{-(1+1./\sqrtTwo)*\dt})circle(1.25pt);
    \draw[thick] ({0*\dt},{3*\dt}) -- ({2*\dt},{3*\dt});
    \draw[thick] ({-1*\dt},{2*\dt}) -- ({3*\dt},{2*\dt});
    \draw[thick] ({-2*\dt},{1*\dt}) -- ({4*\dt},{1*\dt});
    \draw[thick] ({-3*\dt},{0*\dt}) -- ({3*\dt},{0*\dt});
    \draw[thick] ({-2*\dt},{-1*\dt}) -- ({2*\dt},{-1*\dt});
    \draw[thick] ({-1*\dt},{-2*\dt}) -- ({1*\dt},{-2*\dt});
    \draw[thick] ({-2*\dt},{-1*\dt}) -- ({-2*\dt},{1*\dt});
    \draw[thick] ({-1*\dt},{-2*\dt}) -- ({-1*\dt},{2*\dt});
    \draw[thick] ({0*\dt},{-3*\dt}) -- ({0*\dt},{3*\dt});
    \draw[thick] ({1*\dt},{-2*\dt}) -- ({1*\dt},{4*\dt});
    \draw[thick] ({2*\dt},{-1*\dt}) -- ({2*\dt},{3*\dt});
    \draw[thick] ({3*\dt},{0*\dt}) -- ({3*\dt},{2*\dt});
    \draw[thick,fill=colUt] ({0*\dt},{-2*\dt}) circle (3pt);
    \draw[thick,fill=colUt] ({-1*\dt},{-1*\dt}) circle (3pt);
    \draw[thick,fill=colUt] ({1*\dt},{-1*\dt}) circle (3pt);
    \draw[thick,fill=colUt] ({-2*\dt},0) circle (3pt);
    \draw[thick,fill=colUt] ({0*\dt},0) circle (3pt);
    \draw[thick,fill=colUt] ({2*\dt},0) circle (3pt);
    \draw[thick,fill=colUt] ({-1*\dt},{1*\dt}) circle (3pt);
    \draw[thick,fill=colUt] ({1*\dt},{1*\dt}) circle (3pt);
    \draw[thick,fill=colUt] ({3*\dt},{1*\dt}) circle (3pt);
    \draw[thick,fill=colUt] ({0*\dt},{2*\dt}) circle (3pt);
    \draw[thick,fill=colUt] ({2*\dt},{2*\dt}) circle (3pt);
    \draw[thick,fill=colUt] ({1*\dt},{3*\dt}) circle (3pt);
    \draw[thick,fill=colUt] ({0*\dt},{-3*\dt}) circle (1.25pt);
    \draw[thick,fill=colUt] ({-1*\dt},{-2*\dt}) circle (1.25pt);
    \draw[thick,fill=colUt] ({1*\dt},{-2*\dt}) circle (1.25pt);
    \draw[thick,fill=colUt] ({-2*\dt},{-1*\dt}) circle (1.25pt);
    \draw[thick,fill=colUt] ({0*\dt},{-1*\dt}) circle (1.25pt);
    \draw[thick,fill=colUt] ({2*\dt},{-1*\dt}) circle (1.25pt);
    \draw[thick,fill=colUt] ({-3*\dt},{0*\dt}) circle (1.25pt);
    \draw[thick,fill=colUt] ({-1*\dt},{0*\dt}) circle (1.25pt);
    \draw[thick,fill=colUt] ({1*\dt},{0*\dt}) circle (1.25pt);
    \draw[thick,fill=colUt] ({3*\dt},{0*\dt}) circle (1.25pt);
    \draw[thick,fill=colUt] ({-2*\dt},{1*\dt}) circle (1.25pt);
    \draw[thick,fill=colUt] ({0*\dt},{1*\dt}) circle (1.25pt);
    \draw[thick,fill=colUt] ({2*\dt},{1*\dt}) circle (1.25pt);
    \draw[thick,fill=colUt] ({4*\dt},{1*\dt}) circle (1.25pt);
    \draw[thick,fill=colUt] ({-1*\dt},{2*\dt}) circle (1.25pt);
    \draw[thick,fill=colUt] ({1*\dt},{2*\dt}) circle (1.25pt);
    \draw[thick,fill=colUt] ({3*\dt},{2*\dt}) circle (1.25pt);
    \draw[thick,fill=colUt] ({0*\dt},{3*\dt}) circle (1.25pt);
    \draw[thick,fill=colUt] ({2*\dt},{3*\dt}) circle (1.25pt);
    \draw[thick,fill=colUt] ({1*\dt},{4*\dt}) circle (1.25pt);
    \obs{2.6}{0}{0.5*\dx}{colObs}
    \obs{1.6}{1}{0.5*\dx}{colObs}
    \obs{0.6}{0}{0.5*\dx}{colObs}
    \obs{-0.4}{1}{0.5*\dx}{colObs}
    \obs{-1.4}{0}{0.5*\dx}{colObs}
    \obs{-2.4}{1}{0.5*\dx}{colObs}
    \obs{-(3+0.5/\sqrtTwo)}{(-0.5/\sqrtTwo)}{0.5*\dx}{colObs}
    \obs{-4.5}{1}{0.5*\dx}{colObs}
  \end{tikzpicture}
  =
  \begin{tikzpicture}[baseline={([yshift=-0.6ex]current bounding box.center)},
    scale=1.75]
    \def\sqrtTwo{1.41421}
    \draw[thick] ({-3*\dt},{0*\dt}) -- ({-4*\dt},{0*\dt});
    \draw[thick] ({-2*\dt},{1*\dt}) -- ({-(2+1./\sqrtTwo)*\dt},{(1+1./\sqrtTwo)*\dt});
    \draw[thick] ({-1*\dt},{2*\dt}) -- ({-(1+1./\sqrtTwo)*\dt},{(2+1./\sqrtTwo)*\dt});
    \draw[thick] ({0*\dt},{3*\dt}) -- ({-(1./\sqrtTwo)*\dt},{(3+1./\sqrtTwo)*\dt});
    \draw[thick] ({1*\dt},{4*\dt}) -- ({1*\dt},{5*\dt});
    \draw[thick] ({2*\dt},{3*\dt}) -- ({(2.+1./\sqrtTwo)*\dt},{(3+1./\sqrtTwo)*\dt});
    \draw[thick] ({3*\dt},{2*\dt}) -- ({(3.+1./\sqrtTwo)*\dt},{(2+1./\sqrtTwo)*\dt});
    \draw[thick] ({4*\dt},{1*\dt}) -- ({5*\dt},{1*\dt});
    \draw[thick] ({3*\dt},{0*\dt}) -- ({(3+1./\sqrtTwo)*\dt},{-1./\sqrtTwo*\dt});
    \draw[thick] ({2*\dt},{-1*\dt}) -- ({(2+1./\sqrtTwo)*\dt},{-(1+1./\sqrtTwo)*\dt});
    \draw[thick] ({1*\dt},{-2*\dt}) -- ({(1+1./\sqrtTwo)*\dt},{-(2+1./\sqrtTwo)*\dt});
    \draw[thick] ({0*\dt},{-3*\dt}) -- ({0*\dt},{-4*\dt});
    \draw[thick] ({-1*\dt},{-2*\dt}) -- ({-(1+1./\sqrtTwo)*\dt},{-(2+1./\sqrtTwo)*\dt});
    \draw[thick] ({-2*\dt},{-1*\dt}) -- ({-(2+1./\sqrtTwo)*\dt},{-(1+1./\sqrtTwo)*\dt});
    \draw[thick,gray,fill=gray] ({-4*\dt},{0*\dt}) circle (1.25pt);
    \draw[thick,gray,fill=gray] ({-(2+1./\sqrtTwo)*\dt},{(1+1./\sqrtTwo)*\dt})circle (1.25pt);
    \draw[thick,gray,fill=gray] ({-(1+1./\sqrtTwo)*\dt},{(2+1./\sqrtTwo)*\dt})circle (1.25pt);
    \draw[thick,gray,fill=gray] ({-(1./\sqrtTwo)*\dt},{(3+1./\sqrtTwo)*\dt})circle (1.25pt);
    \draw[thick,gray,fill=gray] ({\dt},{5*\dt})circle (1.25pt);
    \draw[thick,gray,fill=gray] ({(2+1./\sqrtTwo)*\dt},{(3+1./\sqrtTwo)*\dt})circle (1.25pt);
    \draw[thick,gray,fill=gray] ({(3+1./\sqrtTwo)*\dt},{(2+1./\sqrtTwo)*\dt})circle (1.25pt);
    \draw[thick,gray,fill=gray] ({5*\dt},{1*\dt}) circle (1.25pt);
    \draw[thick,gray,fill=gray] ({(3+1./\sqrtTwo)*\dt},{(-1./\sqrtTwo)*\dt}) circle (1.25pt);
    \draw[thick,gray,fill=gray] ({(2+1./\sqrtTwo)*\dt},{-(1+1./\sqrtTwo)*\dt})circle (1.25pt);
    \draw[thick,gray,fill=gray] ({(1+1./\sqrtTwo)*\dt},{-(2+1./\sqrtTwo)*\dt})circle (1.25pt);
    \draw[thick,gray,fill=gray] (0,{-4*\dt}) circle (1.25pt);
    \draw[thick,gray,fill=gray] ({-(1+1./\sqrtTwo)*\dt},{-(2+1./\sqrtTwo)*\dt})circle(1.25pt);
    \draw[thick,gray,fill=gray] ({-(2+1./\sqrtTwo)*\dt},{-(1+1./\sqrtTwo)*\dt})circle(1.25pt);
    \draw[thick] ({0*\dt},{3*\dt}) -- ({2*\dt},{3*\dt});
    \draw[thick] ({-1*\dt},{2*\dt}) -- ({3*\dt},{2*\dt});
    \draw[thick] ({-2*\dt},{1*\dt}) -- ({4*\dt},{1*\dt});
    \draw[thick] ({-3*\dt},{0*\dt}) -- ({3*\dt},{0*\dt});
    \draw[thick] ({-2*\dt},{-1*\dt}) -- ({2*\dt},{-1*\dt});
    \draw[thick] ({-1*\dt},{-2*\dt}) -- ({1*\dt},{-2*\dt});
    \draw[thick] ({-2*\dt},{-1*\dt}) -- ({-2*\dt},{1*\dt});
    \draw[thick] ({-1*\dt},{-2*\dt}) -- ({-1*\dt},{2*\dt});
    \draw[thick] ({0*\dt},{-3*\dt}) -- ({0*\dt},{3*\dt});
    \draw[thick] ({1*\dt},{-2*\dt}) -- ({1*\dt},{4*\dt});
    \draw[thick] ({2*\dt},{-1*\dt}) -- ({2*\dt},{3*\dt});
    \draw[thick] ({3*\dt},{0*\dt}) -- ({3*\dt},{2*\dt});
    \draw[thick,fill=colUt] ({0*\dt},{-2*\dt}) circle (3pt);
    \draw[thick,fill=colUt] ({-1*\dt},{-1*\dt}) circle (3pt);
    \draw[thick,fill=colUt] ({1*\dt},{-1*\dt}) circle (3pt);
    \draw[thick,fill=colUt] ({-2*\dt},0) circle (3pt);
    \draw[thick,fill=colUt] ({0*\dt},0) circle (3pt);
    \draw[thick,fill=colUt] ({2*\dt},0) circle (3pt);
    \draw[thick,fill=colUt] ({-1*\dt},{1*\dt}) circle (3pt);
    \draw[thick,fill=colUt] ({1*\dt},{1*\dt}) circle (3pt);
    \draw[thick,fill=colUt] ({3*\dt},{1*\dt}) circle (3pt);
    \draw[thick,fill=colUt] ({0*\dt},{2*\dt}) circle (3pt);
    \draw[thick,fill=colUt] ({2*\dt},{2*\dt}) circle (3pt);
    \draw[thick,fill=colUt] ({1*\dt},{3*\dt}) circle (3pt);
    \draw[thick,fill=colUt] ({0*\dt},{-3*\dt}) circle (1.25pt);
    \draw[thick,fill=colUt] ({-1*\dt},{-2*\dt}) circle (1.25pt);
    \draw[thick,fill=colUt] ({1*\dt},{-2*\dt}) circle (1.25pt);
    \draw[thick,fill=colUt] ({-2*\dt},{-1*\dt}) circle (1.25pt);
    \draw[thick,fill=colUt] ({0*\dt},{-1*\dt}) circle (1.25pt);
    \draw[thick,fill=colUt] ({2*\dt},{-1*\dt}) circle (1.25pt);
    \draw[thick,fill=colUt] ({-3*\dt},{0*\dt}) circle (1.25pt);
    \draw[thick,fill=colUt] ({-1*\dt},{0*\dt}) circle (1.25pt);
    \draw[thick,fill=colUt] ({1*\dt},{0*\dt}) circle (1.25pt);
    \draw[thick,fill=colUt] ({3*\dt},{0*\dt}) circle (1.25pt);
    \draw[thick,fill=colUt] ({-2*\dt},{1*\dt}) circle (1.25pt);
    \draw[thick,fill=colUt] ({0*\dt},{1*\dt}) circle (1.25pt);
    \draw[thick,fill=colUt] ({2*\dt},{1*\dt}) circle (1.25pt);
    \draw[thick,fill=colUt] ({4*\dt},{1*\dt}) circle (1.25pt);
    \draw[thick,fill=colUt] ({-1*\dt},{2*\dt}) circle (1.25pt);
    \draw[thick,fill=colUt] ({1*\dt},{2*\dt}) circle (1.25pt);
    \draw[thick,fill=colUt] ({3*\dt},{2*\dt}) circle (1.25pt);
    \draw[thick,fill=colUt] ({0*\dt},{3*\dt}) circle (1.25pt);
    \draw[thick,fill=colUt] ({2*\dt},{3*\dt}) circle (1.25pt);
    \draw[thick,fill=colUt] ({1*\dt},{4*\dt}) circle (1.25pt);
    \obs{3.5}{0}{0.5*\dx}{colObs}
    \obs{2}{0.6}{0.5*\dx}{colObs}
    \obs{1}{0.4}{0.5*\dx}{colObs}
    \obs{0}{0.6}{0.5*\dx}{colObs}
    \obs{-1}{0.4}{0.5*\dx}{colObs}
    \obs{-2}{0.6}{0.5*\dx}{colObs}
    \obs{-3}{0.4}{0.5*\dx}{colObs}
    \obs{-4.5}{1}{0.5*\dx}{colObs}
  \end{tikzpicture}.
\end{equation}
This equality follows from the fact that the observables can
be represented by diagonal one-site operators and can be therefore freely
moved around the small circle,
\begin{equation}
  \begin{tikzpicture}[baseline={([yshift=-0.6ex]current bounding box.center)},
    scale=1.75]
    \draw[thick] (-\dt,0) -- (\dt,0);
    \draw[thick] (0,-\dt) -- (0,\dt);
    \draw[thick,fill=colUt] (0,0) circle (1.25pt);
    \obs{-0.6}{0}{0.75*\dx}{colObs}
  \end{tikzpicture}=
  \begin{tikzpicture}[baseline={([yshift=-0.6ex]current bounding box.center)},
    scale=1.75]
    \draw[thick] (-\dt,0) -- (\dt,0);
    \draw[thick] (0,-\dt) -- (0,\dt);
    \draw[thick,fill=colUt] (0,0) circle (1.25pt);
    \obs{0}{0.6}{0.75*\dx}{colObs}
  \end{tikzpicture}=
  \begin{tikzpicture}[baseline={([yshift=-0.6ex]current bounding box.center)},
    scale=1.75]
    \draw[thick] (-\dt,0) -- (\dt,0);
    \draw[thick] (0,-\dt) -- (0,\dt);
    \draw[thick,fill=colUt] (0,0) circle (1.25pt);
    \obs{0.6}{0}{0.75*\dx}{colObs}
  \end{tikzpicture}=
  \begin{tikzpicture}[baseline={([yshift=-0.6ex]current bounding box.center)},
    scale=1.75]
    \draw[thick] (-\dt,0) -- (\dt,0);
    \draw[thick] (0,-\dt) -- (0,\dt);
    \draw[thick,fill=colUt] (0,0) circle (1.25pt);
    \obs{0}{-0.6}{0.75*\dx}{colObs}
  \end{tikzpicture}.
\end{equation}

\section{Factorization of the equilibrium state and the local few-site relations}
\label{sec:factorSol}
We start by explicitly expressing coefficients~$\alpha^{(\prime)}_{s_1 s_2 s_3}$,
$\beta^{(\prime)}_{s_1 s_2 s_3}$ that satisfy factorization condition in
equation~\eqref{eq:factCond}. Solving the first two factorization relations we
obtain the following solution,
\begin{equation}
  \begin{aligned}
    \alpha_{000}&=\alpha_{100}=\alpha_{111}=1,& \qquad
    \alpha^{\prime}_{000}&=\alpha^{\prime}_{001}=\alpha^{\prime}_{111}=1,\\
    \alpha_{001}&=\alpha_{101}=\alpha_{110}=\frac{\xi(\lambda+\omega-\xi\omega)}{\lambda+\xi-\xi\omega},&
    \alpha^{\prime}_{011}&=\alpha^{\prime}_{100}=\alpha^{\prime}_{101}=
    \frac{\omega(\lambda+\xi-\xi\omega)}{\lambda+\omega-\xi\omega},\\
    \alpha_{010}&=\frac{\lambda+\xi-\xi\omega}{\lambda+\omega-\xi\omega}, &
    \alpha^{\prime}_{010}&=\frac{\lambda+\omega-\xi\omega}{\lambda+\xi-\xi\omega},\\
    \alpha_{011}&=\frac{\omega(\lambda+\xi-\xi\omega)^2}{(\lambda+\omega-\xi\omega)^2},&
    \alpha^{\prime}_{110}&=\frac{\xi(\lambda+\omega-\xi\omega)^2}{(\lambda+\xi-\xi\omega)^2},
  \end{aligned}
\end{equation}
where we can immediately see that one set of parameters is transformed into
another one by exchanging $\xi$ and $\omega$, and reversing the order of indices,
\begin{equation}
  \alpha^{\prime}_{s_1 s_2 s_3}=\left. \alpha_{s_3 s_2 s_1}\right|_{\xi\leftrightarrow\omega}.
  \end{equation}
Similarly, solving the bottom two equations we obtain 
\begin{equation}
  \begin{aligned}
    \beta_{s_1 s_2 s_3}&=\alpha_{s_1 s_2 s_3},&\qquad
    \beta^{\prime}_{000}&=\beta^{\prime}_{001}=\beta^{\prime}_{110}=\alpha^{\prime}_{000}, &\qquad
    \beta^{\prime}_{011}&=\alpha^{\prime}_{010},\\
    & &
    \beta^{\prime}_{010}&=\beta^{\prime}_{100}=\beta^{\prime}_{101}=\alpha^{\prime}_{100}, &
    \beta^{\prime}_{111}&=\alpha^{\prime}_{110}.
  \end{aligned}
\end{equation}

To demonstrate how the factorization property of the equilibrium state enables us to formulate
the few-site relations~\eqref{eq:genLeftMPS} and~\eqref{eq:genRightMPS}, we first introduce
the following notation for the basis vectors from~$\mathbb{R}^{2^5}$,
\begin{equation}
  \vec{e}_{s_1 s_2 s_3 s_4 s_5} =
  \vec{e}_{s_1} \otimes \vec{e}_{s_2} \otimes \vec{e}_{s_3} \otimes \vec{e}_{s_4} \otimes \vec{e}_{s_5},
  \qquad \vec{e}_0=\begin{bmatrix}1\\0\end{bmatrix},\qquad
  \vec{e}_1=\begin{bmatrix}0\\1\end{bmatrix}.
\end{equation}
Now we can express the first identity from Eq.~\eqref{eq:genLeftMPS} in explicit component form as,
\begin{equation}
  \begin{aligned}
    &\smashoperator{\sum_{s_1,s_2,s_3,s_4,s_5}}
    \vec{e}_{s_1 s_2 s_3 s_4 s_5} \hat{U}_3 P_2 P_4\cdot
    \bra{l} W^{\prime}_{s_3}S W^{\prime}_{s_2} S W^{\prime}_{s_1}\\
    =&\smashoperator{\sum_{s_1,s_2,s_3,s_4,s_5}}
    \vec{e}_{s_1 s_2 s_3 s_4 s_5} \hat{U}_3 P_2 P_4\cdot
    \alpha^{\prime}_{s_3 s_2 s_1} \bra{l} W^{\prime}_{s_2} S W^{\prime}_{s_1}\\
    =&\smashoperator{\sum_{s_1,s_2,s_3,s_4,s_5}}
    \vec{e}_{s_1 s_2 s_3 s_4 s_5} P_2 P_4\cdot
    \bbra{l}A_{s_2}B^{\prime}_{s_3}A_{s_4}\kket{r} \bra{l} W^{\prime}_{s_2} S W^{\prime}_{s_1},
  \end{aligned}
\end{equation}
where to get from the first to the second line, we used the first of the factorization
conditions~\eqref{eq:factCond}. Note that~$\vec{B}$ and~$\vec{B}^{\prime}$ satisfy
an even stronger condition, where we can remove the sum over~$s_1$ and $s_2$. Namely,
\begin{equation}
  \smashoperator{\sum_{s_3,s_4,s_5}}
  \vec{e}_{s_1 s_2 s_3 s_4 s_5} \hat{U}_3 P_2 P_4\cdot
  \alpha^{\prime}_{s_3 s_2 s_1}
  =\smashoperator{\sum_{s_3,s_4,s_5}}
  \vec{e}_{s_1 s_2 s_3 s_4 s_5} P_2 P_4\cdot
  \bbra{l}A_{s_2}B^{\prime}_{s_3}A_{s_4}\kket{r}.
\end{equation}

\section{Matrix-product form of multi-time correlation functions}\label{sec:tsMPSequiv}
To see that the MPS representation~\eqref{eq:MPStimestate} is equivalent
to time-states introduced in~\cite{vMPA2019} we first explicitly
spell out the matrices~$\tilde{A}_s$, $\tilde{A}^{\prime}_s$ which
by definition~\eqref{eq:defTildeA} take the following form,
\begin{equation}
  \begin{aligned}
    \tilde{A}_0&=\begin{bmatrix}
      \alpha_{000}&\alpha_{001}& 0 & 0\\
      \alpha_{000}&\alpha_{001}& 0 & 0\\
      0 & 0 & 0 & 0 \\
      0 & 0 & 0 & 0
    \end{bmatrix},&\qquad
    \tilde{A}_1&=\begin{bmatrix}
      0 & 0 & 0 & 0\\
      0 & 0 & 0 & 0\\
      0 & 0 & 0 & \alpha_{000}+\alpha_{001} \\
      0 & 0 & \alpha_{000}+\alpha_{001} & 0
    \end{bmatrix},\\
    \tilde{A}^{\prime}_0&=\begin{bmatrix}
      \alpha^{\prime}_{000} & 0 & \alpha^{\prime}_{100} &0 \\
      0 & 0 & 0 & 0 \\
      \alpha^{\prime}_{000} & 0 & \alpha^{\prime}_{100} &0 \\
      0 & 0 & 0 & 0
    \end{bmatrix},&
    \tilde{A}^{\prime}_1&=\begin{bmatrix}
      0 & 0 & 0 & 0\\
      0 & 0 & 0 & \alpha^{\prime}_{000}+\alpha^{\prime}_{100}\\
      0 & 0 & 0 & 0\\
      0 & 0 & \alpha^{\prime}_{000}+\alpha^{\prime}_{100} & 0
    \end{bmatrix},
  \end{aligned}
\end{equation}
while the boundary vectors~$\nbbra{\tilde{L}}$ and~$\nkket{\tilde{R}}$
that solve equation~\eqref{eq:defTildeBoundary} can be after some straightforward
algebraic manipulation expressed as
\begin{equation}
  \begin{aligned}
    \nbbra{L}&=\frac{\alpha^{\prime}_{000}+\alpha^{\prime}_{100}}
    {1+\frac{\alpha_{001}}{\alpha_{000}+\alpha_{001}}
    +\frac{\alpha^{\prime}_{100}}{\alpha^{\prime}_{000}\alpha^{\prime}_{100}}}
    \begin{bmatrix}
      \frac{\alpha^{\prime}_{000}}{\alpha^{\prime}_{000}+\alpha^{\prime}_{100}} &
      \frac{\alpha^{\prime}_{100}}{\alpha^{\prime}_{000}+\alpha^{\prime}_{100}} &
      \frac{\alpha^{\prime}_{100}}{\alpha^{\prime}_{000}+\alpha^{\prime}_{100}} &
      \frac{\alpha_{001}}{\alpha_{000}+\alpha_{001}}
    \end{bmatrix},\\
    \nkket{R}&=\frac{1}{\alpha^{\prime}_{000}+\alpha^{\prime}_{100}} \begin{bmatrix}1&1&1&1\end{bmatrix}^T.
  \end{aligned}
\end{equation}
Additionally, we note that the product~$(\alpha_{000}+\alpha_{001})(\alpha^{\prime}_{000}+\alpha^{\prime}_{100})$
is equal to the leading eigenvalue $\lambda$ of~$(W_0^{\prime}+W_1^{\prime})(W_0+W_1)$,
\begin{equation}
  \lambda=(\alpha_{000}+\alpha_{001})(\alpha^{\prime}_{000}+\alpha^{\prime}_{100}).
\end{equation}
Equipped by these relations, it is easy to see that it is possible to introduce
linear maps~$Q$, $U$, $V$,
\begin{equation}
  U=\begin{bmatrix}
    1&0&0&-\frac{\alpha_{001}}{\alpha_{000}}\\
    1&0&0&1\\
    0&0&1&0\\
    0&1&0&0
  \end{bmatrix},\qquad
  V=\begin{bmatrix}
    1&0&0&-\frac{\alpha^{\prime}_{100}}{\alpha_{000}}\\
    0&0&1&0\\
    1&0&0&1\\
    0&1&0&0
  \end{bmatrix},\qquad
  Q=\begin{bmatrix}
    1&0&0&0\\
    0&1&0&0\\
    0&0&1&0
  \end{bmatrix},
\end{equation}
so that the following holds for any $s_1,s_2\in\{0,1\}$,
\begin{equation}
  \begin{aligned}
    \tilde{A}_{s_1} U Q^{T} Q U^{-1} \tilde{A}^{\prime}_{s_2} &= \tilde{A}_{s_1}\tilde{A}^{\prime}_{s_2},&\qquad
    \tilde{A}^{\prime}_{s_1} V Q^{T} Q V^{-1} \tilde{A}_{s_2} &= \tilde{A}^{\prime}_{s_1}\tilde{A}_{s_2},\\
    \nbbra{\tilde{L}} V Q^T Q V^{-1} \tilde{A}_{s_1} &= \nbbra{\tilde{L}}\tilde{A}_{s_1},&
    \tilde{A}^{\prime}_{s_1} V Q^TQ V^{-1}\nkket{\tilde{R}} &= \tilde{A}^{\prime}_{s_1}\nkket{\tilde{R}}.
  \end{aligned}
\end{equation}
This implies that the state~\eqref{eq:MPStimestate} can be equivalently represented by
an MPS with a~$3$-dimensional auxiliary space
\begin{equation}
  \vec{q}=\bra{x_L} \vec{X}_1\vec{X}^{\prime}_2\vec{X}_3\cdots\vec{X}^{\prime}_{2m} \ket{x_R},
\end{equation}
where the new matrices~$X_s$, $X^{\prime}_s$ and boundary vectors~$\bra{x_L}$, $\ket{x_R}$
are defined as
\begin{equation}
  \begin{aligned}
    X_s&=\frac{1}{\alpha_{000}+\alpha_{001}} Q V^{-1} \tilde{A}_s U Q^{T}, &\qquad
    X^{\prime}_s&=\frac{1}{\alpha^{\prime}_{000}+\alpha^{\prime}_{001}} Q V^{-1} \tilde{A}^{\prime}_s U Q^{T},\\
    \bra{x_L}&=\frac{1}{\alpha^{\prime}_{000}+\alpha^{\prime}_{100}}
    \nbbra{\tilde{L}} V Q^T,&
    \ket{x_R}&=(\alpha^{\prime}_{000}+\alpha^{\prime}_{100})  Q V^{-1}\nkket{\tilde{R}},
  \end{aligned}
\end{equation}
which implies the following explicit form,
\begin{equation}
  \begin{gathered}
    \begin{aligned}
      X_0&=\begin{bmatrix}
        \frac{\alpha^{\prime}_{000}}{\alpha^{\prime}_{000}+\alpha^{\prime}_{100}}&0&0\\
        0&0&0\\
        1&0&0
      \end{bmatrix},&\qquad
      X_1&=\begin{bmatrix}
        0&\frac{\alpha^{\prime}_{100}}{\alpha^{\prime}_{000}+\alpha^{\prime}_{100}}&0\\
        0&0&1\\
        0&0&0
      \end{bmatrix},\\
      X^{\prime}_0&=\begin{bmatrix}
        \frac{\alpha_{000}}{\alpha_{000}+\alpha_{001}}&0&0\\
        0&0&0\\
        1&0&0
      \end{bmatrix},&
      X^{\prime}_1&=\begin{bmatrix}
        0&\frac{\alpha_{100}}{\alpha_{000}+\alpha_{001}}&0\\
        0&0&1\\
        0&0&0
      \end{bmatrix},\\
    \end{aligned}\\
    \begin{aligned}
      \bra{x_L}&=\frac{1}{1+\frac{\alpha_{001}}{\alpha_{000}+\alpha_{001}}
      +\frac{\alpha^{\prime}_{100}}{\alpha^{\prime}_{000}+\alpha^{\prime}_{100}}}
      \begin{bmatrix}
        1&\frac{\alpha_{001}}{\alpha_{000}+\alpha_{001}} & 
        \frac{\alpha^{\prime}_{100}}{\alpha^{\prime}_{000}+\alpha^{\prime}_{100}}
      \end{bmatrix},\\
      \ket{x_R}&=\begin{bmatrix}1&1&1\end{bmatrix}^T.
    \end{aligned}
  \end{gathered}
\end{equation}
Finally, to see that this parametrization coincided with the MPS from
Ref.~\cite{vMPA2019}, one needs to only express parameters
$\frac{\alpha_{001}}{\alpha_{000}+\alpha_{001}}$ and
$\frac{\alpha^{\prime}_{100}}{\alpha^{\prime}_{000}+\alpha^{\prime}_{100}}$ in
terms of~$\xi$, $\omega$ and $\lambda$.

\end{appendix}

\bibliography{bobenko.bib}

\begin{thebibliography}{10}
\providecommand{\url}[1]{\texttt{#1}}
\providecommand{\urlprefix}{URL }
\expandafter\ifx\csname urlstyle\endcsname\relax
  \providecommand{\doi}[1]{doi:\discretionary{}{}{}#1}\else
  \providecommand{\doi}{doi:\discretionary{}{}{}\begingroup
  \urlstyle{rm}\Url}\fi
\providecommand{\eprint}[2][]{\url{#2}}

\bibitem{baxter2016exactly}
R.~J. Baxter,
\newblock \emph{Exactly solved models in statistical mechanics},
\newblock Elsevier (2016).

\bibitem{sutherland2004beautiful}
B.~Sutherland,
\newblock \emph{Beautiful models},
\newblock World Scientific Publishing Company (2004).

\bibitem{JSTAT2016}
P.~Calabrese, F.~H.~L. Essler and G.~Mussardo,
\newblock \emph{Introduction to `quantum integrability in out of equilibrium
  systems'},
\newblock J. Stat. Mech. \textbf{2016}(6), 064001 (2016),
\newblock \doi{10.1088/1742-5468/2016/06/064001}.

\bibitem{Calabrese}
P.~Calabrese and J.~Cardy,
\newblock \emph{Quantum quenches in 1+1 dimensional conformal field theories},
\newblock J. Stat. Mech. \textbf{2016}(6), 064003 (2016),
\newblock \doi{10.1088/1742-5468/2016/06/064003}.

\bibitem{Doyon}
O.~A. Castro-Alvaredo, B.~Doyon and T.~Yoshimura,
\newblock \emph{Emergent hydrodynamics in integrable quantum systems out of
  equilibrium},
\newblock Phys. Rev. X \textbf{6}, 041065 (2016),
\newblock \doi{10.1103/PhysRevX.6.041065}.

\bibitem{Bertini}
B.~Bertini, M.~Collura, J.~De~Nardis and M.~Fagotti,
\newblock \emph{Transport in out-of-equilibrium $xxz$ chains: Exact profiles of
  charges and currents},
\newblock Phys. Rev. Lett. \textbf{117}, 207201 (2016),
\newblock \doi{10.1103/PhysRevLett.117.207201}.

\bibitem{Nahum1}
A.~Nahum, J.~Ruhman, S.~Vijay and J.~Haah,
\newblock \emph{Quantum entanglement growth under random unitary dynamics},
\newblock Phys. Rev. X \textbf{7}, 031016 (2017),
\newblock \doi{10.1103/PhysRevX.7.031016}.

\bibitem{Nahum2}
A.~Nahum, S.~Vijay and J.~Haah,
\newblock \emph{Operator spreading in random unitary circuits},
\newblock Phys. Rev. X \textbf{8}, 021014 (2018),
\newblock \doi{10.1103/PhysRevX.8.021014}.

\bibitem{Pollmann}
C.~W. von Keyserlingk, T.~Rakovszky, F.~Pollmann and S.~L. Sondhi,
\newblock \emph{Operator hydrodynamics, otocs, and entanglement growth in
  systems without conservation laws},
\newblock Phys. Rev. X \textbf{8}, 021013 (2018),
\newblock \doi{10.1103/PhysRevX.8.021013}.

\bibitem{Chalker}
A.~Chan, A.~De~Luca and J.~T. Chalker,
\newblock \emph{Solution of a minimal model for many-body quantum chaos},
\newblock Phys. Rev. X \textbf{8}, 041019 (2018),
\newblock \doi{10.1103/PhysRevX.8.041019}.

\bibitem{bertini2019exact}
B.~Bertini, P.~Kos and T.~Prosen,
\newblock \emph{Exact correlation functions for dual-unitary lattice models in
  1+ 1 dimensions},
\newblock Phys. Rev. Lett. \textbf{123}, 210601 (2019),
\newblock \doi{10.1103/PhysRevLett.123.210601}.

\bibitem{avan2016lagrangian}
J.~Avan, V.~Caudrelier, A.~Doikou and A.~Kundu,
\newblock \emph{Lagrangian and hamiltonian structures in an integrable
  hierarchy and space--time duality},
\newblock Nucl. Phys. B \textbf{902}, 415 (2016),
\newblock \doi{10.1016/j.nuclphysb.2015.11.024}.

\bibitem{findlay2019dual}
I.~Findlay,
\newblock \emph{A dual construction of the isotropic landau--lifshitz model},
\newblock Physica D \textbf{398}, 13 (2019),
\newblock \doi{10.1016/j.physd.2019.06.003}.

\bibitem{doikou2019time}
A.~Doikou, I.~Findlay and S.~Sklaveniti,
\newblock \emph{Time-like boundary conditions in the nls model},
\newblock Nucl. Phys. B \textbf{941}, 361 (2019),
\newblock \doi{10.1016/j.nuclphysb.2019.02.022}.

\bibitem{gutkin2020local}
B.~Gutkin, P.~Braun, M.~Akila, D.~Waltner and T.~Guhr,
\newblock \emph{Local correlations in dual-unitary kicked chains},
\newblock arXiv:2001.01298  (2020).

\bibitem{bertini2019entanglement}
B.~Bertini, P.~Kos and T.~Prosen,
\newblock \emph{Entanglement spreading in a minimal model of maximal many-body
  quantum chaos},
\newblock Phys. Rev. X \textbf{9}, 021033 (2019),
\newblock \doi{10.1103/PhysRevX.9.021033}.

\bibitem{piroli2019exact}
L.~Piroli, B.~Bertini, J.~I. Cirac and T.~Prosen,
\newblock \emph{Exact dynamics in dual-unitary quantum circuits},
\newblock Phys. Rev. B \textbf{101}, 094304 (2020),
\newblock \doi{10.1103/PhysRevB.101.094304}.

\bibitem{gopalakrishnan4}
S.~Gopalakrishnan and A.~Lamacraft,
\newblock \emph{Unitary circuits of finite depth and infinite width from
  quantum channels},
\newblock Phys. Rev. B \textbf{100}(6) (2019),
\newblock \doi{10.1103/physrevb.100.064309}.

\bibitem{bertini2019operator1}
B.~Bertini, P.~Kos and T.~Prosen,
\newblock \emph{{Operator Entanglement in Local Quantum Circuits I: Chaotic
  Dual-Unitary Circuits}},
\newblock SciPost Phys. \textbf{8}, 67 (2020),
\newblock \doi{10.21468/SciPostPhys.8.4.067}.

\bibitem{bertini2019operator2}
B.~Bertini, P.~Kos and T.~Prosen,
\newblock \emph{{Operator Entanglement in Local Quantum Circuits II: Solitons
  in Chains of Qubits}},
\newblock SciPost Phys. \textbf{8}, 68 (2020),
\newblock \doi{10.21468/SciPostPhys.8.4.068}.

\bibitem{claeys2020maximum}
P.~W. Claeys and A.~Lamacraft,
\newblock \emph{Maximum velocity quantum circuits},
\newblock arXiv:2003.01133  (2020).

\bibitem{Ziga1}
\v{Z}. Krajnik and T.~Prosen,
\newblock \emph{Kardar-parisi-zhang physics in integrable rotationally
  symmetric dynamics on discrete space-time lattice},
\newblock J Stat Phys \textbf{179}, 110 (2020),
\newblock \doi{10.1007/s10955-020-02523-1}.

\bibitem{Ziga2}
\v{Z}. Krajnik, E.~Ilievski and T.~Prosen,
\newblock \emph{Integrable matrix models in discrete space-time},
\newblock arXiv:2003.05957  (2020).

\bibitem{bobenko1993two}
A.~Bobenko, M.~Bordemann, C.~Gunn and U.~Pinkall,
\newblock \emph{{On two integrable cellular automata}},
\newblock Commun. Math. Phys. \textbf{158}(1), 127 (1993),
\newblock \doi{10.1007/BF02097234}.

\bibitem{wolfram2002new}
S.~Wolfram,
\newblock \emph{A new kind of science}, vol.~5,
\newblock Wolfram media Champaign, IL (2002).

\bibitem{Takesue}
S.~Takesue,
\newblock \emph{Reversible cellular automata and statistical mechanics},
\newblock Phys. Rev. Lett. \textbf{59}, 2499 (1987),
\newblock \doi{10.1103/PhysRevLett.59.2499}.

\bibitem{prosenMejiaMonasterioCA54}
T.~Prosen and C.~Mej{\'{\i}}a-Monasterio,
\newblock \emph{Integrability of a deterministic cellular automaton driven by
  stochastic boundaries},
\newblock J. Phys. A: Math. Theor. \textbf{49}(18), 185003 (2016),
\newblock \doi{10.1088/1751-8113/49/18/185003}.

\bibitem{inoueTakesueCA54}
A.~Inoue and S.~Takesue,
\newblock \emph{Two extensions of exact nonequilibrium steady states of a
  boundary-driven cellular automaton},
\newblock J. Phys. A: Math. Theor. \textbf{51}(42), 425001 (2018),
\newblock \doi{10.1088/1751-8121/aadc29}.

\bibitem{prosenBucaCA54}
T.~Prosen and B.~Bu{\v{c}}a,
\newblock \emph{Exact matrix product decay modes of a boundary driven cellular
  automaton},
\newblock J. Phys. A: Math. Theor. \textbf{50}(39), 395002 (2017),
\newblock \doi{10.1088/1751-8121/aa85a3}.

\bibitem{bucaetalLargeDev}
B.~{Bu{\v{c}}a}, J.~P. {Garrahan}, T.~{Prosen} and M.~{Vanicat},
\newblock \emph{Exact large deviation statistics and trajectory phase
  transition of a deterministic boundary driven cellular automaton},
\newblock Phys. Rev. E \textbf{100}, 020103 (2019),
\newblock \doi{10.1103/PhysRevE.100.020103}.

\bibitem{TMPA2018}
K.~Klobas, M.~Medenjak, T.~Prosen and M.~Vanicat,
\newblock \emph{Time-dependent matrix product ansatz for interacting reversible
  dynamics},
\newblock Commun. Math. Phys. \textbf{371}(2), 651 (2019),
\newblock \doi{10.1007/s00220-019-03494-5}.

\bibitem{vMPA2019}
K.~Klobas, M.~Vanicat, J.~P. Garrahan and T.~Prosen,
\newblock \emph{Matrix product state of multi-time correlations},
\newblock arXiv:1912.09742  (2019).

\bibitem{gopalakrishnan1}
S.~Gopalakrishnan,
\newblock \emph{Operator growth and eigenstate entanglement in an interacting
  integrable floquet system},
\newblock Phys. Rev. B \textbf{98}, 060302 (2018),
\newblock \doi{10.1103/PhysRevB.98.060302}.

\bibitem{gopalakrishnan2}
S.~Gopalakrishnan, D.~A. Huse, V.~Khemani and R.~Vasseur,
\newblock \emph{Hydrodynamics of operator spreading and quasiparticle diffusion
  in interacting integrable systems},
\newblock Phys. Rev. B \textbf{98}, 220303 (2018),
\newblock \doi{10.1103/PhysRevB.98.220303}.

\bibitem{gopalakrishnan3}
A.~J. Friedman, S.~Gopalakrishnan and R.~Vasseur,
\newblock \emph{Integrable many-body quantum floquet-thouless pumps},
\newblock Phys. Rev. Lett. \textbf{123}, 170603 (2019),
\newblock \doi{10.1103/PhysRevLett.123.170603}.

\bibitem{alba2019RCA54}
V.~Alba, J.~Dubail and M.~Medenjak,
\newblock \emph{Operator entanglement in interacting integrable quantum
  systems: the case of the rule 54 chain},
\newblock Phys. Rev. Lett. \textbf{122}, 250603 (2019),
\newblock \doi{10.1103/PhysRevLett.122.250603}.

\end{thebibliography}

\nolinenumbers

\end{document}